\newif\ifwordcount
\wordcountfalse

\documentclass{aastex631}
\usepackage{multirow,bigdelim}
\usepackage{cprotect}
\usepackage{xcolor}
\usepackage[thinc]{esdiff}
\usepackage{amsmath}
\usepackage{tablefootnote}
\usepackage{booktabs}
\usepackage{hhline}
\usepackage{color}

\usepackage{cprotect}
\usepackage{graphicx}% Include figure files
\usepackage{dcolumn}% Align table columns on decimal point
\usepackage{bm}% bold math
\usepackage{mathtools}
\usepackage{fancyvrb}
\usepackage{hyperref}
\usepackage{graphicx}
\usepackage{caption}
\usepackage{subcaption}
\usepackage{ragged2e}

\DeclareCaptionJustification{justified}{\justifying}
\newcommand{\beq} {\begin{equation}}
\newcommand{\eeq} {\end{equation}}
\newcommand{\bal} {\begin{aligned}}
\newcommand{\eal} {\end{aligned}}

\begin{document}

\renewcommand{\arraystretch}{1.1}

% Only include extra packages if you really need them. Common packages are:
% \usepackage{graphicx}	% Including figure files
% \usepackage{amsmath}	% Advanced maths commands
% \usepackage{amssymb}	% Extra maths symbols
% \setcounter{tocdepth}{5}
%%%%% AUTHORS - PLACE YOUR OWN COMMANDS HERE %%%%%

% Please keep new commands to a minimum, and use \newcommand not \def to avoid
% overwriting existing commands. Example:
%\newcommand{\pcm}{\,cm$^{-2}$}	% per cm-squared

%%%%%%%%%%%%%%%%%%% TITLE PAGE %%%%%%%%%%%%%%%%%%%
% Title of the paper, and the short title which is used in the headers.
% Keep the title short and informative.
\title{
CLASS\_SZ II: Notes and Examples of Fast and Accurate Calculations of Halo Model, Large Scale Structure and Cosmic Microwave Background Observables
}
%Cosmological observable emulators for Stage-IV surveys I: Cosmic Microwave Background \jch{This title is not accurate.  The paper contains high-precision emulators for $P(k)$ and other quantities as well.  Revision needed.}

% The list of authors, and the short list which is used in the headers.
% If you need two or more lines of authors, add an extra line using \newauthor
%\onecolumn
\author{Boris Bolliet}
\affiliation{Kavli Institute for Cosmology, University of Cambridge, Cambridge, United Kingdom}
\author{Aleksandra Kusiak} 
\affiliation{DAMTP, Centre for Mathematical Sciences, Wilberforce Road, Cambridge CB3 0WA, UK}
\affiliation{Kavli Institute for Cosmology, University of Cambridge, Cambridge, United Kingdom}
\author{Fiona McCarthy}
\affiliation{DAMTP, Centre for Mathematical Sciences, Wilberforce Road, Cambridge CB3 0WA, UK}
\affiliation{Center for Computational Astrophysics, Flatiron Institute, New York, NY, USA 10010}
\affiliation{Kavli Institute for Cosmology, University of Cambridge, Cambridge, United Kingdom}
\author{Alina Sabyr}
\affiliation{Department of Physics, Columbia University, New York, NY, USA 10027}
\author{Kristen Surrao}
\affiliation{Department of Physics, Columbia University, New York, NY, USA 10027}
\author{Jens Chluba}
\affiliation{Jodrell Bank Centre for Astrophysics, Alan Turing Building, University of Manchester, Manchester M13 9PL}
\author{Carmen Embil Villagra}\affiliation{DAMTP, Centre for Mathematical Sciences, University of Cambridge, Wilberforce Road, Cambridge CB3 OWA, UK}
\affiliation{Kavli Institute for Cosmology Cambridge, Madingley Road, Cambridge CB3 0HA, UK}
\author{Simone Ferraro}
\affiliation{Lawrence Berkeley National Laboratory, One Cyclotron Road, Berkeley, CA, USA 94720}
\affiliation{Berkeley Center for Cosmological Physics, Department of Physics, University of California, Berkeley, CA, USA 94720}
\author{Boryana Hadzhiyska} 
\affiliation{Berkeley Center for Cosmological Physics, Department of Physics, University of California, Berkeley, CA, USA 94720}
\author{Dongwon Han} 
\affiliation{DAMTP, Centre for Mathematical Sciences, Wilberforce Road, Cambridge CB3 0WA, UK}
\author{J.~Colin Hill}
\affiliation{Department of Physics, Columbia University, New York, NY, USA 10027}
\affiliation{Center for Computational Astrophysics, Flatiron Institute, New York, NY, USA 10010}
\author{Juan Francisco Mac\'{\i}as-P\'erez}
\affiliation{Laboratoire de Physique Subatomique et de Cosmologie, Universit\'e
Grenoble-Alpes,\\ CNRS/IN2P3, 53, 
avenue des Martyrs, 38026 Grenoble
cedex, France}
\affiliation{Perimeter Institute for Theoretical Physics, Waterloo, Ontario, N2L 2Y5, Canada}
\author{Abhishek Maniyar}
\affiliation{SLAC National Accelerator Laboratory 2575 Sand Hill Road Menlo Park, California 94025, USA}
\affiliation{Kavli Institute for Particle Astrophysics and Cosmology, 382 Via Pueblo Mall Stanford, CA  94305-4060, USA}
\author{Yogesh Mehta}
\affiliation{School of Earth and Space Exploration, Arizona State University, Tempe, AZ, USA 85287}
\author{Shivam Pandey}
\affiliation{Department of Physics, Columbia University, New York, NY, USA 10027}
\author{Emmanuel Schaan}
\affiliation{SLAC National Accelerator Laboratory 2575 Sand Hill Road Menlo Park, California 94025, USA}
\affiliation{Kavli Institute for Particle Astrophysics and Cosmology, 382 Via Pueblo Mall Stanford, CA  94305-4060, USA}
\author{Blake Sherwin}
\affiliation{DAMTP, Centre for Mathematical Sciences, Wilberforce Road, Cambridge CB3 0WA, UK}
\affiliation{Kavli Institute for Cosmology, University of Cambridge, Madingley Road, Cambridge CB3 0HA}
\author{Alessio Spurio Mancini}
\affiliation{Department of Physics, Royal Holloway, University of London, Egham Hill, Egham, TW20 0EX, United Kingdom}
\author{\'{I}\~{n}igo Zubeldia}$^{1,2}$ 
\affiliation{Institute of Astronomy, University of Cambridge, Madingley Road, Cambridge CB3 0HA}
\affiliation{Kavli Institute for Cosmology, University of Cambridge, Madingley Road, Cambridge CB3 0HA}
% \affiliation{Institute of Astronomy, University of Cambridge, Madingley Road, Cambridge CB3 0HA}
% \affiliation{Kavli Institute for Cosmology, University of Cambridge, Madingley Road, Cambridge CB3 0HA}
% \affiliation{DAMTP, Centre for Mathematical Sciences, Wilberforce Road, Cambridge CB3 0WA, UK}
% % List of institutions
% $^{1}$DAMTP, Centre for Mathematical Sciences, Wilberforce Road, Cambridge CB3 0WA, UK\\
% $^{3}$ Department of Physics, Columbia University, New York, NY 10027, USA\\
% $^{4}$ Department of Physics and Astronomy, University of Pennsylvania, Philadelphia, PA 19104, USA}

% \author{Authors}

% These dates will be filled out by the publisher
\date{Received 2025}

% Enter the current year, for the copyright statements etc.
% \pubyear{2023}

% Don't change these lines

% \renewcommand{\arraystretch}{1.1}
% \label{firstpage}
% \pagerange{\pageref{firstpage}--\pageref{lastpage}}
% \maketitle

\begin{abstract}

These notes are very much work-in-progress and simply intended to showcase, in various degrees of details (and rigour), some of the cosmology calculations that \textsc{class\_sz}  can do. We describe the \textsc{class\_sz} code (\href{https://github.com/CLASS-SZ}{link}) in C, Python and Jax. Based on the Boltzmann code \textsc{class}, it can compute a wide range of observables relevant to current and forthcoming CMB and Large Scale Structure surveys. This includes galaxy  shear and clustering, CMB lensing, thermal and kinetic Sunyaev and Zeldovich observables, Cosmic Infrared Background, cross-correlations and three-point statistics. Calculations can be done either within the halo model or the linear bias model. For standard $\Lambda$CDM cosmology and extensions, \textsc{class\_sz} uses high-accuracy \textsc{cosmopower} emulators of the CMB and matter power spectrum to accelerate calculations. With this, along with efficient numerical integration routines, most \textsc{class\_sz} output can be obtained in less than 500 ms (CMB $C_\ell$'s or matter $P(k)$ take $\mathcal{O}(1\mathrm{ms})$), allowing for fast or ultra-fast parameter inference analyses. Parts of the calculations are ``\textit{jaxified}", so the  software can be integrated into differentiable pipelines.  
\end{abstract}

% \begin{keywords}
% large-scale structure of Universe – cosmic background radiation – methods: statistical – methods: data analysis
% \end{keywords}

%%%%%%%%%%%%%%%%%%%%%%%%%%%%%%%%%%%%%%%%%%%%%%%%%%

\tableofcontents

%%%%%%%%%%%%%%%%% BODY OF PAPER %%%%%%%%%%%%%%%%%%

\section{Introduction}
The \textsc{class\_sz} code was initially developed for a re-analysis of parameter constraints of the \textit{Planck} Compton-$y$ power spectrum \citep[see][]{Bolliet:2017lha}, built upon the \textsc{szfast} code \citep{KomatsuSeljak2002,Dolag:2015dta}. Subsequently it was extended to accommodate the calculations of a wide range of observables of interest in cosmology today, involving the thermal and kintetic Snyaev Zeldovich effects (tSZ and kSZ), galaxies, gravitational weak lensing, or the Cosmic Infrared Background. The main  purpose of \textsc{class\_sz} is the calculation of observables within the halo-model \citep[e.g.,][]{1991ApJ...381..349S,Mo_1996,Scoccimarro:2000gm,Seljak:2000gq,Cooray:2002dia}, but it can do much more. 

Since \textsc{class\_sz} is built on top of the Boltzmann code \textsc{class} \citep{2011arXiv1104.2932L,Lesgourgues_2011_CLASSIII,classII} (and more specifically, the version v2.9.4 of \textsc{class}), it can also compute all observables readily available in \textsc{class}. Notably, this includes the Cosmic Microwave Background (CMB) temperature and polarization anisotropy power spectra and the linear and non-linear matter power spectrum. Importantly, this also means that the \textsc{class\_sz} predictions are always strictly consistent with the cosmological model computed by \textsc{class}: \textsc{class\_sz} does not use approximations or fitting functions for distances and perturbation transfer functions, but simply uses the \textsc{class} output. Moreover, this means that \textsc{class\_sz} can compute predictions for all the extended cosmologies available in \textsc{class}, like spatial curvature, massive neutrinos, dynamical dark energy, decaying dark matter, and so on. 

The \textsc{class\_sz} code is written in C and has python wrapper, called \textsc{classy\_sz} and coded in \texttt{cython}, so that it can be called from \texttt{python} code or within a jupyter notebook. It relies on OpenMP multithreading paralellization for all embarassingly parallel tasks. In addition, Fourier transforms which do not have analytical expressions are evaluated numerically using the FFTW3 \citep{FFTW05} implementation of the Fast Fourier Transform (FFT) algorithm and its FFTLog \citep{Hamilton_2000} counterpart for logarithmic grids.

Via the python wrapper, \textsc{class\_sz} can use emulators of the matter and CMB power spectra to accelerate calculations. In particular, it has been adapted to use the  \textsc{cosmopower} emulators \citep{Spurio_Mancini_2022} presented in \cite{Bolliet:2023sst}. These emulators cover $\Lambda$CDM, massive neutrinos, $w$CDM and  $\Lambda$CDM+$N_\mathrm{eff}$ and are accurate enough to be used for Stage IV analyses. Hence, within these models, \textsc{class\_sz} can also be used to perform fast Markov Chains Monte Carlo  analysis  on standard cosmological likelihoods such as the ones implemented in \textsc{cobaya} \citep{Torrado:2020dgo} and \textsc{cosmosis} \citep{Zuntz:2014csq}.  The typical evaluation time of \textsc{class\_sz}, using the emulators and requesting high-accuracy CMB power spectra plus a few halo-model predictions is $\lesssim 0.5$ second. This time goes down massively if halo-model predictions are not requested, to become similar to \textsc{cosmopower} time of order 1ms.

\textsc{class\_sz} should be seen as a stand-alone code, independent of \texttt{class} as once can in principle plug emulators for distances, CMB $C_\ell$'s and matter $P(k)$ that are based on \textsc{camb} \citep{Lewis:1999bs}, other Boltzmann codes, or simulations \citep[e.g.,][]{Arico:2020lhq}.

Our goal in this work is to describe the range of observables that \textsc{class\_sz} computes and how it does it so that the computations are as efficient as possible. The code is public, available online on GitHub \footnote{\href{https://github.com/CLASS-SZ}{https://github.com/CLASS-SZ/class\_sz}} with documentation on ReadTheDocs \footnote{\href{https://class-sz.readthedocs.io/en/latest/index.html}{https://class-sz.readthedocs.io}}. Along with this paper, we release a series of legacy notebooks to show how to run \textsc{class\_sz}\footnote{\href{https://github.com/CLASS-SZ/notebooks}{https://github.com/CLASS-SZ/notebooks} \label{fn:nbs}}. Note that a first official version release of \textsc{class\_sz} was described recently in \cite{Bolliet:2022pze}.

The paper is organized as follows. In section \ref{sec:computing} we explain how to compute with \textsc{class\_sz} and summarize the main numerical routines that it uses. In section \ref{sec:cmb} we describe how to compute CMB anisotropy and matter power spectra using cosmopower emulators  implemented in \textsc{class\_sz}. In section \ref{sec:linbias}, we show how to compute linearly biased observables, which essentially consists of the non-linear power spectrum scaled by a bias factor. In section \ref{sec:hmf} we describe our implementation of the halo mass function and mass conversion routines.  In section \ref{sec:tracers}, we describe the Large Scale Structure (LSS) tracers as implemented \textsc{class\_sz}. In this section we also show their power spectra and cross-power spectra, several bispectra, position space correlations and beyond Limber predictions. In section \ref{sec:szcounts} we describe different implementations of sz cluster counts predictions available in \textsc{class\_sz}. In section \ref{sec:lkls} we describe how \textsc{class\_sz} is interfaced with \textsc{cobaya} and \textsc{cosmosis} and explain how to run standard likelihood analysis. In section \ref{sec:modify} we explain how to modify \textsc{class\_sz}, including how to modify tracer profiles our implementing a new tracer.  We conclude in section \ref{sec:conclusion}.

\section{Computing with \textsc{class\_sz}}\label{sec:computing}

%The code borrows special functions and an integrator for oscillatory functions from \verb|gsl| \citep{galassi2018scientific}, interpolation and root finding routines from J.~Burkardt's scientific library,\footnote{\href{https://people.math.sc.edu/Burkardt/index.html}{https://people.math.sc.edu/Burkardt/index.html}}

The \textsc{class\_sz} code and its python wrapper \textsc{classy\_sz}  can be called exactly like \textsc{class} and \textsc{classy}. The code should be primarily run via its Python wrapper. Nonetheless, it is possible to run the C code from a terminal console, from inside the \textsc{class\_sz} repository, one can run a calculations with the command:

\texttt{\$ ./class explanatory.ini}

\noindent where \texttt{explanatory.ini} is an input parameter file. Alternatively, from inside a python code, one can import the python wrapper: \texttt{import classy\_sz}; set the parameter values with the \texttt{set(\{...\})} method; and run a calculation with the \texttt{compute()} method. Instructions are given in the README file of the repository and example of calculations in python can be found in the tutorial notebooks (see footnote \ref{fn:nbs}).

A major improvement of \textsc{class\_sz} over \textsc{class}, even for CMB and matter power spectra calculations, is that it can use the high-accuracy \textsc{cosmopower} emulators released by \cite{Bolliet:2023sst}. Currently these emulators cover $\Lambda$ Cold Dark Matter (CDM) cosmology and the $\Lambda$CDM+$\Sigma m_\nu$,  $w$CDM and $\Lambda$CDM+$\Sigma N_\mathrm{eff}$ extensions. To run \textsc{class\_sz} using the emulators, the calculations must be done via the python wrapper \textsc{classy\_sz}. The procedure is the same as the usual one, except that the relevant method to call is not \texttt{compute()} but \texttt{compute\_class\_szfast()}. 

When many calculations have to be done using a fixed cosmology, it is possible to run \textsc{class\_sz} again without redoing the calculation of the standard \textsc{class} quantities. For this one simply has to call the method \texttt{compute\_class\_sz()} after updating the parameter dictionary. 

The background part of class, take $\mathcal{O}(50\mathrm{ms})$. Computing this plus all CMB power spectra with the emulators takes $\mathcal{O}(80\mathrm{ms})$.

By default, calculations of correlators in \textsc{class\_sz} are done within the Limber approximation \citep{Limber, loverde&Afshordi2008} such that wavenumbers $k$'s are mapped to angular multipoles $\ell$'s via
\begin{equation}
k=(\ell+1/2)/\chi:=k_\ell,\label{eq:limber}
\end{equation}
where $\chi$ is the comving distance and where we introduced the notation $k_\ell$ as a shorthand to refer to the wavenumber expressed in terms of $\ell$. (See subsection \ref{ssec:beyondlimber} for calculations without Limber aproximations.) 

There are three classes of integrals performed by \textsc{class\_sz}:
\begin{itemize}
    \item \textbf{Fourier and Hankel transforms.} Hankel transforms (i.e., Fourier transforms of radial functions) are performed either using \textsc{gsl}'s implementation of the \texttt{QAWO} algorithm\footnote{See this \href{https://www.gnu.org/software/gsl/doc/html/integration.html}{link} and \cite{galassi2018scientific}.}, or FFTLog as implemented by A.~Slosar\footnote{See \href{https://github.com/slosar/FFTLog}{https://github.com/slosar/FFTLog} and the C++ code \textsc{copter} \citep{2009PhRvD..80d3531C}.}.  Using FFTLog is generally a factor of $\approx 10$ faster than \texttt{QAWO}. Hence FFTLog is the default \textsc{class\_sz} choice for Fourier and Hankel transform. In the Python wrapper, the FFTLog operations are carried out with \texttt{mcfit} \citep{2019ascl.soft06017L}.
     \item \textbf{Integrals over masses.} Integrals over halo, sub-halo or galaxy masses are performed using an adaptive Patterson scheme \citep{Patterson_1968} imported from  {\tt CosmoTherm} \citep{ct2011}. This scheme is based on fully nested quadrature rules, which ensure none of the integrand evaluations are wasted.  The default \textsc{class\_sz} integration variable is $\ln m$ where $m$ is the mass. 
      \item \textbf{Integrals over redshift.}  Integrals over redshifts and comoving volume are also performed using the Patterson scheme. The default \textsc{class\_sz} integration variable is $\ln(1+z)$ (or equivalently, $\ln(a)$), where $z$ denotes redshift and $a$ is the scale factor. 
\end{itemize}

For root-finding, useful for converting between masses definitions for example, are done using Brent's method \citep{brent2002algorithms} from J.~Burkardt's scientific library\footnote{\href{https://people.math.sc.edu/Burkardt/index.html}{https://people.math.sc.edu/Burkardt/index.html}}. Linear interpolation of 1D and 2D arrays are also done using Burkardt's library. 

Default units in \textsc{class\_sz} are $\mathrm{Mpc}/h$ for distances and  $\mathrm{M}_\odot/h$ for masses.

If the requested observable is a 3D bispectrum or power spectrum, which depend on wavenumbers $k$'s, \textsc{class\_sz} is parallelized with respect to $k$ values. 

If the requested observable is an angular power spectrum, which depends on multipoles $\ell$'s, \textsc{class\_sz} is parallelized with respect to $\ell$ values.

All embarrassingly parallel tasks are done with \textsc{OpenMP} multi-threading. This means that, when possible, ``for loops" over multipoles, wavenumbers, masses or redshifts are parallelized.  

Most redshift integrals correspond to integrals over comoving volume $\mathrm{d}\mathrm{v}$, which we evaluate as follows, using Patterson's scheme,  according to 
\begin{equation}
    \int [\cdot\cdot\cdot]\mathrm{dv}= \int [\cdot\cdot\cdot]\frac{\mathrm{dv}}{\mathrm{d}z}\mathrm{d}z=\int [\cdot\cdot\cdot]\chi^2\mathrm{d}\chi=\int_{\ln(1+z_\mathrm{min})}^{\ln(1+z_\mathrm{max})}[\cdot\cdot\cdot]\frac{c\chi^2}{H}(1+z)\mathrm{d}\ln(1+z).\label{eq:zint}  
\end{equation}
where $H$ is the time dependent Hubble parameter and $c$ is the speed of light. The differential comoving volume is therefore given by
\begin{equation}
\frac{\mathrm{dv}}{\mathrm{d}z} = \frac{c\chi^2}{H}.
\end{equation}

\begin{figure*}
    \includegraphics[width=1.\columnwidth]{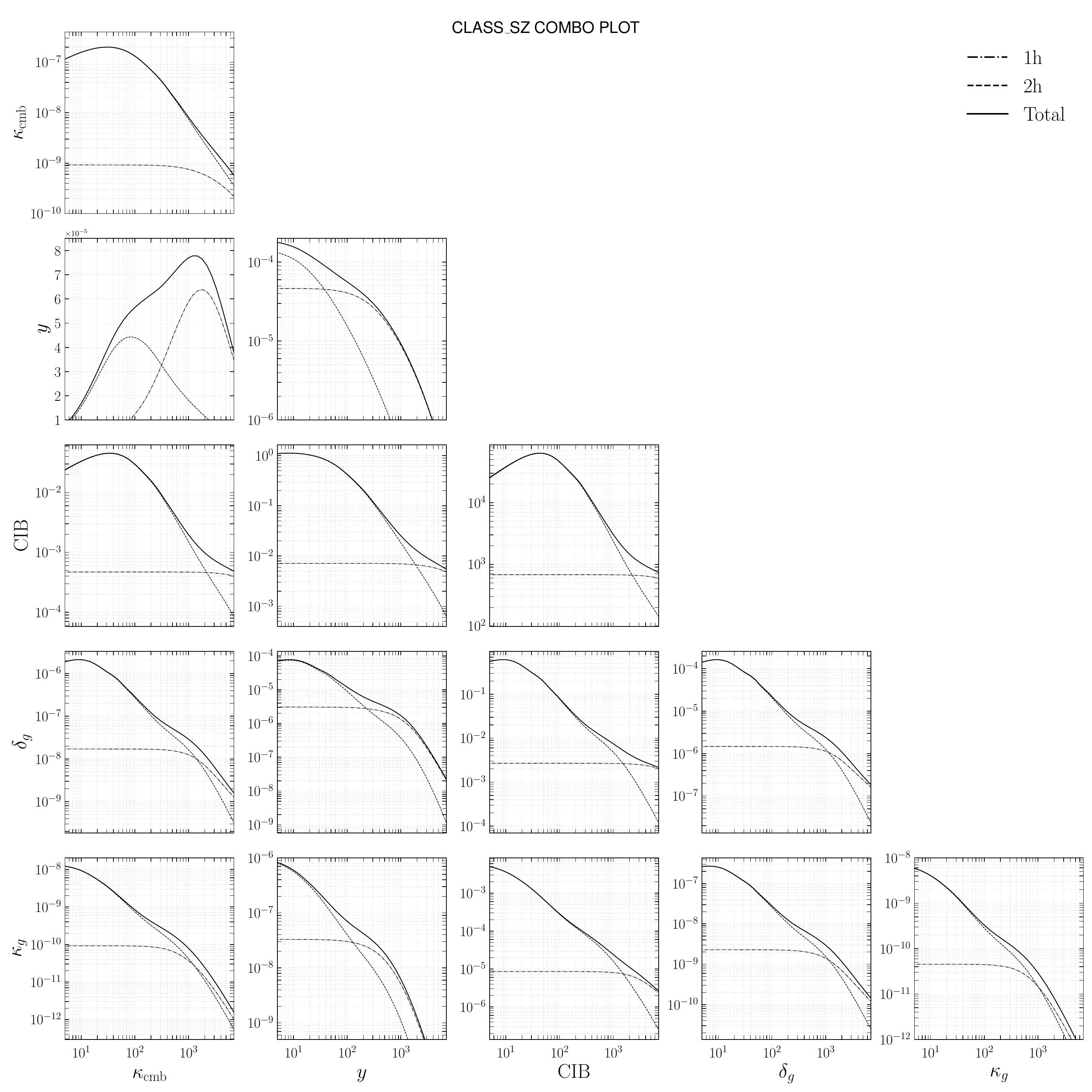}
    \vspace{-0.2cm}
    \caption{\textit{Combo} plot showing the main angular power spectra between different tracers implemented in \textsc{class\_sz}. The notebook to generate this figure is avaialble online (see footnote \ref{fn:combo}).}
    \label{fig:combo}
\end{figure*}

In figure \ref{fig:combo} we show the main cross and auto power spectra computed by the code. The notebook to replicate this calculation is available\footnote{\href{https://class-sz.readthedocs.io/en/latest/notebooks/class_sz_combo.html}{https://class-sz.readthedocs.io/en/latest/notebooks/class\_sz\_combo.html}\label{fn:combo}}.

\section{CMB Anisotropy and Matter Power Spectra}\label{sec:cmb}

All quantities computed by \textsc{class} can also be computed using \textsc{class\_sz}, since the latter is built on top of the former. For instance, to compute CMB and matter power spectra one can pass \textsc{class} inpute files to \textsc{class\_sz} and compute with the same command. 

The most standard calculations performed by \textsc{class} are the CMB temperature and polarization power spectra for E and B-modes, the lensing potential power spectrum, and the linear and non-linear matter power spectrum. For the non-linear modelling of matter perturbations, \textsc{class} can use both \textsc{halofit} and \textsc{hmcode}. All of this is avalaible in \textsc{class\_sz}. 

\begin{figure*}
    \includegraphics[width=1.\columnwidth]{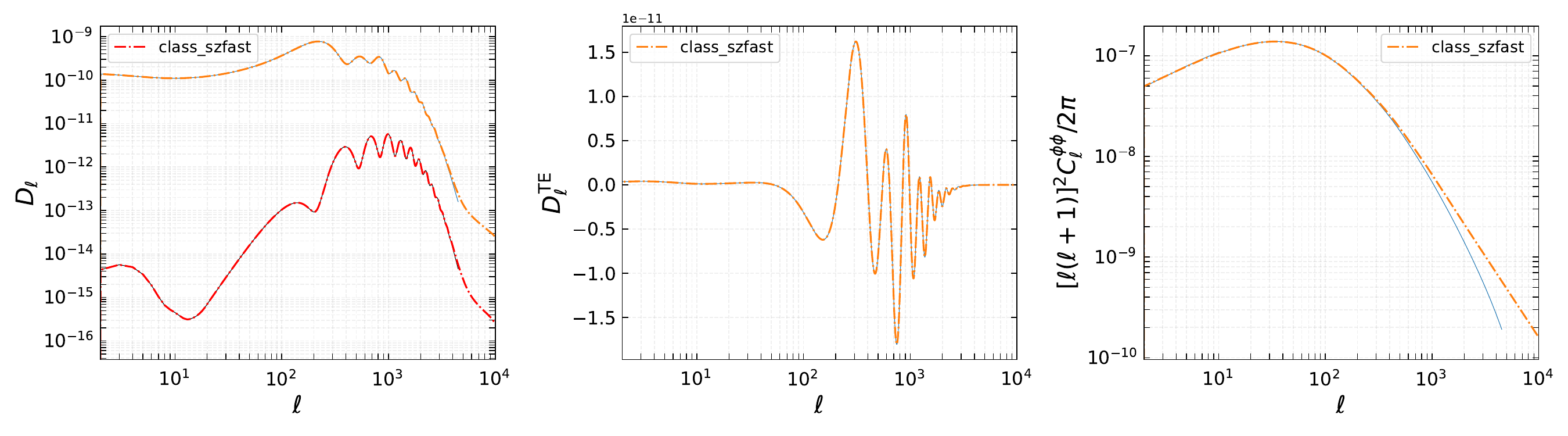}
    \vspace{-0.2cm}
    \caption{The cmb power spectra computed with \textsc{class\_sz}. The dashed-dotted line show the calculation using the \textit{fast mode} of  \textsc{class\_sz} which is based on \textsc{cosmopower} emulators.}
    \label{fig:cmb_cls}
\end{figure*}

On Figure \ref{fig:cmb_cls}, we show the CMB power spectra computed with \textsc{class\_sz}. The standard calculation, identical to \textsc{class} is shown as the thin solid lines and takes seconds, or minutes depending on accuracy requirements, to evaluate. The dashed-dotted lines show the same spectra computed with the \textit{fast mode} of \textsc{class\_sz}, based on \textsc{cosmopower} emulators \citep{Bolliet:2023sst}.  The fast-mode takes $\approx 0.03$ seconds to evaluate. The notebook to compute these predictions is available online\footnote{\href{https://class-sz.readthedocs.io/en/latest/notebooks/classy\_szfast\_cmb\_cls.html}{https://class-sz.readthedocs.io/en/latest/notebooks/classy\_szfast\_cmb\_cls.html}}.

 \begin{figure*}
    \includegraphics[width=1.\columnwidth]{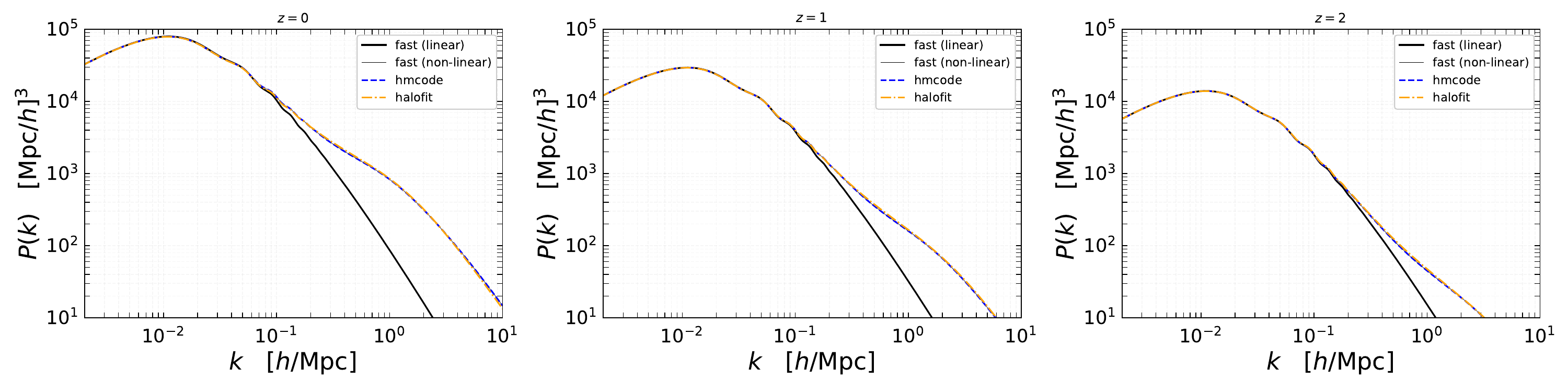}
    \vspace{-0.2cm}
    \caption{The matter power spectrum at 3 different redshifts using the  \textit{fast mode} of  \textsc{class\_sz} which is based on \textsc{cosmopower} emulators.}
    \label{fig:mpk}
\end{figure*}

On Figure \ref{fig:mpk}, we show the linear and non-linear matter power spectrum. As for the CMB, the \textsc{class\_sz} predictions for the matter power spectrum are strictly identical to those of \textsc{class}, or the high-accuracy \textsc{cosmopower} emulators if \textsc{class\_sz} is called with the fast-mode. As we will see in the next sections, the matter power spectrum is a central piece of LSS calculations. One of the advantages of \textsc{class\_sz} is that it does not rely on analytical approximations for the matter power spectrum but uses the exact \textsc{class} calculation. 

Unbiased and linearly-biased (see next Section) are based on the non-linear matter power spectrum (see Figure \ref{fig:mpk}). For the non-linear regime, one can request the same models as in \textsc{class}. Namely, \textsc{hmcode} \citep{Mead15,Mead_2021} or \textsc{halofit} \citep{Takahashi:2012em}.

Note that the matter power spectrum can also be computed within the halo-model using the Navarro-Frenk-White density profile, see Section  \ref{sssec:nfw}.

\section{Unbiased and Linearly Biased Observables}\label{sec:linbias}

Unbiased and linearly biased and weak lensing observables do not generally require an integration over masses (at least not explicitely). They can be obtained by integrating the matter power spectrum over comoving volume. 
Linearly biased observables refer to tracers of the matter field that are proportional to a bias. This is the case for the galaxy number density field. 

\subsection{Galaxies}

In the linear bias approximation, the galaxy overdensity $\delta_g$ is related to the matter overdensity $\delta_m$ by  
\begin{equation}
    \delta_g = b_g \delta_m,
\end{equation}
where $b_g$ is the galaxy bias. The galaxy power spectrum $P_{gg}$ can then be approximated as 
\begin{equation}
P_{gg}(k) = b_g^2 P_{NL}(k)\label{eq:pkgg_hf}
\end{equation}
where $P_{NL}$ is the non-linear matter power spectrum computed with, e.g., \textsc{halofit} or \textsc{hmcode}. With this, the galaxy angular power spectrum can be obtained as 
\begin{equation}
    C^{gg}_\ell = \int \mathrm{d}\mathrm{v}  W^g(z)W^g(z)P_{gg}\left(k_\ell\right)=b_g^2\int \mathrm{d}\mathrm{v}  W^g(z)W^g(z)P_{NL}\left(k_\ell\right)\label{eq:clgg}
\end{equation}
 Furthermore, the redshift dependent kernel is given by 
\begin{equation}
    W^g(z) = \frac{H}{\chi^2c }\varphi_\mathrm{g}^\prime(z)\quad\mathrm{with}\quad\varphi_\mathrm{g}^\prime(z) = \frac{1}{N_\mathrm{g}^\mathrm{tot}}\frac{\mathrm{d}N_\mathrm{g}}{\mathrm{d}z}\quad \mathrm{where}\quad N_\mathrm{g}^\mathrm{tot}=\int \mathrm{d}z\frac{\mathrm{d}N_\mathrm{g}}{\mathrm{d}z}.\label{eq:wg}
\end{equation}
The normalization by $N_g^\mathrm{tot}$ ensures that $\int \varphi_\mathrm{g}^\prime(z)\mathrm{d}z=1$. 

In \textsc{class\_sz}, this calculation can be done by adding \texttt{gal\_gal\_hf} to the \texttt{output} entry in the parameter dictionnary. The subscript $\texttt{hf}$ is a reference to the \textsc{halofit} power spectrum that can be used for the matter power spectrum, rather than the halo model. If \texttt{halofit} is added to the \texttt{non\_linear} entry of the parameter dictionnary, it is indeed the  \textsc{halofit} power spectrum that will be used for $P_{NL}$ in Eq.~\ref{eq:pkgg_hf}. If \texttt{hmcode} is used instead, is the \textsc{hmcode} power spectrum that will be used for $P_{NL}$. 
Once $P_{NL}$ is computed by \textsc{class}, within \textsc{class\_sz}, and given normalized galaxy redshift distribution $\phi^\prime_g(z)$ that is passed to the code (see \href{https://github.com/borisbolliet/class_sz/blob/master/notebooks/class_szfast_plots_and_tutorial_ngal.ipynb}{link}), the integral over redshift are done in parallel for as many multipoles as requested using \textsc{openmp} multi-threading. If one uses the \texttt{galn\_galn\_hf} option as an output, \textsc{class\_sz} will compute as many galaxy samples as requested, in paralell. For example for 4 galaxy samples and $\mathcal{O}(100)$ multipoles, the full calculation takes $\approx 0.3$ seconds.

Another observable that can be computed in a very similar way is the lensing convergence power spectrum, which is an unbiased tracer of the matter field. \textsc{class\_sz} can compute predictions for galaxy and CMB weak lensing.

\subsection{Galaxy Weak Lensing}

For galaxy weak lensing the angular power spectrum is computed as 
\begin{equation}
    C^{\kappa_g\kappa_g}_\ell = \int \mathrm{d}\mathrm{v}  W^{\kappa_g}(z)W^{\kappa_g}(z)P_{NL}\left(k_\ell\right)\label{eq:clgg}
\end{equation}
where $W^{\kappa_g}(z)$ is the galaxy weak lensing kernel given by 
\begin{equation}
 W^\mathrm{\kappa_{_{g}}}(z)=\frac{3}{2}\frac{\Omega_m H_0^2}{c^2\chi^2}(1+z) \chi I_s(\chi)\quad\mathrm{with}\quad I_s(\chi)=\int_z^{+\infty}\mathrm{d}z_\mathrm{s} \varphi^\prime_s(z_\mathrm{s})\frac{\chi_s-\chi}{\chi_s}\label{eq:wkg}
\end{equation}
where $\varphi^\prime_s$ is the normalized redshift distribution of source galaxies, and we used the notation $\chi_s=\chi(z_\mathrm{s})$ for the comoving distance to redshift $z_\mathrm{s}$, and where $\Omega_m$ and $H_0$ denote the matter fraction and Hubble parameter values today, respectively. To request this calculation in \textsc{class\_sz} we add \texttt{gallens\_gallens\_hf} as an output, or \texttt{ngallens\_ngallens\_hf} if we want to compute many galaxy samples at once.

\subsection{CMB Weak Lensing}

We can compute CMB weak lensing power spectra with \textsc{class\_sz} using this approach too. This is done according to: 

\begin{equation}
C^{\kappa_\mathrm{cmb}\kappa_\mathrm{cmb}}_\ell = \int \mathrm{d}\mathrm{v}  W^{\kappa_\mathrm{cmb}}(z)W^{\kappa_\mathrm{cmb}}(z)P_{NL}\left(k_\ell\right)\label{eq:clkk}
\end{equation}
where $W^{\kappa_\mathrm{cmb}}(z)$ is the CMB weak lensing kernel
\begin{equation}
 W^\mathrm{\kappa_{_{CMB}}}(\chi)=\frac{3}{2}\frac{\Omega_\mathrm{m}H_0^2}{c^2\chi^2}(1+z)\chi \frac{\chi_\star-\chi}{\chi_\star}.\label{eq:wkcmb}
\end{equation}
where $\chi_\star$ is the comoving distance to last scattering, i.e., \texttt{chi\_star} in \textsc{class}. Note that formally, this is the same as Eq.~\ref{eq:wkg} for $\varphi^\prime_s(z_\mathrm{s})=\delta^{D}(z-z_\star)$ where $\delta^D$ is Dirac's delta function and $z_\star$ is the redshift of the last scattering surface. To compute CMB weak lensing power spectrum using this approach in \textsc{class\_sz} we request \texttt{lens\_lens\_hf} as an output. 

\subsection{Cross-Correlations Between Galaxies and Gravitational Weak Lensing}

Cross correlations between galaxy density, galaxy lensing and CMB lensing are also available using the same approach.

For galaxy - galaxy lensing cross-power spectrum we request \texttt{gal\_gallens\_hf} as an output, for galaxy - CMB lensing we request \texttt{gal\_lens\_hf}  and for galaxy lensing - CMB lensing we request \texttt{gallens\_lens\_hf} as an output. These are computed as:
\begin{align}
C^{g\kappa_g}_\ell &= b_g\int \mathrm{d}\mathrm{v}  W^{g}(z)W^{\kappa_g}(z)P_{NL}\left(k_\ell\right)\label{eq:clkk} \\
C^{g\kappa_\mathrm{cmb}}_\ell &= b_g\int \mathrm{d}\mathrm{v}  W^{g}(z)W^{\kappa_\mathrm{cmb}}(z)P_{NL}\left(k_\ell\right)\label{eq:clkk}\\
C^{\kappa_g\kappa_\mathrm{cmb}}_\ell &= \int \mathrm{d}\mathrm{v}  W^{\kappa_g}(z)W^{\kappa_\mathrm{cmb}}(z)P_{NL}\left(k_\ell\right)\label{eq:clkk}
\end{align}
respectively, using the redshift dependent kernels defined above in Eq.~\eqref{eq:wg}, \eqref{eq:wkg} and \eqref{eq:wkcmb}.

\subsection{Scale Dependent Bias from Non-Gaussianity}
On large scale, the bias can become strongly scale-dependent when initial conditions are non-Gaussian. Departure from Gaussianity, parameterized by $f_{NL}$ \citep[see][]{Komatsu:2001rj}, contribute to a scale-dependent bias of the form \citep{Dalal:2007cu}:
\begin{equation}
    \Delta b (k,m) = 3 f_{NL}(b(m)-1)\delta_c \frac{\Omega_m H_0^2}{k^2 T(k)D(z)}
\end{equation}
where $\delta_c$ is the critical overdensity for spherical collapse, $D$ is the normalized growth factor, and $T$ is the transfer function. In \textsc{class}, $T$ is the sum of the transfer functions for the metric perturbations $\phi$ and $\psi$. Here $b$ is the bias of the tracer.

% \bb{fiona's plot}
%%% Fiona plot below::
\begin{figure}
\includegraphics[width=\textwidth]{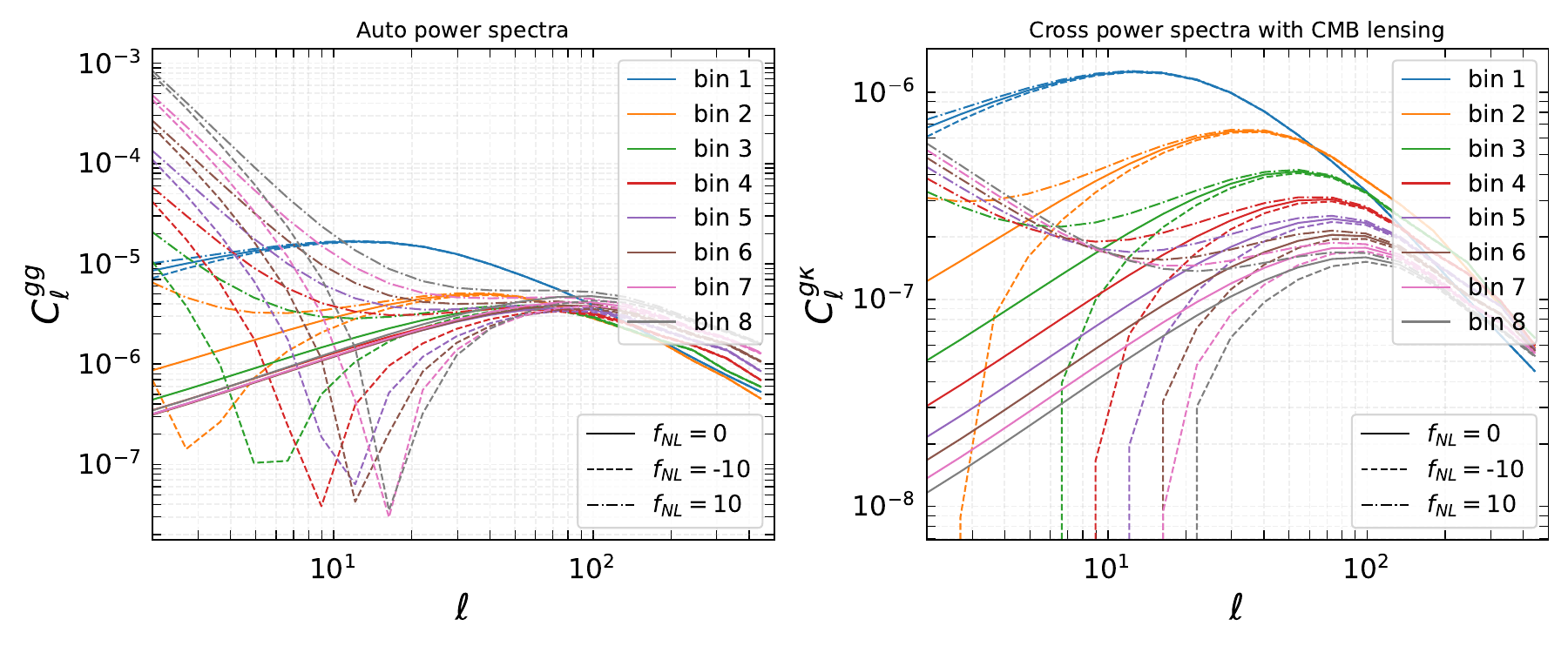}
\caption{\textit{Left:}The auto power spectra of an LSST-like sample of galaxies, with Gaussian redshift bins (of equal extent in redshift) and no photometric errors. \textit{Right:} The cross power spectra of the same galaxy bins with CMB lensing. The power spectra are computed for various values of $f_{\mathrm{NL}}$.} 
\end{figure}
%%% end Fiona plot

\subsection{Matter Bispectrum}
The expression of the tree-level matter bispectrum in Eulerian perturbation theory, for an Einstein de-Sitter Universe \citep{1984ApJ...279..499F}
\begin{equation}
    B_\mathrm{TL}(k_1,k_2,k_3) = 2F_2(k_1,k_2,k_3)P_{L}(k_1)P_{L}(k_2)+2\,\mathrm{cyc.}\label{eq:bt}
\end{equation}
where we did not write explicitly the permutations between modes and where the  $F_2$ kernel is given by \citep{1984ApJ...279..499F,Goroff:1986ep}\footnote{See Chapter 12 of \cite{dodelson2020modern} for a presentation of second-order cosmological perturbation theory and \cite{bernardeau2002} for further details.}
\begin{equation}
    F_2(k_1,k_2,k_3)=F_2(\mathbf{k}_1,\mathbf{k}_2)=\frac{5}{7}+\frac{1}{2}\cos\theta_{12}\left(\frac{k_1}{k_2}+\frac{k_2}{k_1}\right)+\frac{2}{7}(\cos\theta_{12})^2\quad\mathrm{with}\quad \cos\theta_{12}=\frac{\mathbf{k}_1\cdot\mathbf{k}_2}{k_1 k_2}=\frac{k_3^2-k_2^2-k_1^2}{2k_1 k_2}\label{eq:f2s}.
\end{equation}
(This is the expression as implemented in \verb|class_sz|, which takes the three wavenumber moduli as an input.) The \cite{Gil_Mar_n_2012} bispectrum fitting formula has the same form as Eq.~\eqref{eq:bt}, except that the linear matter power spectrum is replaced by its non-linear counterpart and that it includes extra scale- and redshift-dependent coefficients in front of the terms in the expression of the $F_2$ kernel. There are nine parameters that control the scale and redshift dependence of those coefficients, whose values are found by fitting the data from $N$-body simulations. Thus, the non-linear matter bispectrum is written as
 \begin{equation}
    B_{\mathrm{eff}}^{NL}(k_1,k_2,k_3) = 2F_2^\mathrm{eff}(k_1,k_2,k_3)P_{NL}(k_1)P_{NL}(k_2)+2\,\mathrm{cyc}.\label{eq:b_gilma}
\end{equation}
For $F_2^\mathrm{eff}$ we implemented the \cite{Scoccimarro:2000ee} and \cite{Gil_Mar_n_2012} formulas and fitting parameters.  The main difference between these formulas is that the \cite{Gil_Mar_n_2012} formula corrects the unphysical oscillations associated with the BAOs in the power spectrum which are visible in the \cite{Scoccimarro:2000ee} prediction \citep[see, e.g.,][for details]{Gil_Mar_n_2012}. On large scales, both formulas match with the tree-level bispectrum of Eq.~\eqref{eq:bt}. Note that the \cite{Gil_Mar_n_2012} formula is calibrated on a fairly restricted $k$- and $z$-range, namely:  $0.03\,h/\mathrm{Mpc}\leq k \leq 0.4 h/\mathrm{Mpc}$, and $0\leq z\leq 1.5$. Recent  matter bispectrum fitting formulas have been derived by \cite{Takahashi_2020} on a broader $k$- and $z$-range ($k\lesssim 3 h/\mathrm{Mpc}$ and $z<3$) but we do not discuss them here as they have not been used in the context of the kSZ effect.

The \cite{Scoccimarro:2000ee} formula reads
\begin{eqnarray}
    F_2^\mathrm{eff}(\mathbf{k}_i,\mathbf{k}_j) & = &\frac{5}{7}a(n_i,k_i)a(n_j,k_j)\nonumber\\
                     & +&\frac{1}{2}\cos (\theta_{ij})\left(\frac{k_i}{k_j}+\frac{k_j}{k_i}\right)b(n_i,k_i)b(n_j,k_j)+\frac{2}{7}\cos^2 (\theta_{ij})c(n_i,k_i)c(n_j,k_j)\label{eq:socci}
\end{eqnarray}
where 
\begin{eqnarray}
 \nonumber \label{abc} a(n,k)&=&\frac{1+\sigma_8^{a_6}(z)[0.7Q_3(n)]^{1/2}(q a_1)^{n+a_2}}{1+(q a_1)^{n+a_2}}, \\
 b(n,k)&=&\frac{1+0.2a_3(n+3)q^{n+3}}{1+q^{n+3.5}}, \\
\nonumber c(n,k)&=&\frac{1+4.5a_4/[1.5+(n+3)^4](q a_5)^{n+3}}{1+(q a_5)^{n+3.5}}.
\end{eqnarray}
where 
\begin{equation}
n \equiv \frac{d\log P_L(k)}{d\log k}    
\end{equation}
is the slope of the linear matter power spectrum at $k$. To accelerate calculations, we pre-tabulate $n$ on a $(z,k)$-grid, which we then interpolate when necessary. 
Note that since BAO is present in $P_L(k)$ it creates oscillations in n(k). In the \cite{Scoccimarro:2000ee} approach, these oscillations are kept.

And 
\begin{equation}
    Q_3(n) = \frac{4-2^n}{1+2^{n+1}}
\end{equation}
and 
\begin{equation}
q = \frac{k}{k_{NL}}\quad\mathrm{with}\quad k_{NL}\quad
     \mathrm{such}\,\,\mathrm{that}\quad \frac{k_{NL}^3P_L(k_{NL})}{2\pi^2} = 1
\end{equation}
where the coefficients are
\begin{equation}
\nonumber a_1=0.25,\, a_2=3.5,\, a_3=2,\,  a_4=1,\, a_5=2,\, a_6=-0.2 \,.
\end{equation}

\begin{table}
\begin{center}
\begin{tabular}{l|l|l}
\hline
$a_1=0.484$ & $a_2=\,3.740$ & $a_3=-0.849$ \\
$a_4=0.392$ & $a_5=\,1.013$ & $a_6=-0.575$ \\
$a_7=0.128$ & $a_8=-0.722$ & $a_9=-0.926$ \\
\hline
\end{tabular}
\end{center}
\caption{Best-fit parameters from \protect\cite{Gil_Mar_n_2012}, to compute the effective $F_2$ kernel of Eq.~\eqref{eq:socci} with the modified functions of Eq.~\eqref{eq:gilmar}.}
\label{tab:tabla_fit}
\end{table}
%See Eqs.~2.6-2.12 of \cite{Gil_Mar_n_2012} for the expression of the effective kernel $F_2^\mathrm{eff}$, as well as the values of the fitting parameters.  

The \cite{Gil_Mar_n_2012} formula is the similar but with modified functions:
\begin{eqnarray}
 \nonumber \label{abc_new} \tilde{a}(n,k)&=&\frac{1+\sigma_8^{a_6}(z)[0.7Q_3(n)]^{1/2}(q a_1)^{n+a_2}}{1+(q a_1)^{n+a_2}}, \\
 \tilde{b}(n,k)&=&\frac{1+0.2a_3(n+3)(q a_7)^{n+3+a_8}}{1+(q a_7)^{n+3.5+a_8}}, \\
\nonumber \tilde{c}(n,k)&=&\frac{1+4.5a_4/[1.5+(n+3)^4](q a_5)^{n+3+a_9}}{1+(q a_5)^{n+3.5+a_9}}.\label{eq:gilmar}
\end{eqnarray}

The values of the $a_i$'s parameters are reported in table \ref{tab:tabla_fit}.

In the \cite{Gil_Mar_n_2012} approach, the oscillations in $n$ due to the BAO are removed. To do so,  we simply interpolate $n(k)$ through the mean. To find the mean, we locate the extrema of the oscillations and define interpolating nodes at middle between the extrema (in $\ln k$).

\begin{figure*}
    \includegraphics[width=1.\columnwidth]{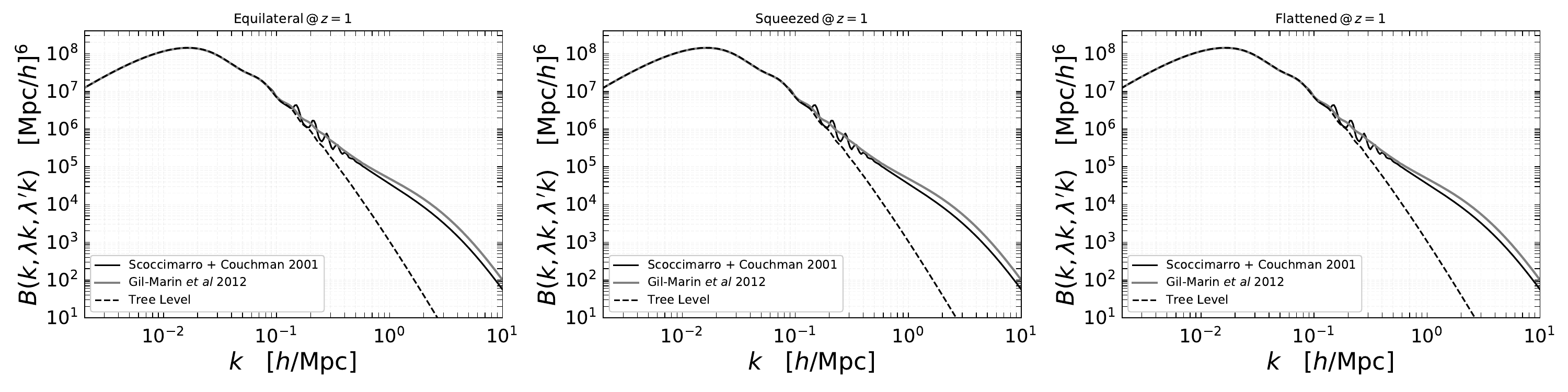}
    \vspace{-0.2cm}
    \caption{The matter bispectrum for the equilateral, squeezed and flattened configuration using the Tree-Level approximation and the  \cite{Scoccimarro:2000ee} and \cite{Gil_Mar_n_2012} formulas.}
    \label{fig:Bkz_effective}
\end{figure*}

The  matter bispectrum computed with the \cite{Gil_Mar_n_2012} and \cite{Scoccimarro:2000ee} fitting formulas are plotted in  Figure~\ref{fig:Bkz_effective} for $z=1$ and for three different configurations.  A notebook is available online to reproduce these calculations\footnote{\href{https://github.com/CLASS-SZ/notebooks/blob/main/class_sz_matterbispectrum_at_z.ipynb}{https://github.com/CLASS-SZ/notebooks/blob/main/class\_sz\_matterbispectrum\_at\_z.ipynb}}.

\section{Halo Abundance}\label{sec:hmf}

\subsection{Halo Mass Function}

The model assumes that matter is distributed within distinct spherical halos whose abundance is determined by the linear matter power spectrum through the halo mass function (HMF). The HMF determines the comoving number density of haloes of mass $M$ at redshift $z$ via \citep[e.g,][]{1974ApJ...187..425P,1991ApJ...379..440B,Tinker_2008,Tinker2010}
\begin{equation}
    \frac{\mathrm{d}n}{\mathrm{d}m} = \nu f(\nu)\frac{\rho_\mathrm{m,0}}{m}\frac{\mathrm{d}\ln\sigma^{-1}}{\mathrm{d}m}\label{eq:hmf}
\end{equation}
where 
\begin{equation}
\nu(m,z)=\delta_c/\sigma(m,z)\label{eq:peak}
\end{equation}
\footnote{Note that \cite{Tinker2010} use the \textit{peak-height} definition $\nu \equiv \delta_c/\sigma(m,z)$ while {\texttt{class\symbol{95}sz}} uses $\nu \equiv (\delta_c/\sigma)^2$ as in E. Komatsu's {\texttt{szfast}} code. Also,  \cite{Tinker_2008} do not use the peak height explicitly, but  $\sigma^{-1}$ instead.} is the \textit{peak height} in the linear density field with $\delta_c=(3/20)(12\pi)^{2/3}\approx1.686$ the spherical collapse density threshold \citep[see][for the $\Omega_\mathrm{m}$ correction - not used here]{nm1997},  $\rho_\mathrm{m,0}$ is the mean matter density at $z=0$ and 
\begin{equation}
    \sigma^2(m,z) =\frac{1}{2\pi^2}\int\mathrm{d}kk^2\hat{\mathrm{W}}(kR)^2 P_{L}(k,z)\label{eq:sig2}
\end{equation}
is the variance of the matter density field smoothed in region of radius $R=(3m/4\pi\rho_\mathrm{m,0})^{1/3}$ using the Fourier transform of the real-space top-hat window function $\hat{\mathrm{W}}(x)=3j_1(x)/x$ where $j_1(x) =  \left[\sin(x)-x\cos(x)\right]/x^2$ is the first-order spherical Bessel function. Here, $P_L$ is the linear matter power spectrum.

We have implemented four different parameterisations
of the HMF: the \cite{Bocquet:2015pva}  fitting formula obtained
from the \verb|Magneticum| simulation with the impact of baryons; the \cite{Tinker_2008} formula; 
the \cite{Tinker2010} formula (see their Eq. 3), an updated version of the former;
and the \cite{Jenkins:2000bv} formula. 

The \cite{Bocquet:2015pva} and \cite{Tinker_2008} HMFs are expressed as 
\begin{equation}
f\left(\sigma,z\right)=  A\left[\left(\frac{\sigma}{b}\right)^{-a}+1\right]\exp\left(-\frac{c}{\sigma^{2}}\right).\label{eq:MF}
\end{equation}
The  \cite{Tinker2010}  HMF is parameterized as
\begin{equation}
f\left(\nu,z\right)=  \alpha\left[\left(\beta^2 \nu\right)^{-\phi}+1\right]\nu^{\eta}\exp\left(-\gamma\frac{\nu}{2}\right)\sqrt{\nu},\label{eq:MF-T10}
\end{equation}
where $\nu$ is defined via $\sigma=1.685/\sqrt{\nu}$. The fitting parameters of these HMFs depend on redshift and are reported in table \ref{tab:HMFparams}. The \cite{Jenkins:2000bv} formula for the HMF evaluated at  $M_{\mathrm{180m}}$ (over-density mass of 180 times the mean matter density) does not have an explicit redshift dependence and reads as
%\begin{equation}
$f\left(\sigma\right)= 0.301\exp\left(-|0.64-\ln\sigma|^{3.82}\right)$.\\%\label{eq:MF-J01}
%\end{equation}

\begin{table}
\begin{center}
%\setcellgapes{2pt}\makegapedcells
\begin{tabular}{l|cccccccccc}
 &$A_0$ & $a_0$ & $b_0$ & $c_0$&$A_z$ & $a_z$ & $b_z$ & $c_z$&&\tabularnewline
\hline 
\cite{Bocquet:2015pva}   & 0.228 & 2.15 & 1.69 &1.30 & 0.285 & -0.058 & -0.366 & -0.045&&\tabularnewline
\cite{Tinker_2008}   & 0.186 & 1.47 & 2.57 &1.19 & -0.14 & -0.06 & -0.011 & 0&&\tabularnewline
\hline
\tabularnewline
 &$\alpha_0$ & $\beta_0$ & $\gamma_0$ & $\eta_0$&$\phi_0$ & $\alpha_z$ & $\beta_z$ & $\gamma_z$&$\eta_z$&$\phi_z$\tabularnewline
 \hline
\cite{Tinker2010}  & 0.368 & 0.589 & 0.864 & -0.243 & -0.729 & 0 & 0.2 & -0.01&0.27&-0.08
\end{tabular}
\end{center}
\caption{Parameters for the halo mass functions (HMF). Note that these parameters values are relevant for \protect\cite{Bocquet:2015pva}, \protect\cite{Tinker_2008} and  \protect\cite{Tinker2010}  HMFs evaluated at the over-density mass $M_{200m}$ (for the \protect\cite{Tinker_2008} formula at $M_{1600m}$, the value of $b_z$ has to be replaced by  $b_z=-0.314$). Given a parameter $p=A,b,..,$ the redshift dependence is obtained as $p=p_0 (1+z)^{p_z}$.\label{tab:HMFparams}}
\end{table}

By consistency, the HMF must be such that
\begin{equation}
    \int \mathrm{d}\nu f(\nu)=1,\quad\int \mathrm{d}\nu b^{(1)}(\nu)f(\nu)=1,\quad\int \mathrm{d}\nu b^{(n)}(\nu)f(\nu)=0\,\,\mathrm{for}\,\,n>1,\label{eq:hmconsist}
\end{equation}
where $b^{(1)}$ is the linear bias (see Eq.~\ref{eq:b1tink}) and $b^{(n)}$ are higher order biases (e.g., Eq.~\ref{eq:b2pbs}). These constraints enforce that all matter is within halos and that it is not biased with respect to itself \citep[e.g.,][]{Tinker2010}.

One should keep in mind that these fitting formulas are calibrated on simulations with a limited mass and redshift range. Namely $0.25\lesssim \sigma^{-1}\lesssim 2.5$, which corresponds to masses $\sim 10^{10}-10^{15}\,\mathrm{M}_\odot /h$ at $z=0$) and $0<z\lesssim 2$, for the \cite{Tinker_2008} and \cite{Tinker2010} functions. Note also that \cite{Tinker_2008} suggests to use $f(\sigma,z=2.5)$ for all $z>2.5$, while \cite{Tinker2010} suggests to use $f(\sigma,z=3)$ for all $z>3$\footnote{Note that ccl does not have the $z>3$ condition.}. This may be important for quantities that have contribution from high redshift, like CIB or CMB lensing. 

 \begin{figure*}
 \includegraphics[width=1.\columnwidth]{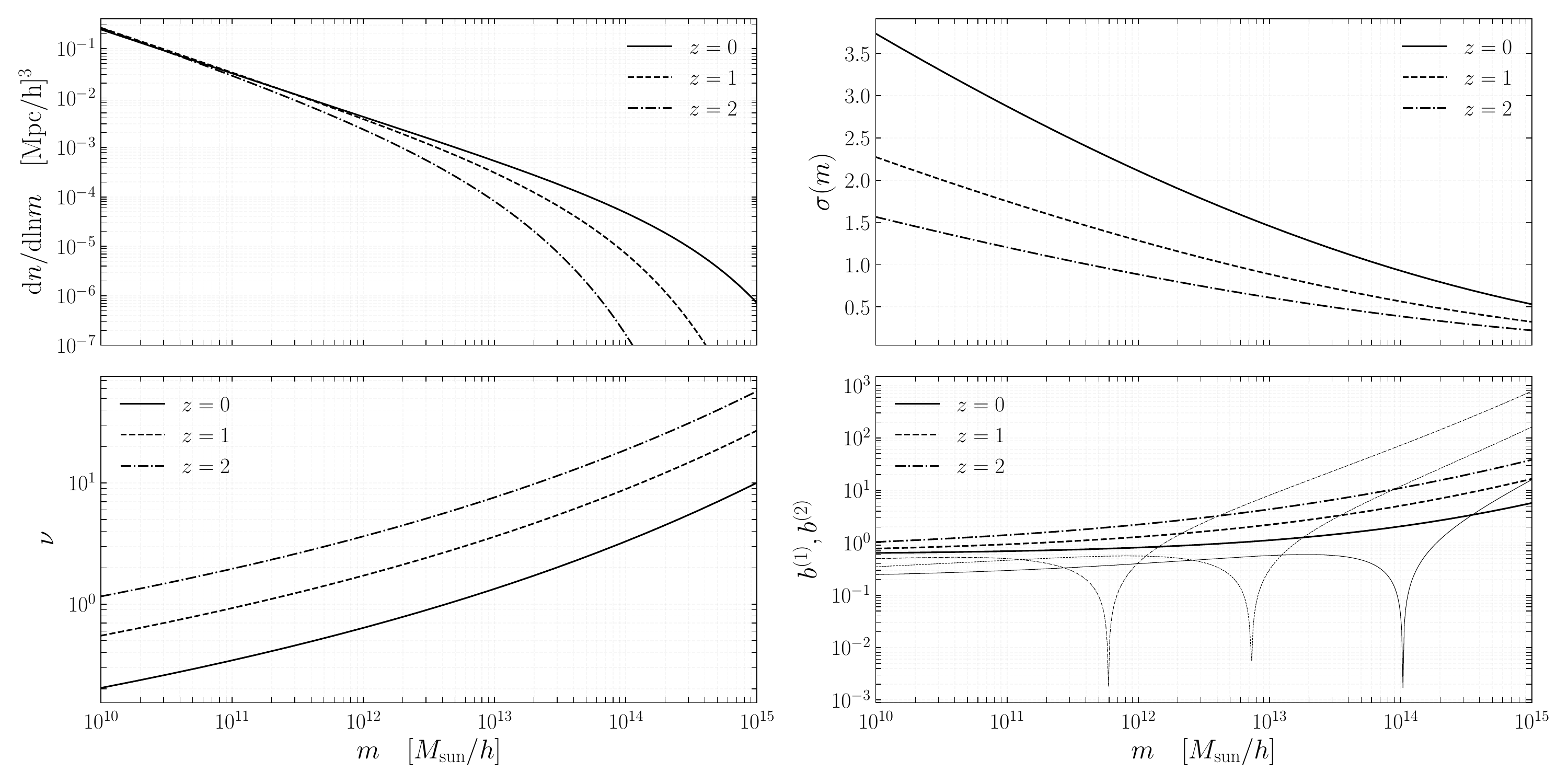}
    \vspace{-0.2cm}
     \caption{\textit{Top left:} Differential halo number density (Eq.~\ref{eq:hmf}). \textit{Top right:} Smoothed matter density field variance (Eq.~\ref{eq:sig2}). \textit{Bottom left:} Peak height (Eq.~\ref{eq:peak}). \textit{Bottom right:} First and second order biases (Eq.~\ref{eq:b1tink} and~\ref{eq:b2pbs}). These calculations can be replicated using the online notebook.\footnote{\href{https://github.com/CLASS-SZ/notebooks/blob/main/class\_sz\_hmf\_and\_sigma.ipynb}{https://class-sz.readthedocs.io/en/latest/notebooks/classy\_szfast\_hmf\_and\_sigma.html}}}
    \label{fig:hmf}
\end{figure*}

\subsection{Halo Bias}\label{ssec:bias}

In \textsc{class\_sz} the fiducial first-order bias is the \cite{Tinker2010} formula:
\begin{equation}
    b^{(1)}(\nu)=1-A\frac{\nu^a}{\nu^a+\delta_c^a}+B\nu^b+C\nu^c\label{eq:b1tink}
\end{equation}
with parameters fixed to the values in Table \ref{tab:tinkerbias1}, which are from \cite{Tinker2010}.  It is a power-law at high-mass, flattens out at low-mass and is 1 at $\nu=0$. Recall, $\nu=\delta_c/\sigma(m)$ with $\sigma(m)$ from Eq.~\eqref{eq:sig2}.

We compute the second-order bias $b^{(2)}$ with the peak background split formula using Eq.~(8) of \cite{10.1093/mnras/stv702}, which follows \cite{Scoccimarro:2000gm}. Namely,

\begin{equation}
    b_2(\nu) = 2 (1+a_2)(\epsilon_1+E_1)+\epsilon_2+E_2\label{eq:b2pbs}
\end{equation}
where $a_2=-17/21$ is from the spherical collapse model and where $\epsilon_i$'s and $E_i$'s are computed from the \cite{Tinker2010} HMF parameters (see Table \ref{tab:HMFparams}) as given in Table \ref{tab:b2}. See also \cite{phils2020} for details on higher order biases.

\begin{table}
\begin{center}
\begin{tabular}{l|c}
\hline
$A$ & $1.0+0.24 y\exp[-(4/y)^4]$\\
$a$ & $0.44y -0.88$\\
$B$ & $0.183$\\ 
$b$ & $1.5$\\ 
$C$ & $0.019+0.107y+0.19\exp[-(4/y)^4]$\\
$c$ & 2.4\\ 
\hline
\end{tabular}
\end{center}
\caption{Best-fit parameters from Table 2 of \protect\cite{Tinker2010} first-order bias coefficients. With $y =\log_{10}\Delta$ and $\Delta$ is with respect to mean density. This means that if we work with critical density masses, we need to set $\Delta=\Delta_c/\Omega_m(z)$. When doing that, note that we do not include neutrinos in $\Omega_m$.}
\label{tab:tinkerbias1}
\end{table}

\begin{table}
\begin{center}
\begin{tabular}{l|c}
\hline
$\epsilon_1$ & $(c\nu-2d)/\delta_c$ \\
$\epsilon_2$ & $[c\nu(c\nu-4d-1)+2d(2d-1)]/\delta_c^2$ \\
$E_1$ & $2a(\delta_c[(b\nu)^{-a}+1])^{-1}$\\ 
$E_2/E_1$ & $(-2a+c\nu -4d+1)/\delta_c$ \\
\hline
\end{tabular}
\end{center}
\caption{Coefficients for the second order bias, from \protect\cite{10.1093/mnras/stv702}. With parameters $a=\phi$, $b=\beta^2$,$c=\gamma$ and $d=\eta+0.5$ computed from the \protect\cite{Tinker2010} parameters. Note that $\nu=\delta_c^2/\sigma^2$.}
\label{tab:b2}
\end{table}

\subsection{Halo Model Consistency}

%\bb{see the nice discussin in appendix A of https://arxiv.org/pdf/2005.00009.pdf}

In the halo model, we routinely assume that all the mass in the Universe is in the form of halos.
This results in a number of consistency relations:
\beq
\left\{
\bal
&\int dm \ n(m) \frac{m}{\bar{\rho}} = 1\\
&\int dm \ n(m) b_1(m) \frac{m}{\bar{\rho}} = 1\\
&\forall i>2, \int dm \ n(m) b_i(m) \frac{m}{\bar{\rho}} = 0\\
\eal
\right.
\eeq
The first constraint encode that the mean matter density, expressed as a sum over halos, is equal to the mean matter density in the Universe.
The other constraints indicate that the matter overdensity field, expressed as a sum over halos, is equal to the matter overdensity field.
In practice, the mass function $n(m)$ is only known over a finite range $[m_\text{min}, m_\text{max}]$.
The convergence of the integral for the power spectrum 1-halo term is usually not an issue.
But the 2-halo term is often very sensitive to the minimum mass $m_\text{min}$, and does not converge until extremely low halo masses, for which the notion of halo may even be ill-defined.
In practice, other halo model observables (e.g., higher point functions) may be sensitive to either integration bound.

%\bb{in fact 1h at very high k can be an issue... see comment in cmb lens plot}

As discussed in \cite{Schmidt_2016}, we address this issue by implementing the halo model consistency conditions of Eq.~\eqref{eq:hmconsist} as follows:
\begin{align}
\int_0^{+\infty} \mathrm{d}n\hat{u}^X(m,z)\hat{u}^Y(m,z) &= \int_{m_\mathrm{min}}^{m_\mathrm{max}}\mathrm{d}n \hat{u}^X(m,z)\hat{u}^Y(m,z)+N_\mathrm{min}(z)\hat{u}^X(m_\mathrm{min},z)\hat{u}^Y(m_\mathrm{min},z)\label{eq:hmct1}\\
\int_0^{+\infty} \mathrm{d}nb^{(i)}(m,z)\hat{u}^X(m,z) &=\int_{m_\mathrm{min}}^{m_\mathrm{max}}\mathrm{d}nb^{(i)}(m,z)\hat{u}^X(m,z) + b^{(i)}_{m_\mathrm{min}}(z)[\rho_\mathrm{m,0}/m_\mathrm{min}]\hat{u}^X(m_\mathrm{min},z)\label{eq:hmct2}
\end{align}
with $i=1,2$ for the first and second order bias of Eq.~\eqref{eq:b1tink} and \eqref{eq:b2pbs}, respectively \citep[see also][]{phils2020,mead2021}. The \textit{counter-terms} on the RHS account for the low-mass part of HMF that cannot be parameterized using current N-body simulations. With this implementation,  ``halo model predictions do not depend on any properties of low-mass halos that are smaller than the scales of interest" \citep{Schmidt_2016}. The counter-terms require a mass integral at each redshift that we pretabulate as 
\begin{align}
N_\mathrm{min}(z) &= [1-I_0(z)]\rho_\mathrm{m,0}/m_\mathrm{min}\quad\mathrm{with}\quad I_0(z)=\int_{m_\mathrm{min}}^{m_\mathrm{max}}\mathrm{d}n m/\rho_\mathrm{m,0}\label{eq:ct1}\\
b^{(1)}_\mathrm{min}(z) &= 1-I_1(z)\quad\mathrm{with}\quad I_1(z)=\int_{m_\mathrm{min}}^{m_\mathrm{max}}\mathrm{d}nb^{(1)}(m,z)m/\rho_\mathrm{m,0}\label{eq:ct2}\\
b^{(2)}_\mathrm{min}(z) &= -I_2(z)\quad\mathrm{with}\quad I_2(z)=\int_{m_\mathrm{min}}^{m_\mathrm{max}}\mathrm{d}nb^{(2)}(m,z)m/\rho_\mathrm{m,0}.\label{eq:ct3}
\end{align}
One can check that Eq. \eqref{eq:hmct1}-\eqref{eq:hmct2} amounts to substituting the HMF $\mathrm{d}n/\mathrm{d}m$ with $\mathrm{d}n/\mathrm{d}m+N_\mathrm{min}\delta(m-m_\mathrm{min})$ in all mass integrals and setting a cut-off at $m_\mathrm{min}$. Eq. \eqref{eq:ct1}-\eqref{eq:ct3} are then equivalent to the consistency conditions
\begin{equation}
\int \mathrm{d}n m =\rho_\mathrm{m,0},\quad
\int \mathrm{d}n m b^{(1)}(m,z) = \rho_\mathrm{m,0}, \quad
\int \mathrm{d}nm b^{(2)}(m,z) = 0 ,\label{eq:hmc}
\end{equation}
ensuring that all mass is within halos and that matter is unbiased with respect to itself. 

These consistency conditions can have a significant contribution to power spetcra of tracers, especially weak lensing tracers which receive contribution from low-mass halos. 

\subsection{Concentration-Mass Relations}\label{ref:ssec}

Halo-concentration is a fundamental parameter of dark matter halos. . Moreoever, assuming that dark matter follows NFW, to convert from the virial mass to the over-density mass, one needs concentration-mass  relations (see Subsection \ref{ref:ssec}). 
 
 We have implemented various concentration-mass relations \textsc{CLASS\_SZ}, including \cite{Duffy_2008},  \cite{Zhao_2009},  \cite{Klypin_2011}, \cite{Bhattacharya2013}, \cite{S_nchez_Conde_2014}. 

 The \cite{Duffy_2008} relations assumes that the virial concentration is 
 \begin{equation}
 c_{\mathrm{vir}} = 7.85\times(M_{\mathrm{vir}}/2\times10^{12})^{-0.81}\times(1+z)^{-0.71}\label{eq:duffy}
\end{equation}
where $M_{\mathrm{vir}}$ is in units $h^{-1}\mathrm{M_\odot}$. 

\cite{Klypin_2011} uses 
\begin{equation}
c_{\mathrm{vir}} = c_0\times(M_{\mathrm{vir}}/10^{12})^{-0.075}\times[1+(M_{\mathrm{vir}}/M_{\mathrm{0}})^{-0.26}],
\end{equation}
where $c_0$ and $M_0$ are functions of redshift and whose tabulation can be found in Table~3 of that reference. 

\cite{S_nchez_Conde_2014} uses concentrations at the over-density mass
$M_{{\mathrm{200c}}}$ of 200 times the critical
density of the universe instead of $M_{\mathrm{vir}}$, i.e., 
\begin{equation}
c_{\mathrm{200}} = \sum_{i=0}^{5} c_i \times [\ln(M_{\mathrm{200c}})]^i \times(1+z)^{-1}
\end{equation}
where the values for the coefficients $c_i$ are given below Eq.~1 of the reference. 

Finally \cite{Zhao:2008wd}  does not give an explicit concentration-mass relation but computes it numerically at every redshift with the \verb|mandc| code\footnote{website: \href{http://202.127.29.4/dhzhao/mandc.html}{http://202.127.29.4/dhzhao/mandc.html}}. In this case, we ran \verb|mandc| for the \textit{Planck 2015} best-fit cosmological parameters and for several redshift values and tabulated the concentration for subsequent interpolation. The tabulation is available in the \textsc{class\_sz} repository.

Another standard choice of the concentration-mass relation for dark matter halos is the one introduced in \cite{Bhattacharya2013}:
\begin{equation}
    c_\mathrm{vir}  = 7.7 D^{0.9}\nu^{-0.29}\label{eq:dutton_macio}
\end{equation}
where $D$ is the growth factor.

We also have implemented the \cite{Dutton:2014xda} formula:
\begin{equation}
    \log_{10}c_\mathrm{vir} = a + b \log_{10}(M_\mathrm{vir}/10^{12})
\end{equation}
with $m_\mathrm{vir}$ in $M_\mathrm{sun}/h$ and 
\begin{eqnarray}
    a & =& 0.537 + (1.025-0.537)\exp(-0.718 z^{1.08})\nonumber\\
    b & = &-0.097 + 0.024 z\nonumber
\end{eqnarray}

We also have the option of passing a fixed value for the concentration, which is useful for testing purposes. 

More physical models, based explicitly on the matter power spectrum as \cite{Diemer:2014gba,Diemer:2018vmz} will become available in \textsc{class\_sz} in the near future.

\subsection{Conversion Between Mass Definitions}
 Although here we exclusively used the $m_{200\mathrm{c}}$ mass definition, we explain how to convert between  mass definitions as it can be useful for comparison with other analyses or to implement different mass functions, HOD's and tracer profiles. To convert between $m_{\Delta}$ and $m_{\Delta^{\prime}}$, we  compute $m_{\Delta^{\prime}}=\int_0^{r_{\Delta^\prime}}\mathrm{d}r4\pi r^2 \rho_{_\mathrm{NFW}}(r)$ with the NFW profile defined in terms of $r_s=r_\Delta/c_\Delta$. Its expression is equivalent to
\begin{equation}
\frac{m_{\Delta^{\prime}}}{m_{\Delta}}-\frac{f_{_\mathrm{NFW}}(c_{\Delta})}{f_{_\mathrm{NFW}}(c_{\Delta}r_{\Delta^\prime}/r_{\Delta})}=0\quad\mathrm{with}\quad r_{\Delta^\prime} = \left[3 m_\Delta/(4\pi  \Delta^\prime \rho_\mathrm{crit}(z)) \right]^{1/3}
\end{equation}
and 
\begin{equation}
    f_{_\mathrm{NFW}}(x)=[\ln(1+x)-x/(1+x)]^{-1}
\end{equation}
which can be solved for $m_{\Delta^{\prime}}$ with a root-finding algorithm. In \verb|class_sz| we use Brent's method \citep{brent2002algorithms}.

\begin{figure*}
    \includegraphics[width=1.\columnwidth]{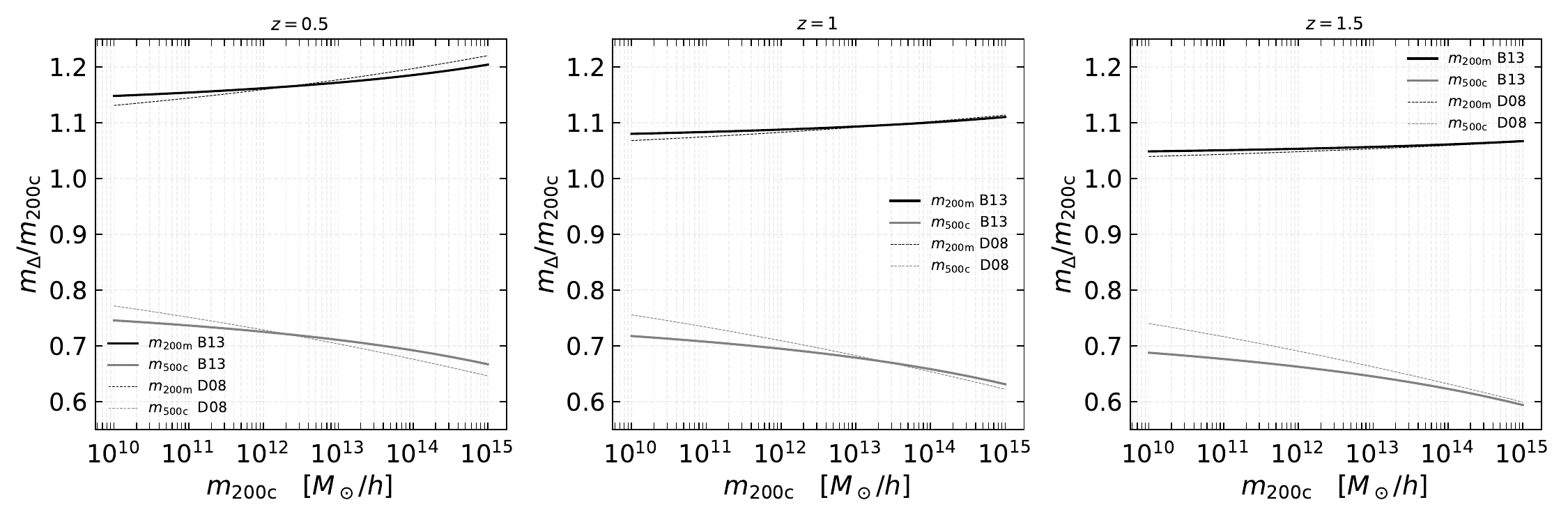}
    \vspace{-0.2cm}
    \caption{mass conversion.}
    \label{fig:mconv}
\end{figure*}

For reference, we show the conversions between $m_\mathrm{200c}$, $m_\mathrm{200m}$ and $m_\mathrm{500c}$ at three redshifts for the \cite{2013ApJ...766...32B} and \cite{Duffy_2008} concentration-mass relations in Figure~\ref{fig:mconv}. Overall, $m_\mathrm{200m}$ is $\approx 5-20\%$ larger than  $m_\mathrm{200c}$, while $m_\mathrm{500c}$ is $\approx 20-40\%$ lower than  $m_\mathrm{200c}$. The \cite{2013ApJ...766...32B} and \cite{Duffy_2008} agree well at high masses but differ substantially at low masses. 

In the \textsc{Websky} simulations \citep{websky_Stein_2020}, the mass conversion between 200m and 200c is simplified:
\begin{equation}
    m_{200c} = \Omega_m^{0.35} m_{200m}
\end{equation}
we also have this implemented. 

\subsection{Sub-halo Mass Function}\label{ss:subhmf}

We implemented two version of the subhalo mass function. These are currently used in \textsc{class\_sz} for CIB calculation. The \cite{2010ApJ...719...88T} function is 
\begin{equation}
    \frac{dN}{d\ln m_\mathrm{sub}} = 0.30\mu^{-0.7}\exp\left[-9.9\mu^{2.5}\right]\label{eq:subhmf1}
\end{equation}
where $\mu=m_\mathrm{host}/m_\mathrm{sub}$, $m_\mathrm{host}$ is the host halo mass and  $m_\mathrm{sub}$ is the sub-halo mass (see their Eq.~12).
The second subhalo mass function currently available in \textsc{class\_sz} is the \cite{Jiang:2013kza} formula, which reads (see their Eq.~21):
\begin{equation}
    \frac{dN}{d\ln m_\mathrm{sub}} = \left[ \gamma_1 \mu^{\alpha_1}+\gamma_2 \mu^{\alpha_2} \right]\exp[-\beta \mu^\zeta]\label{eq:subhmf2}
\end{equation}
where parameter values are $(\gamma_1,\alpha_1,\gamma_2,\alpha_2,\beta,\zeta)=(0.13, -0.83, 1.33,
-0.02, 5.67, 1.19)$. Note that this formula is the one used in the \textsc{websky} simulations \citep{websky_Stein_2020} for their CIB model.

\section{LSS Tracers and Halo-Model Power Spectra}\label{sec:tracers}

\subsection{Power Spectra}\label{ss:ps}

Let $X$ and $Y$ be two LSS tracers with radial profiles  $u^X$ and $u^Y$.  Their 3D power spectrum is defined via
\begin{equation}
    \langle X(\bm{k}_1)Y(\bm{k}_2)\rangle=(2\pi)^2\delta(\bm{k}_1+\bm{k}_2)P_{XY}(k_1)
\end{equation}
The halo model power spectrum for the RHS is $P_{XY}^\mathrm{hm} = P_{XY}^\mathrm{1h}+P_{XY}^\mathrm{2h}$ where the  1-halo term, $P_{XY}^\mathrm{1h}$, accounts for correlations between points within the same halo, and the 2-halo term, $P_{XY}^\mathrm{2h}$, accounts for correlations between points in distinct halos. Each term can be expressed  using the 3D Fourier transforms of the profiles. All the profiles we consider are radially symmetric, therefore Fourier transforms reduce to Hankel transforms. The Hankel transform of a radial profile is given by
\begin{equation}
    \hat{u}(k)=4\pi \int_0^\infty \mathrm{d}r r^2 j_0(kr)\mathrm{H}(r_\mathrm{out}-r)u(r)\quad\mathrm{where}\quad j_0(x)=\sin(x)/x\label{eq:hatu}
\end{equation}
is the spherical Bessel function of order 0 and where we added the Heaviside step function $\mathrm{H}$ in order to truncate the profile at some radius $r_\mathrm{out}$. 

In \textsc{class\_sz} we have two methods to evaluate Eq.~\eqref{eq:hatu}. The QAWO implementation of GSL and the FFTLog implementation. FFTLog is much faster.

Note that in the $k\rightarrow 0$ limit, $\hat{u}$ is the volume average of $u$ within a sphere of radius $r_\mathrm{out}$.
Explicitly, the 1- and 2-halo terms are 
\begin{equation}
    P_{XY}^\mathrm{1h} = \langle \hat{u}^X\hat{u}^Y\rangle_n\quad\mathrm{with}\quad
    P_{XY}^\mathrm{2h} = \langle b^{(1)}\hat{u}^X\rangle_n\langle b^{(1)}\hat{u}^Y\rangle_n P_{L}
\end{equation}
where $b^{(1)} = b^{(1)}(m,z)$ is the first-order halo bias of Eq.~\eqref{eq:b1tink}.

  In general, for two fields $X$ and $Y$ there is a contribution to the 1-halo power spectrum coming from  correlated fluctuations so that $\langle\hat{u}^X\hat{u}^Y\rangle =(1+r) \langle\hat{u}^X\rangle\langle\hat{u}^Y\rangle$ with $r\neq0$ (here the angle brackets are to be understood as ensemble-average at fixed mass and redshift). Although, we can often assume $r\ll1$ since it is unlikely that two different fields $X$ and $Y$ would fluctuate in a correlated way. See e.g. \cite{Koukoufilippas:2019ilu} for an analysis where they take this  into account. This can be tested on simulations. 

\subsection{Profiles}

In this subsection we describe the profiles of the matter field tracers. Here, by profile we refer to 3d spherically symmetric, i.e., radial functions that describe how a given tracer is distributed around a halo. In \ref{sssec:nfw} we describe the NFW profile, in \ref{ref:sssicmdens} the ICM gas density profile, in \ref{sss:icmpress} the ICM gas pressure profile, in \ref{sss:te}  the gas temperature, in \ref{sss:galhod} galaxie counts, in \ref{sss:gallens} galaxy lensing, in \ref{sss:cmblens} CMB lensing and in \ref{sss:cib} the CIB.

\subsubsection{Dark Matter Density}\label{sssec:nfw}

The Navarro-Frenk-White density profile is defined as $\rho_{_\mathrm{NFW}}(r)=\rho_{\mathrm{m},0} u^\mathrm{_{NFW}}(r)$ with 
\begin{equation}
    u^\mathrm{_{NFW}}(r) =\frac{\rho_s}{\rho_{\mathrm{m},0}} \frac{1}{\frac{r}{r_s}\left(1+\frac{r}{r_s}\right)^2}\quad\mathrm{where}\quad r_s = r_\Delta/c_\Delta.
\end{equation}
Here, the scale radius $r_s$ is defined in terms of characteristic radius and concentrations $r_\Delta$ and $c_\Delta$. These depend on the halo mass $m_\Delta$. The concentration is often computed with a relation calibrated on simulations \citep[e.g.,][]{Duffy_2008,2013ApJ...766...32B}. In this paper, we use the \cite{2013ApJ...766...32B} relation. It is common to take $r_\Delta$ as the radius of the spherical region of mass $m_\Delta$ within which the density is $\Delta$ times the critical density,
at redshift $z$. Thus,
\begin{equation}
    r_\Delta = \left[3 m_{\Delta}/(4\pi  \Delta_\mathrm{crit} \rho_\mathrm{crit}(z)) \right]^{1/3}.\label{eq:rd}
\end{equation}
Common values for $\Delta_\mathrm{crit}$ are $180$, $200$ and $500$. Instead of using the critical density as a reference, one can use the matter density which means replacing $\Delta_\mathrm{crit}$ by $\Delta_m =\Delta_\mathrm{crit}\Omega_\mathrm{m}(z)$, where $\Omega_\mathrm{m}(z)=\rho_\mathrm{m}(z)/\rho_\mathrm{crit}(z)$. Another common choice for these definitions are the virial mass and radius, which amount to replacing $\Delta_\mathrm{crit}$ by  
\begin{equation}
\Delta_c(z) = 18\pi^2 +82 x - 39 x ^2
\end{equation}
with $x=\Omega_m(z)-1$ as given in Eq.~6 of \cite{1998} (case $\Omega_r=0$). 

By definition, we  have $m_\Delta = \int_0^{r_\Delta}\mathrm{d}r4\pi r^2 \rho_{_\mathrm{NFW}}(r)$, which yields 
\begin{equation}
    \rho_s = \frac{m_\Delta}{4\pi r_s^3 }f_{_\mathrm{NFW}}(c_\Delta).\label{eq:rho_s}
\end{equation}

The Fourier transform of $u^{_\mathrm{NFW}}$ truncated at $r_\mathrm{cut}=\lambda r_\Delta$ (see Eq. \ref{eq:hatu}) has an analytical expression given by \citep{Scoccimarro:2000gm}:
\begin{equation}
    \hat{u}^\mathrm{_{NFW}}(k) = \frac{m_{\lambda r_\Delta}}{\rho_{\mathrm{m},0}}    \left([\mathrm{Ci}((1+\lambda c_\Delta)q)-\mathrm{Ci}(q)] \cos q
    +    [\mathrm{Si}((1+\lambda c_\Delta) q)-\mathrm{Si}(q)]\sin q
    -\frac{\sin(\lambda c_\Delta q)}{(1+\lambda c_\Delta)q}\right)f_{_\mathrm{NFW}}(\lambda c_\Delta)\label{eq:nfwtrunc}
\end{equation}
where $m_{\lambda r_\Delta}$ is the mass within $\lambda r_\Delta$ (i.e., $m_\Delta$ for $\lambda=1$) and where $\mathrm{Ci}(x)=\int_x^{\infty}\mathrm{d}t\cos(t)/t$ and $\mathrm{Si}(x)=\int_0^x\mathrm{d}t\sin (t)/t$ are the cosine and sine integrals, and $q = (1+z)kr_s = \ell/\ell_s$.\footnote{In the last equality we defined  $\ell_s=d_A/r_s$ with angular diameter distance $d_A = \chi/(1+z)$, and traded wavenumber for multipole according to $k\chi=\ell+1/2$.} Note the $(1+z)$ factor. \footnote{see also appendix of https://arxiv.org/pdf/1505.07833.pdf for the analytical formula}

Noting that $q\propto m_\Delta^{1/3}$, the asymptotic behaviors of $\hat{u}^\mathrm{_{NFW}}$ when $k\rightarrow0$ or $m_\Delta\rightarrow0$ are the same, namely
\begin{equation}
    \lim_{q\rightarrow 0}\hat{u}^\mathrm{_{NFW}}_k = \frac{m_{\lambda r_\Delta}}{\rho_\mathrm{m,0}}.\label{eq:q0limnfw}
\end{equation}
This is an important property which implies that in the low-$k$ regime $\langle \hat{u}^\mathrm{_{NFW}}_k\rangle_n\approx1$ (when $\lambda =1$), crucial to maintaining the halo model consistency conditions of Eq.~\eqref{eq:hmconsist}.

 When CDM density is assumed to follow the NFW profile, the halo model matter power spectrum at $z$ is $P^\mathrm{hm}_{\delta_\mathrm{m}\delta_\mathrm{m}}=P_{\delta_\mathrm{m}\delta_\mathrm{m}}^\mathrm{1h}+P_{\delta_\mathrm{m}\delta_\mathrm{m}}^\mathrm{2h}$ with 
 \begin{equation}
      P_{\delta_\mathrm{m}\delta_\mathrm{m}}^\mathrm{1h}(k,\chi) = \langle \hat{u}^{_\mathrm{NFW}}_{k}\hat{u}^{_\mathrm{NFW}}_{k}\rangle_n\quad\mathrm{and}\quad P_{\delta_\mathrm{m}\delta_\mathrm{m}}^\mathrm{2h}(k,\chi) = \langle b^{(1)}\hat{u}^{_\mathrm{NFW}}_{k} \rangle_n^2 P_{L}(k,\chi)\label{eq:pkhm}
 \end{equation}
where $P_L$ is the linear matter power spectrum. In the low-$k$ limit we have $\langle b^{(1)}\hat{u}^{_\mathrm{NFW}}_{k} \rangle_n\rightarrow 1$ (by construction and consistency) so that $P_{\delta_\mathrm{m}\delta_\mathrm{m}}^\mathrm{2h}\sim P_L$, whereas  $P_{\delta_\mathrm{m}\delta_\mathrm{m}}^\mathrm{1h}\sim \langle m_\Delta^2/\rho_\mathrm{m,0}^2\rangle$ which is independent of $k$.

On ultra-large scales the power spectrum should grow as $k^4$. Hence, we follow \cite{mead2015}  and add an exponential damping to the 1-halo term, of the form:
\begin{equation}
   f(k) = 1-\exp(-k^2/k_\mathrm{damp}^2)\label{eq:damp}
\end{equation}
with $k_\mathrm{damp}=0.01 \mathrm{Mpc}^{-1}$.

In principle, the 2-halo term should also be damped in the non-linear regime due to perturbative effects. We do not  account for this subtlety yet and refer to \cite{mead2021} and \cite{phils2020} for details on these aspects, for now.

\begin{figure*}
    \includegraphics[width=1.\columnwidth]{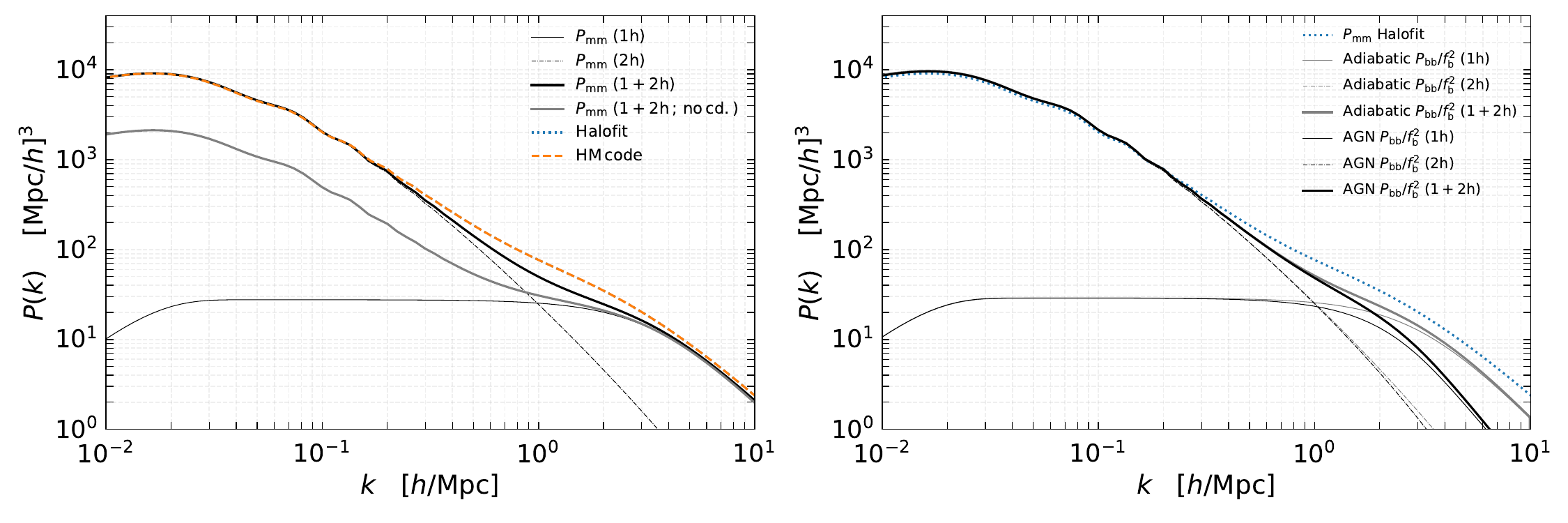}
    \vspace{-0.5cm}
    \caption{The matter porwer spectrum in the halo model. At high k on the left the match is roughly ok, but difficult to get better than that. See notebook \href{https://github.com/CLASS-SZ/notebooks/blob/main/class_sz_baryonspk.ipynb}{https://github.com/CLASS-SZ/notebooks/class\_sz\_baryonspk.ipynb}.}
    \label{fig:pkshm}
\end{figure*}

At low-$k$ the 2-halo term dominates and we have $P^\mathrm{hm}_{\delta_\mathrm{m}\delta_\mathrm{m}}\sim P_L$. The halo model matter power spectrum is plotted on the left panel of Figure~\ref{fig:pkshm} against the \verb|halofit| and \verb|hmcode| (which are nearly identical). The mismatch between the halo model power spectrum and the N-body calibrated formulas (\verb|halofit| and \verb|hmcode|) in the transition regime between the 2-halo and 1-halo term is a well-known short-coming of the halo model. This issue has been addressed in several manners. For instance, \cite{mead2015} suggest using $P^{\mathrm{hm}}_{\delta_\mathrm{m}\delta_\mathrm{m}}=[(P^{\mathrm{1h}}_{\delta_\mathrm{m}\delta_\mathrm{m}})^\alpha+(P^{\mathrm{2h}}_{\delta_\mathrm{m}\delta_\mathrm{m}})^\alpha)]^{1/\alpha}$ where $\alpha$ is a free parameter. Another approach, proposed in \cite{phils2020} is to use perturbation theory at one-loop order in the modeling of the 2-halo term, amounting to replace $P_{L}$ in Eq.~\eqref{eq:pkhm} by %\begin{equation}
    $P_{NL}=P_{L}+P_\mathrm{spt}+P_\mathrm{ct}$
%\end{equation}
where $P_\mathrm{spt}=P_{22}+P_{13}$ with $P_{22}$ and $P_{13}$ resulting from higher-order terms associated with the $F_2$ and $F_3$ coupling kernels  \citep[see, e.g.,][]{bernardeau2002} and $P_\mathrm{ct}(k)=-c_\mathrm{s}^2k^2P_{L}(k)$ with $c_\mathrm{s}$ a free parameter of the model. What these approaches have in common is inclusion of the  \textit{nuisance} parameters to the model. The extra nuisance parameters then need to be calibrated on simulations or marginalized over.  The current version of \textsc{class\_sz} does not have this extra pieces of modelling for the transition regimes. This will be implemented soon.

{\bf{Dark Matter Density 2-halo Term.}}  We compute the 2-halo term of the density profile in real space using 

\begin{equation}
    I_{2h}(k,z) = \langle b^{(1)}\hat{u}_{k}\rangle_n 
\end{equation}
which we tabulate on a $(z,k)$ grid, in parallel. We then FFTLog the product with linear Pk, as
\begin{equation}
    \rho_{2h}(r,z;m)(k,z) = \frac{1}{2\pi^2}b^{(1)}(m,z)\int d k k^2 P_L(k,z) I_{2h}(k,z)j_0(kr)\label{eq:dens2hterm}
\end{equation}
This is done at each $z$, in parallel.  Importantly, at low-$k$, the integral $I_{2h}$ is constant and should be 1. In the limit $r$ goes to 0, $\rho_{2h}$ is is  $\int dk k^2P_L(k,z)I_{2h}(k,z)$. For the NFW profile, this integral converges because at high-$k$ the integral falls of as $k^{-2}$.

\subsubsection{ICM Density}\label{ref:sssicmdens}

For the gas electron density profile, $\rho_\mathrm{gas,free}$, we consider two parameterizations. First, the NFW formula \citep{Navarro_1997} rescaled by the baryon fraction $f_\mathrm{b}$, i.e.,
\begin{equation}
    \rho_{\mathrm{gas,free}}(r)=f_\mathrm{b}f_\mathrm{free}\rho_{_\mathrm{NFW}}(r)
\end{equation}
where $\rho_{_\mathrm{NFW}}(r)$ is the usual NFW profile (see previous subsection). Second, a generalized NFW (gNFW) formula, following   \cite{Battaglia_2016}:
\begin{equation}
    \rho_{\mathrm{gas,free}}(r)=f_\mathrm{b}f_\mathrm{free}\rho_\mathrm{crit}(z) C \left(\frac{r}{x_\mathrm{c}r_\mathrm{200c}}\right)^\gamma \left[1+\left(\frac{r}{x_\mathrm{c}r_\mathrm{200c}}\right)^\alpha\right]^{-\frac{\beta+\gamma}{\alpha}}, \label{eq:ugnfw}
\end{equation}
where $r_\mathrm{200c}$ is the characteristic radius associated with the overdensity mass $m_\mathrm{200c}$ (see Eq.~\ref{eq:rd}), with $x_c=0.5$ and $\gamma=-0.2$ kept fixed throughout the paper and with mass and redshift dependent parameters $C,\alpha,\beta,\gamma$, such that 
\begin{equation}
    p= A_0\left(\frac{m_\mathrm{200c}}{10^{14}M_\odot}\right)^{A_m}\left(1+z\right)^{A_z}\quad\mathrm{for}\quad p \in \{C,\alpha,\beta,\gamma\}.
\end{equation}
For $A_0,A_m,A_z$ we use the best-fit values from \cite{Battaglia_2016} reported in Table \ref{tab:gnfwb16}, corresponding to either the \textit{AGN feedback} model (that is our fiducial assumption) or the \textit{Adiabatic} model. Note that the NFW profile is a subcase of the gNFW formula, when parameters are set to $x_c=1/c_{200c}$, $\gamma=-1$, $\alpha=1$, $\beta=3$, and $C=\rho_{s}/\rho_\mathrm{crit}(z)$, where $c_{200c}$ is the concentration computed with the \cite{2013ApJ...766...32B} relation and $\rho_s$ is the normalization of the NFW profile defined in Eq.~\eqref{eq:rho_s}.
\begin{table}
\begin{centering}
\begin{tabular}{c|ccc|ccc}
 & \multicolumn{3}{c}{\textit{AGN feedback}} & \multicolumn{3}{c}{\textit{Adiabatic}}\tabularnewline
$p$ & $A_0$ & $A_{m}$ & $A_{z}$ & $A_0$ & $A_{m}$ & $A_{z}$\tabularnewline
\hline 
$C$ & $4\times10^{3}$ & $0.29$ & $-0.66$ & $1.9\times10^{4}$ & $0.09$ & $-0.95$\tabularnewline
$\alpha$ & $0.88$ & $-0.03$ & $0.19$ & $0.70$ & $-0.017$ & $0.27$\tabularnewline
$\beta$ & $3.83$ & $0.04$ & $-0.025$ & $4.43$ & $0.005$ & $0.037$\tabularnewline
\end{tabular}
\par\end{centering}
\caption{Best-fit values of the parameters of the generalized NFW gas density profile formula fit to simulations from  \cite{Battaglia_2016}. \textit{Adiabatic} corresponds to simulations whose sub-grid model has only gravitational heating. \textit{AGN feedback} corresponds to a sub-grid model with radiative cooling, star formation, supernova feedback, cosmic rays, and AGN feedback. The gas density profile is computed using these parameters in Eq.~\eqref{eq:ugnfw}. \citep[See][for details.]{Battaglia_2016}}
\label{tab:gnfwb16}
\end{table} 
With this, we compute the Fourier transform of the profile as 
\begin{equation}
\hat{u}^\mathrm{e}_k=4\pi \int_0^\infty \mathrm{d}r r^2 j_0(kr)\mathrm{H}(r_\mathrm{cut}-r)u^\mathrm{e}(r)\quad\mathrm{with}\quad j_0(x)=\frac{\sin(x)}{x}\quad\mathrm{and}\quad u^\mathrm{e}(r)=\frac{\rho_\mathrm{gas,free}(r)}{\rho_\mathrm{m,0}},\label{eq:hatub}
\end{equation}
where $\mathrm{H}$ is the Heaviside step function (which truncates the profile at $r_\mathrm{cut}$) and where we used the fact that the profiles are radially symmetric to write the Fourier transform as a Hankel transform. In general, it is necessary to truncate the density profiles because their volume integrals do not converge or may have support at unphysically large radii. For the NFW profile, we set the truncation radius to $r_\mathrm{cut}=r_\mathrm{200c}$. For the gNFW profile we require $r_\mathrm{cut}$ to be such that the enclosed gas mass is the same as in the NFW case, i.e., $f_\mathrm{b}m_{200c}$. We then find $r_\mathrm{cut}$ numerically with Brent's method \citep{brent2002algorithms}, solving
\begin{equation}
F(r_\mathrm{out};m_\mathrm{200c},z)=0\quad\mathrm{with}\quad F(r_\mathrm{out};m_\mathrm{200c},z)=4\pi \int_0^\mathrm{r_\mathrm{cut}}\mathrm{d}r r^2\rho_\mathrm{gas}(r;m_\mathrm{200c},z)-f_\mathrm{b}m_\mathrm{200c},\label{eq:brent}
\end{equation}
where we wrote the mass and redshift dependence explicitly to emphasize the fact that this operation is done at each mass and redshift. The method is illustrated in \cite{Bolliet:2022pze}.

We note that halo models based on the \cite{Battaglia_2016} gas density profile parameterization have been used in multiple previous analyses \citep[e.g.,][]{Smith:2018bpn, Munchmeyer:2018eey,Cayuso:2021ljq,Roy:2022muv},  which computed Fourier and harmonic space two-point functions. In principle, our results could be checked against these studies. One notable difference is that previous works often truncate the gas density profile at $r_\mathrm{200c}$, and rescale its amplitude by a factor such that the enclosed mass is $m_\mathrm{200c}$.  We argue that our truncation method is more consistent, as it preserves the total gas mass but does not alter the density as a function of radius.

We also have implemented the Baryon Correction Model \citep{Schneider:2015wta} with the updated model presented in \cite{Schneider:2018pfw} and \cite{Giri:2021qin}. This model is based on the following parameterization \citep{Giri:2021qin}:

\begin{equation}
    \rho_\mathrm{gas} = \rho_\mathrm{gas,0}\frac{
    f_b-f_\mathrm{star}(m)}{[1+10(\frac{r}{r_\mathrm{vir}})]^{\beta(m)}[1+(\frac{r}{\theta_\mathrm{ej}r_\mathrm{vir}})^\gamma]^\frac{\delta-\beta(m)}{\gamma}}
\end{equation}
where $f_\mathrm{star}$ is the stellar mass fraction, parameterized as 
\begin{equation}
    f_\mathrm{star}=0.055\left(\frac{m}{m_s}\right)^{-\eta_\mathrm{star}}
\end{equation}
with $m_s = 2.5\times 10^{11}M_\mathrm{sun}/h$ and $\eta_\mathrm{star}$ a free parameter of the model. The power law is 
\begin{equation}
    \beta(m) = \frac{3\left(m/m_c\right)^\mu}{1+(m/m_c)^\mu}
\end{equation}
with $m_c$ and $\mu$ free parameters of the model. Also $\gamma,\theta_{ej}$ and $\delta$ are free parameters. The mass here is $m_\mathrm{200c}$. For $m_c$ there is also a redshift dependent parameterization:
\begin{equation}
    \log_{10}m_c(z)= \log_{10}m_{c,0}(1+z)^{\nu_{\log_{10}m_c}}
\end{equation}

Our fiducial values are $\log_{10}(M_{c,0}) = 13.25$, $\theta_{\textrm{ej}}= 4.711$, $\eta_\mathrm{star} = 0.2$,  $\delta=7.0$, $\mu=1$, $\gamma=2.5, \nu_{\log_{10}m_c} = -0.038$ which reproduce the BAHAMAS \citep{McCarthy2017} simulation power spectrum\footnote{see the README of \href{https://github.com/sambit-giri/BCemu}{https://github.com/sambit-giri/BCemu}} at $z=0$.

The electron power spectrum is computed in the halo model using the Fourier transform of the gas density profile, $\hat{u}^\mathrm{e}$ of Eq.~\eqref{eq:hatub}, as 
 \begin{equation}
      P_{\delta_\mathrm{e}\delta_\mathrm{e}}^\mathrm{1h}(k,\chi) = \langle \hat{u}^{\mathrm{e}}_{k}\hat{u}^{\mathrm{e}}_{k}\rangle_n\quad\mathrm{and}\quad P_{\delta_\mathrm{e}\delta_\mathrm{e}}^\mathrm{2h}(k,\chi) = \langle b^{(1)}\hat{u}^{\mathrm{e}}_{k} \rangle_n^2 P_{L}(k,\chi).\label{eq:pkb}
 \end{equation}
The gas density profile is normalized such that 
\begin{equation}
\lim_{k\rightarrow 0}  \hat{u}^{\mathrm{e}}_{k} =  f_\mathrm{b} f_\mathrm{free}\frac{m_\Delta}{\rho_\mathrm{m,0}},
\end{equation}
and there for in the low-$k$ limit we have $P_{\delta_\mathrm{e}\delta_\mathrm{e}}^\mathrm{hm}\approx P_{\delta_\mathrm{e}\delta_\mathrm{e}}^\mathrm{2h}\approx f_\mathrm{b}^2P_L$, irrespective of the gas density profile assumption. In the high-$k$ regime, the difference between the gas density profile and the NFW profile can be significant and therefore the scale dependence of the baryon power spectrum can depart from that of the non-linear matter power spectrum. This is illustrated in the middle panel of Figure~\ref{fig:pkshm}.

{\bf{Gas Density 2-halo Term.}} We compute the 2-halo term of the density profile in real space using 
the same formulas as for the NFW profile, except we trade NFW for baryons. For the baryon profiles, the $k$-integral integral converges because at high-$k$ the integrand falls of more rapidly than $k^{-2}$.

\begin{figure*}
\includegraphics[width=1.\columnwidth]{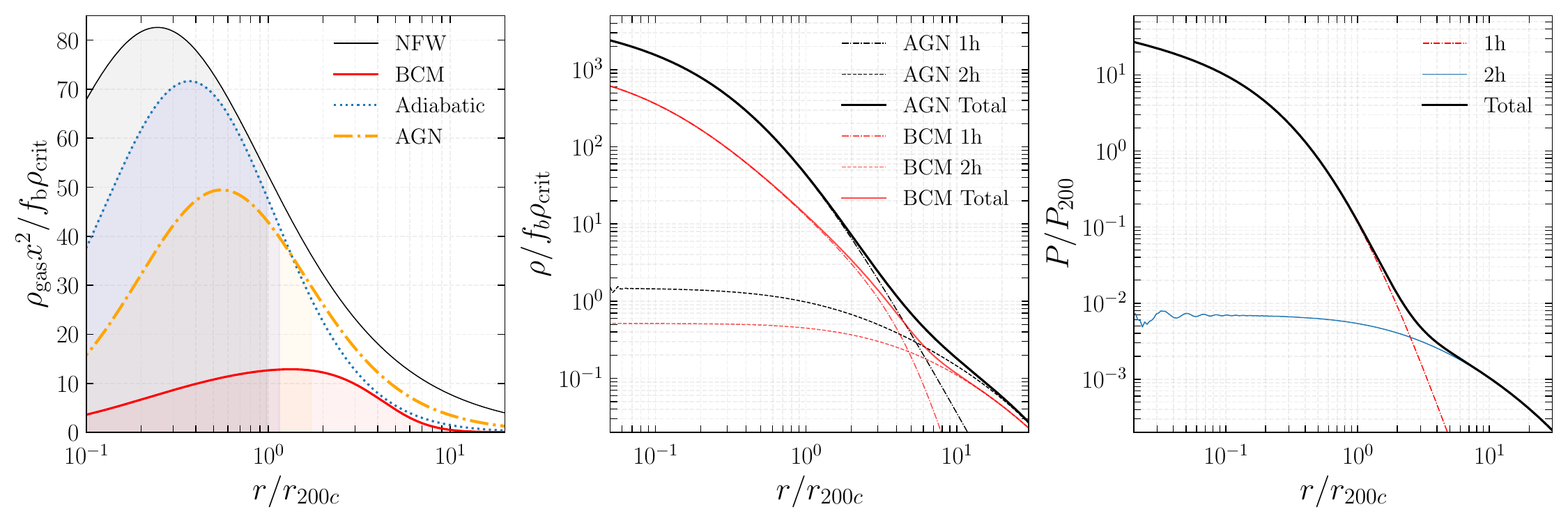}
    \centering
    \vspace{-0.2cm}
    \caption{Example of ICM pressure and density profiles. Note that the battglia et al profile fit has 2-halo contribution.}
    \label{fig:profilesrealspace}
\end{figure*}

\subsubsection{ICM Pressure}\label{sss:icmpress}
Defining $\sigma_\mathrm{T}$ as the Thomson cross-section and $m_e$ as the electron mass, the electron pressure, \textit{i.e.}, Compton-$y$, multipole-space kernel $ u_\ell^y (M,z)$ as a function of halo mass $M$ and redshift $z$ is given by~(\textit{e.g.},~\cite{KomatsuSeljak2002,HillPajer2013})
\begin{equation}
    u_\ell^y(M,z) = \frac{\sigma_\mathrm{T}}{m_\mathrm{e} c^2} \frac{4\pi r_s}{\ell_s^2} \int_{x_{\rm{min}}}^{x_\mathrm{max}} \mathrm{d}x \, x^2 \, \mathrm{sinc}(w_\ell x)  P_e(x,M,z),
    \quad \mathrm{with}\quad w_\ell =\frac{\ell+\tfrac{1}{2}}{\ell_s} \,, 
\label{eq:uy}
\end{equation}
where the mass-dependent $r_s$ and $\ell_s$ are the characteristic radius and the characteristic multipole of the pressure profile, related via:
\begin{equation}
    \ell_s=\frac{d_A}{r_s}=\frac{1}{(1+z)}\frac{\chi}{r_s} \,,
\end{equation}
where $d_A = \chi /(1+z)$ is the angular diameter distance to redshift $z$. The integration variable $x=r/r_s$ is the ratio of the distance from the center of the halo $r$ and its characteristic radius $r_s$. The pressure profile $P_e(x,M,z)$ is a quantity that parameterizes the radial pressure, and there exist various choices for $P_e$ in the literature. We set $x_{\rm{min}} = 10^{-5}$ and $x_{\rm{max}} = 4$. 

There are several options for the pressure profile in \verb|class_sz|, including the generalized Navarro-Frenk-White (GNFW), \cite{Battaglia_2012}, \cite{Planck2013SZ}, and \cite{Arnaud2010} profiles. The GNFW profile (\cite{NFW:1996, NFW1997, Nagai:2007mt}) has the formula for the pressure profile $P_e(x,M,z)$
\begin{equation}
     P_e(x,M,z) = P_{\Delta} P_0 \left(\frac{x}{x_c}\right)^{\gamma^y} \left[1+\left(\frac{x}{x_c}\right)^{\lambda^y}\right]^{-\beta^y} \,,
\end{equation}
where
\begin{equation}
     P_{\Delta} = \frac{G \Delta M_{\Delta}\rho_c(z) \Omega_b}{2 R_{\Delta} \Omega_m}
\end{equation}
for any spherical overdensity definition $\Delta$ relative to the critical density $\rho_c$. In the \cite{Battaglia_2012} profile, and the literature after \cite{pandey2021crosscorrelation, pandey2020},    the generalized NFW formula is used, setting $\lambda^y = 1.0$ and $\gamma^y = -0.3$. They parameterize $P_0$, $x_c$, and $\beta^y$ according to a scaling relation. Letting $X$ denote any of the parameters $P_0$, $x_c$, and $\beta^y$, this parameter $X$ can then be written as
\begin{equation}
    X (M_{\Delta})  = X_{\Delta} \left(\frac{M_{\Delta}}{10^{14} M_\odot}\right)^{\alpha^y} (1+z)^{\omega^y}, 
\end{equation}
where $X_{\Delta}$ is the value of that parameter at $M_{\Delta} = 10^{14} M_\odot$ at $z = 0$, and $\alpha^{y}$ and $\omega^y$ are free parameters. We set those parameters to the standard \cite{Battaglia_2012} values (the ``AGN feedback model at $\Delta=200$'' from their Table~1), which are also summarized in Table~\ref{table:pp_b12} here.
\begin{table}
\centering
\setlength{\tabcolsep}{10pt}
\renewcommand{\arraystretch}{1.5}
    \begin{tabular}{c|c|c|c|c}

         Parameter &  Parameter description & $X_{\Delta}$ & $\alpha^y$ & $\omega^y$ \\
         \hline \hline
         $P_0$ &   Amplitude of the pressure profile &18.1 & 0.154 & -0.758 \\
         \hline
          $x_c$ & Core scale of the pressure profile& 0.497 &  -0.00865 & 0.731 \\
        \hline
        $\beta^y$ & Shape of the pressure profile & 4.35 & 0.0393 & 0.415 \\

    \end{tabular}
    \caption{Values of the AGN feedback \cite{Battaglia_2012} pressure profile parameters.}
    \label{table:pp_b12}
\end{table}

Other pressure profiles include the \cite{Planck2013SZ} and \cite{Arnaud2010} pressure profiles. By combining SZ and X-ray profiles into a joint fit to the generalized NFW profile, \cite{Planck2013SZ} found best fit parameters  [$P_0$,$c_{500}$,$\gamma$,$\alpha$,$\beta$ ]  =  [6.41, 1.81, 0.31, 1.33, 4.13], where $P_0$ defines the normalization, $c_{500}$ is the concentration parameter at $R_{500}$, and $\gamma$, $\alpha$, and $\beta$ are the slopes in the central, intermediate, and outer regions, respectively. Specifically, these parameters are related to the pressure profile via \citep{Nagai:2007mt}
\begin{equation}
    \frac{P(r)}{P_{500}} = \frac{P_{0} } { (c_{500}x)^{\gamma}\left[1+(c_{500}x)^\alpha\right]^{(\beta-\gamma)/\alpha}}
\end{equation}
with $x \equiv \frac{r}{r_s}$ and $r_s = \frac{r_{500}}{c_{500}}$. In that work, $\gamma$ was fixed to 0.31 and the other four parameters were free parameters. In \cite{Arnaud2010} these best fit parameters were found to be  [$P_{0}$,$c_{500}$,$\gamma$,$\alpha$,$\beta$] = [$8.403~{h_{70}^{-3/2}}$, 1.177, 0.3081, 1.0510, 5.4905].

% \begin{figure}[t]
%     \centering
%     \includegraphics[scale=0.6]{yy.pdf}
%     \caption{ The caption is broken} %Comparison of the Compton-$y$ auto-correlation measurements from 
%     %\textit{Planck} 2015 
    
%     %\citep{planck2015_tsz} (orange dots with error bars), ACT 2020 \cite{Choi_2020} (at $\ell=3000$), and SPT 2021 along with the theoretical predictions computed using the Battaglia} 
    
%     %\textit{et al.}~(2012) }%\cite{Battaglia_2012} pressure profile (solid lines) used in this work}
    
%     %\cite{Choi_2020} (at $\ell=3000$), and SPT 2021 \cite{Reichardt_2021} (at $\ell=3000$), along with the theoretical predictions computed using the Battaglia \textit{et al.}~(2012) \cite{Battaglia_2012} pressure profile (solid lines) used in this work.}
%     \label{fig:yy}
% \end{figure}

To obtain the auto- and cross-correlations of the tSZ field at some frequency, we multiply the auto- and cross-correlations of the Compton-$y$ field by the standard tSZ spectral function at each frequency $\nu$, \textit{i.e.}:
\begin{equation}
    g(\nu) = T_{\rm{CMB}} \left(x \, {\rm{coth}}\left(\frac{x}{2}\right) -4\right) , 
    \label{eq:g_nu_tsz} \,
\end{equation}
where $x = h_P \nu /(k_B T_{\rm{CMB}} )$, with $h_P$ the Planck constant, $k_B$ the Boltzmann constant, and $T_{\rm{CMB}}$ the CMB temperature today, for which we take $T_{\rm{CMB}}=2.726$ K.

{\bf{Gas Pressure 2-halo Term.}} We compute the 2-halo term of the pressure profile the same way as for the density profile (see Subsection \ref{ref:sssicmdens} and Eq.~\ref{eq:dens2hterm}), except we trade density for pressure.

\subsubsection{ICM Temperature}\label{sss:te}
We have implemented several electron temperature scaling relations. This includes the hydrostatic equilibrium relation \citep[e.g.,][]{2005A&A...441..893A,2007ApJ...668....1N,Erler:2017dok},
%\jc{\sc [check again from Arnaud paper]}
%---------------------------
\begin{equation}
    k_B T_{\rm e} \approx 5 \,\mathrm{keV} \left[\frac{E(z)m_{500c}}{3\times 10^{14}h^{-1}M_\mathrm{sun}}\right]^{2/3}\label{eq:Temhe}
\end{equation}
%---------------------------
and those given by \cite{Lee:2019tes}, which were fitted to the BAHAMAS \citep{McCarthy2017} and MACSIS \citep{Barnes:2016ret} simulations  \citep[see also][for a multi-simulation comparison]{Lee:2022svd},
%---------------------------
\begin{equation}
    k_B T_{\rm e} \approx A \,\mathrm{keV} \left[\frac{m_{500c}}{3\times 10^{14}h^{-1}M_\mathrm{sun}}\right]^{B+C\log(m_{500c}/3\times 10^{14}h^{-1}M_\mathrm{sun})}E(z)^{2/3}.\label{eq:Temhelee}
\end{equation}
%---------------------------
A self-similar relation can be recovered when $B=2/3$. The values for the parameters $A,B$ and $C$ are given in Table~\ref{tab:abclee} 
 and allow us to estimate the $y$-weighted temperature, $k_B T_{\rm e}=\langle y \,k_B T_{\rm e} \rangle/\langle y\rangle$, of clusters of a given mass. This temperature is directly relevant to the relativistic SZ effect \citep{Wright1979, Itoh98, Sazonov1998, Challinor1999, Chluba2012SZpack} and can exceed the X-ray temperature noticeably. It furthermore exhibits additional redshift-dependence beyond the $E(z)^{2/3}$ scaling \citep{Lee:2019tes, Lee:2022svd}.
%-----------------------------------
\begin{table}
\centering
\setlength{\tabcolsep}{10pt}
\renewcommand{\arraystretch}{1.5}
    \begin{tabular}{c|c|c|c}
           &  A & B & C  \\
        \hline
         $z=0$ &  4.763 & 0.581 & 0.013\\
          $z=0.5$ &  4.353 & 0.571 & 0.008\\
        $z=1$&  3.997 & 0.593 & 0.009 \\
    \end{tabular}
    \caption{Values of the \cite{Lee:2019tes} parameters temperature-mass relation [see Eq.~\eqref{eq:Temhelee}], from their Table~4.}
    \label{tab:abclee}
\end{table}
%-----------------------------------

The SZ intensity power spectrum across the whole sky is given by a weighted average of the tSZ signals from multiple clusters with varying temperatures. This means that the exact frequency scaling of the SZ power spectrum receives relativistic temperature corrections beyond the non-relativistic SZ formula, $g(\nu)$, given by Eq.~\eqref{eq:g_nu_tsz}. This correction can be modeled using the $y^2$-weighted temperature power spectrum \citep{Remazeilles:2018laq}:
%-----------------------------------
\begin{equation}
    C_\ell^{T_e} = \frac{C_\ell^{T_e yy}}{C_\ell^{yy}}\quad \mathrm{with}\quad C_\ell^{T_e yy} = \int \mathrm{dv} \langle T_e \hat{u}^y(k_\ell) \hat{u}^y (k_\ell)\rangle_n \quad \text{and where} \quad C_\ell^{yy} = \int \mathrm{dv} \langle\hat{u}^y(k_\ell) \hat{u}^y (k_\ell)
\end{equation}
%-----------------------------------
is the angular tSZ power spectrum (see section \ref{ss:autospec}). 
This can also be thought of as a simple moment expansion \citep{Chluba2017moments} of the $yy$-power spectrum with relativistic corrections included.
Note that for this calculation we compute the 1-halo term contribution only. 

\begin{figure*}
\includegraphics[width=0.75\columnwidth]{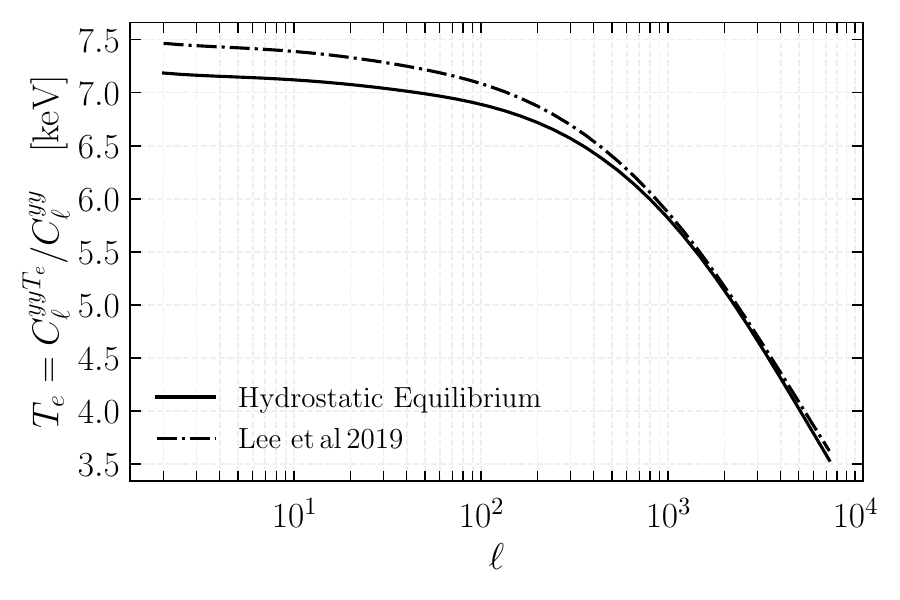}
    \centering
    \vspace{-0.2cm}
    \caption{Example of electron temperature powerspectrum. The solid line show the prediction from hydrostatic equilibrium (Eq.~\ref{eq:Temhe}) and the dotted-dashed line show the prediction from fits to simulation, see Eq.~\eqref{eq:Temhelee} and \cite{Lee:2019tes}.}
    \label{fig:teyy}
\end{figure*}

We show the predictions for $C_\ell^{T_e}$ from the two temperature-mass relations in Figure \ref{fig:teyy}. The calculations are given in a notebook online.\footnote{\href{https://github.com/CLASS-SZ/notebooks/blob/main/class_sz_sztemperature_powerspectrum.ipynb}{https://github.com/CLASS-SZ/notebooks/blob/main/class\_sz\_sztemperature\_powerspectrum.ipynb}}
By using this $\ell$-dependent SZ temperature, one can model the exact SZ (intensity) power-spectrum across frequency as $$C_{\ell}^{\rm SZ}(\nu)\approx g(\nu, C_\ell^{T_e})^2\,C_\ell^{yy},$$
where $g(\nu, C_\ell^{T_e})^2$ includes the exact relativistic temperature corrections that can be reliably computed with {\tt SZpack} \citep{Chluba2012SZpack}. The corrections to the frequency scaling are expected to reach  $\simeq 10-20\%$ at high frequencies for typical temperature of $\simeq 5\,{\rm keV}$. Alternatively, one could modify the $yy$-power spectrum reconstruction to account for the relativistic temperature corrections and measure $C_\ell^{T_e}$ \citep{Remazeilles2020}. This could provide additional insight into the cluster {\it gastrophysics}, and \textsc{class\_sz} now provides the tools for exploring this direction.

\begin{figure*}
\includegraphics[width=0.35\columnwidth]{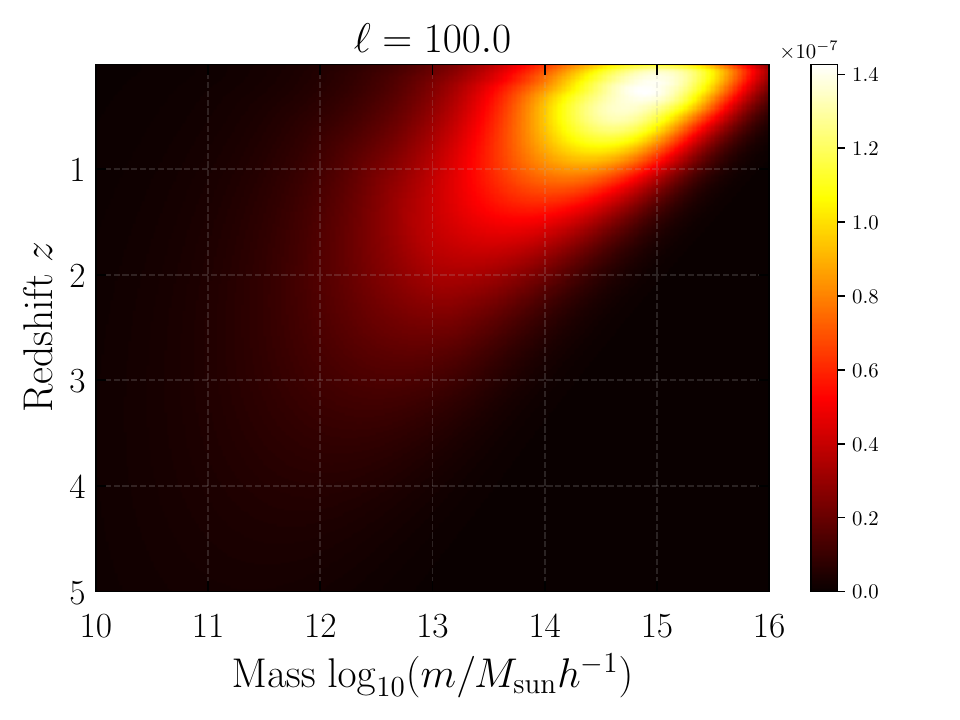}\includegraphics[width=0.35\columnwidth]{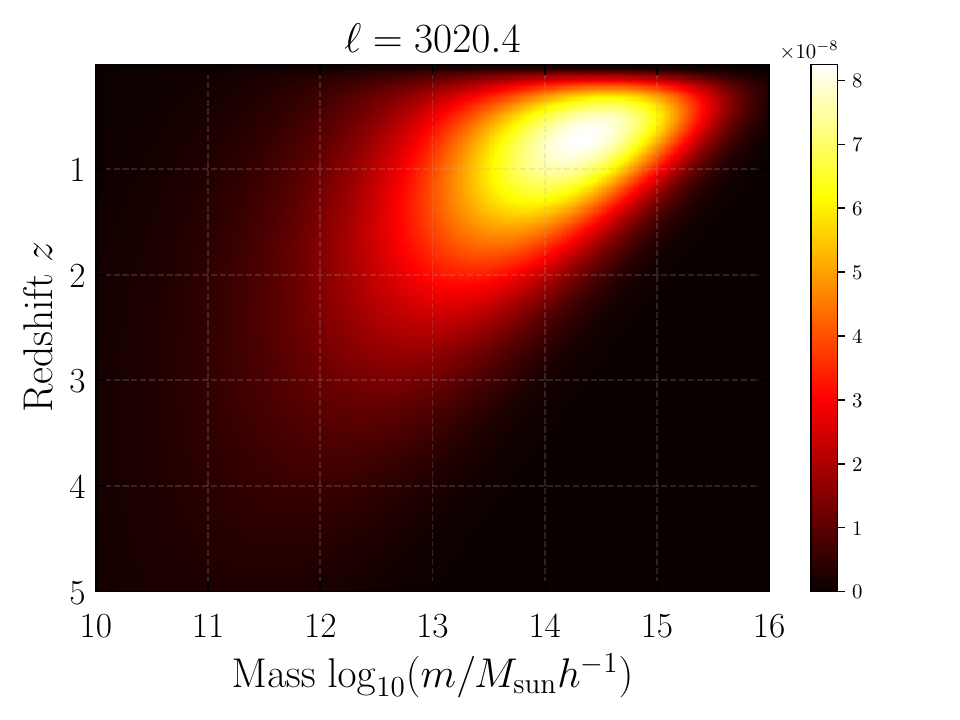}\includegraphics[width=0.35\columnwidth]{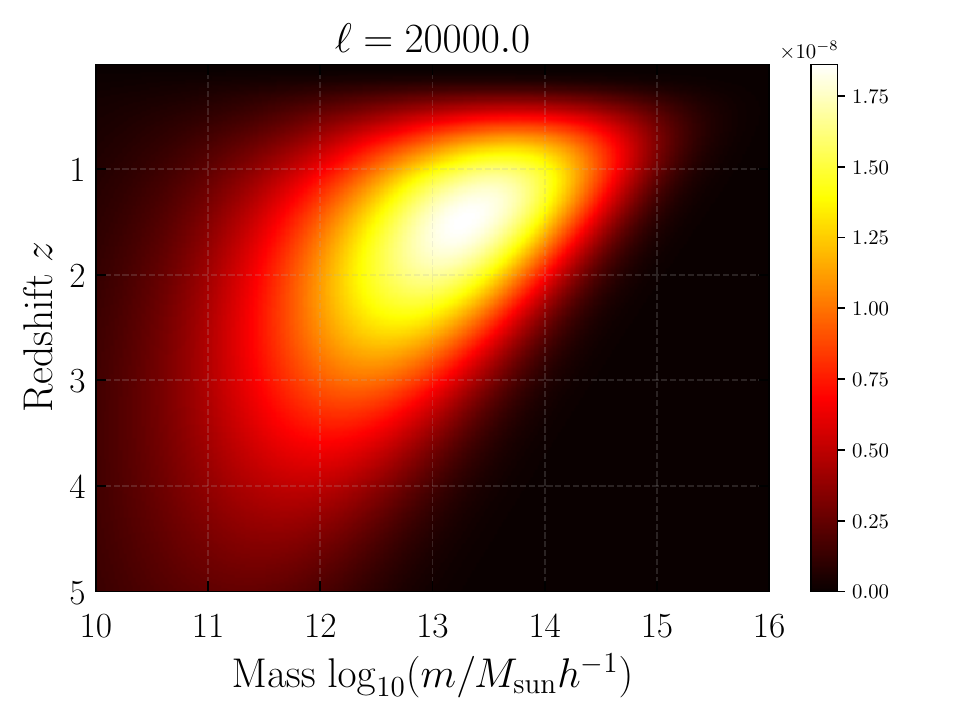}
    \centering
    \vspace{-0.cm}
    \caption{mean-y heatmap.}
    \label{fig:yheatmap}
\end{figure*}

\subsubsection{Galaxy Counts}\label{sss:galhod}

Galaxies populate dark matter halos in complicated ways. Galaxy halo occupation distributions (HOD) is a simple description of galaxy clustering,  \cite[see, e.g.,][]{Berlind:2001xk,Zheng:2004id,zz2007}. In \textsc{class\_sz} we use a model that is general enough to match that of \cite{Zehavi_2011},  \cite{Koukoufilippas:2019ilu} and \cite{za2021}. HOD models generally assume  types of galaxies, central and satellite galaxies. 

The expectation value for the number of central galaxies in a halo of mass $m$ is given by 
\begin{equation}
  N_\mathrm{cent}(m) = \frac{1}{2}\left(1+\mathrm{erf}\left[\frac{\log_{10}(m/m_\mathrm{min})}{\sigma_{\mathrm{log}_{10}m}}\right]\right)\label{eq:ncent}
\end{equation}
where $m_\mathrm{min}$ is a pivot halo mass above which, on average, halos have a central galaxy. Here,  $\sigma_{\mathrm{log}_{10}m}$ controls the steepness of the transition in mass from no galaxy to having at least one galaxy in the host halo of mass $m$. 

The expectation value for the number of satellite galaxies is a power law with an exponent $\alpha_s$
\begin{equation}
 N_\mathrm{sat}(m) =  \mathrm{H}(m-m_0)N_\mathrm{cent}(m)\left(\frac{m-m_0}{m_1}\right)^{\alpha_s},\label{eq:nsat}
\end{equation}
where $m_\mathrm{1}$ is a pivot halo mass above which the number of satellites in the host halo increases steeply, and $m_0$ is a mass threshold.

Given a specific HOD (Eq.~\ref{eq:ncent} and \ref{eq:nsat}) we can compute the galaxy number density and galaxy bias at $z$ as 
\begin{equation}
    \bar{n}_{\mathrm{g}}(z) =\langle  N_\mathrm{cent}+ N_\mathrm{sat} \rangle_n,\quad\mathrm{with}\quad b_\mathrm{g}(z)=\frac{1}{\bar{n}_{\mathrm{g}}(z)}\langle b^{(1)}(N_\mathrm{cent}+ N_\mathrm{sat})\rangle_n,\label{eq:ngbar}
\end{equation}
where $b^{(1)}$ is the linear bias of Eq.~\eqref{eq:b1tink}. As for the spatial distribution, central galaxies are naturally assumed to be located at the center of halos (their \textit{density profile} is a Dirac delta function) and satellite galaxies are assumed to be randomly distributed along an NFW-like radial profile. Thus, the Fourier transform of  the \textit{galaxy density profile} is
\begin{equation}
    \hat{u}^\mathrm{g}_k = \frac{1}{\bar{n}_\mathrm{g}}\left(N_\mathrm{cent}+N_\mathrm{sat}\hat{u}^\mathrm{sat}_k\right)\label{eq:ugk}
\end{equation}
where $\hat{u}^\mathrm{sat}_k$ is the same Eq.~\eqref{eq:nfwtrunc} with $\lambda=1$ and without the ${m_{\lambda r_\Delta}}/{\rho_{\mathrm{m},0}}$ prefactor. In addition, $c_\Delta$ is often replaced by a free parameter $c_\mathrm{sat}$ to allow for more freedom in the radial distribution. Another important HOD quantity is the Fourier transform of the second moment of the satellites galaxy distribution:
\begin{equation}
\hat{u}^\mathrm{gg}_k=\frac{1}{\bar{n}_{\mathrm{g}}^2}\left[N_\mathrm{sat}^2(\hat{u}^{\mathrm{sat}}_k)^2+2N_\mathrm{sat}\hat{u}^{\mathrm{sat}}_k\right],\label{eq:uggk}
\end{equation}
\citep[see, e.g.,][]{van_den_Bosch_2013,Koukoufilippas:2019ilu}, as it determines the 1-halo term of the galaxy power spectrum.

\subsubsection{Galaxy Lensing}\label{sss:gallens}

The tracer relevant to galaxy weak lensing is the matter distribution. Therefore Galaxy weak lensing power spectra are LOS-integrals of the  matter power spectrum. These are often computed using Non-linear matter power spectrum. But halo model is also useful, especially in order to quantify effects of baryons.

\subsubsection{CMB Lensing}\label{sss:cmblens}

This is the same as Galaxy weak lensing except that the redshift kernel is different because the photons come from the last-scattering-surface. Again, this is commonly computed with non-linear fits, but halo model is useful to model effects of baryons.

\begin{figure*}
    \includegraphics[width=0.5\columnwidth]{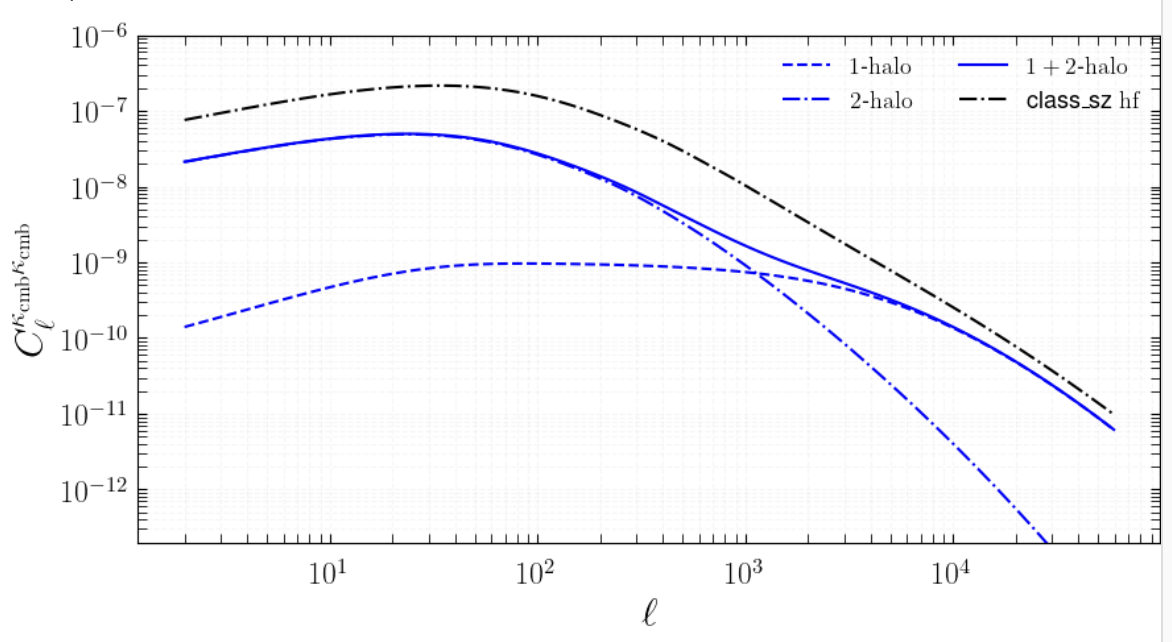}\includegraphics[width=0.5\columnwidth]{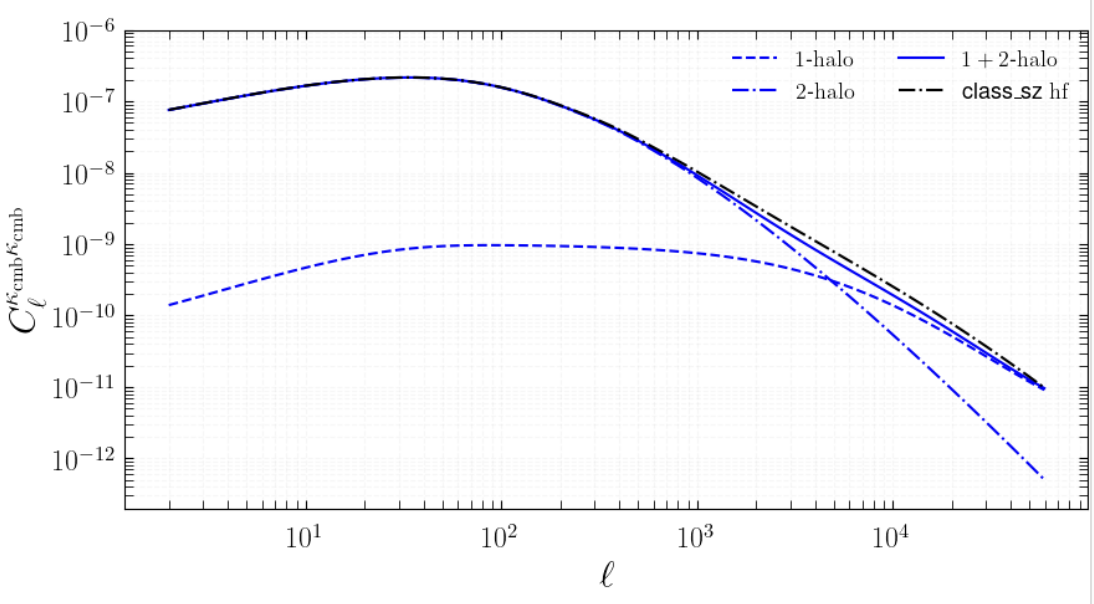}
    \vspace{-0.2cm}
    \caption{CMB lens computed with \textsc{class\_sz} using different models \emph{Left}: Without consistency condition. \emph{right}: with consistency see notebook \href{https://github.com/CLASS-SZ/notebooks/blob/main/class_sz_cmblensing_halomodel.ipynb}{https://github.com/CLASS-SZ/notebooks/class\_sz\_cmblensing\_halomodel.ipynb}.}
    \label{fig:CMBlens_cons}
\end{figure*}

\subsubsection{Cosmic Infrared Background}\label{sss:cib}

\begin{figure*}
    \includegraphics[width=1.\columnwidth]{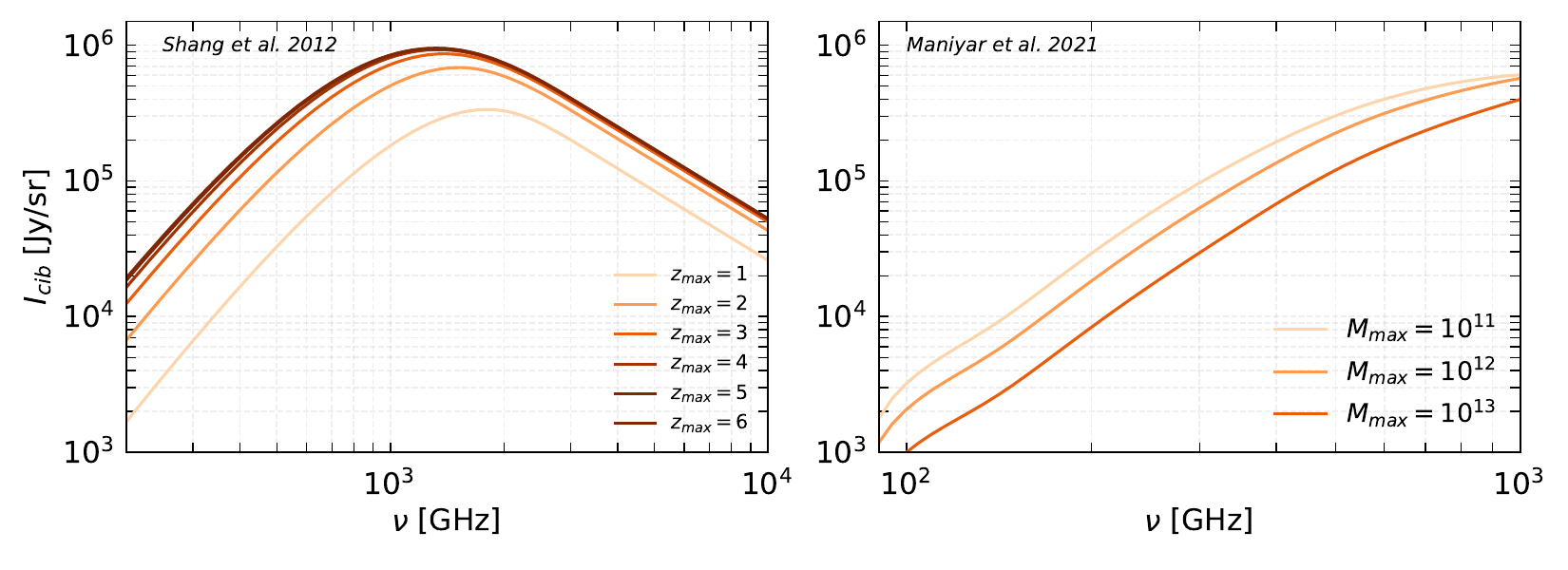}
    \vspace{-0.2cm}
    \caption{CIB monopole computed with \textsc{class\_sz} using two different models \emph{Left}: CIB monopole using halo model prescription from \cite{Shang2012} integrated up to different maximum redshifts. \emph{Right}: Effects of varying the most efficient halo mass sourcing the CIB emission on the CIB model from \cite{Maniyar2021}.}
    \label{fig:CIBmonopole}
\end{figure*}

{\bf{Shang et al. CIB model.}} We give the halo model description of the CIB emission, which is based on the model presented in \cite{Shang2012}, which was further used in many other analyses, including \cite{Viero_2013_hermes, websky_Stein_2020, McCarthy:2021lfp, Sabyr_2022}. Following \cite{Shang2012}, we can define the CIB in the halo model analogously to the galaxy HOD, but with additional prescriptions that describe the infrared emission of each galaxy.

First, we can write the specific intensity of the CIB $I_{\nu}$ at frequency $\nu$ as 
\begin{equation}
    I_{\nu} = \int \frac{\mathrm{d} \chi}{\mathrm{d}z} a(z) \bar{j}_\nu(z) \mathrm{d}z \,, 
\end{equation}
where $\bar{j}_\nu(z)$ is the average emissivity~(\textit{e.g.},~\cite{McCarthy:2021lfp}): 
\begin{equation}
    \bar{j}_\nu(z) = \int_{}^{} \mathrm{d}M \frac{\mathrm{d}n}{\mathrm{d}M} \frac{L_{(1+z)\nu}(M,z)}{4\pi} \,, 
    \label{eq:j_nu}
\end{equation}
where $L_{(1+z)\nu}(M,z)$ is the infrared luminosity of a halo of mass $M$ at redshift $z$ and the factor of $(1+z)$ in the frequency accounts for redshifting of the emitted radiation. 

The luminosity $L_{(1+z)\nu}(M,z)$ is the sum of the contributions from the central and satellite galaxies, defined by $L_{(1+z)\nu}^{c}(M,z)$ and $L_{(1+z)\nu}^{s}(M,z)$, respectively. In this work, we follow the assumption made in \cite{McCarthy:2021lfp} that both the central and satellite luminosity depend on the same galaxy luminosity model $L^{\rm{gal}}_{(1+z)\nu} (M,z)$ (which is only dependent on the mass of the host halo $M$ and redshift $z$), weighted by the number of centrals and satellites, respectively. Thus, the central luminosity, $L_{(1+z)\nu}^{c}(M,z)$ can be defined as
 \begin{equation}
     L_{(1+z)\nu}^{c}(M,z) = N_{c}^{\rm{CIB}} (M,z) L^{\rm{gal}}_{(1+z)\nu} (M,z) \,,
     \label{eq:L_cent}
 \end{equation}
where the number of CIB central galaxies in a halo, $N_{c}^{\rm{CIB}}$, similarly to the galaxy HOD, is either zero or one, depending on whether the halo mass $M$ is smaller or larger than the parameter $M_{\rm{min}}^{\rm{CIB}}$, the minimum mass to host a central galaxy that sources CIB emission. This requirement can be  written as
\begin{equation}
    N_{c}^{\rm{CIB}} (M,z) = 
    \begin{cases}
      0, & \text{if}\ M < M_{\rm{min}}^{\rm{CIB}} \\
      1, &  \text{if}\ M \geq M_{\rm{min}}^{\rm{CIB}} \,.
    \end{cases}
\end{equation}
The satellite luminosity, $L_{(1+z)\nu}^{s}(M,z)$, that is, the luminosity of a halo due to its satellite galaxies, is given by
\begin{equation}
    L_{(1+z)\nu}^{s}(M,z) = \int \mathrm{d} M_s \frac{\mathrm{d} N}{\mathrm{d} M_s}  L^{\rm{gal}}_{(1+z)\nu} (M,z) \,,
    \label{eq:L_sat}
\end{equation}
where $\mathrm{d} N/\mathrm{d} M_s$ is the subhalo mass function. For the subhalo mass function we have implemented several formulas (see Subsection \ref{ss:subhmf}). Note that \cite{McCarthy:2020qjf} uses the \cite{2010ApJ...719...88T} formula while \cite{websky_Stein_2020} uses the \cite{Jiang:2013kza} formula.

The luminosity of galaxies $L^{\rm{gal}}_{(1+z)\nu}$ is governed by the \emph{luminosity-mass relation} (or \emph{L--M relation}), that can be written as 
\begin{equation}
    L^{\rm{gal}}_{(1+z)\nu} = L_0 \Phi(z) \Sigma(M) \Theta((1+z)\nu) \,,
    \label{eq:LM_relation}
\end{equation}
where $L_0$ is a normalization factor, $\Phi(z)$ describes the redshift evolution, $\Sigma(M)$ the mass dependence, and $\Theta$ is the SED of the infrared emission. We describe parametrized functions for the \emph{L--M relation} below.

The redshift evolution of the \emph{L--M relation} is parametrized by a power law index $\delta^{\rm{CIB}}$ in the form of
\begin{equation}
    \Phi(z) = (1+z)^{\delta^{\rm{CIB}}} \,.
    \label{eq:Phi_cib}
\end{equation}
It is well-motivated by observations~\cite{Stark_2009, Gonzalez_2011}, however, to extend the redshift evolution of the \emph{L--M relation}, by including the so-called plateau redshift $z_p$, where $\delta^{\rm{CIB}} = 0$ at $z \geq z_p$:
\begin{equation}
    \Phi(z) = 
    \begin{cases}
      (1+z)^{\delta^{\rm{CIB}}}, & \text{if}\ z < z_p \,, \\
      1, &  \text{if}\ z \geq z_p \,.
    \end{cases}
    \label{eq:Phi_cib_zp}
\end{equation} 
This approach was taken in \cite{websky_Stein_2020} and \cite{Viero_2013_hermes}, where the authors assumed $z_p = 2$, motivated by observations \cite{Stark_2009, Gonzalez_2011}. We also follow this prescription and include $z_p$ as a parameter in our model, to be specified below.

The mass dependence of the \emph{L--M relation} is written as 
\begin{equation}
\Sigma (M) = \frac{M}{\sqrt{2\pi \sigma^2_{L-M}}} e^{-(\mathrm{log}_{10} M - \mathrm{log}_{10} M_{\rm{eff}}^{\rm{CIB}})/2 \sigma^2_{L-M}} \,, 
\end{equation}
with two free parameters, $M_{\rm{eff}}^{\rm{CIB}}$ the peak of the specific IR emissivity, and $\sigma^2_{L-M}$ which controls
the range of halo masses that source the CIB emission. 

The CIB SED is the standard modified blackbody combined with a power law decline at high frequencies 
\begin{equation}
    \Theta = 
    \begin{cases}
       \nu^{\beta^{\rm{CIB}}}B_{\nu}(T_d(z))& \text{if}\ \nu < \nu_0, \\
      \nu^{—\gamma^{\rm{CIB}}} &  \text{if}\  \nu \geq \nu_0,
    \end{cases}
    \label{eq:theta_cib}
\end{equation}
where $B_{\nu}(T)$ is the Planck function at temperature $T$, $\nu_0$ is the break frequency that has to satisfy the continuous derivative requirement
\begin{equation} 
   \frac{\mathrm{d} \ln \Theta(\nu,z) }{\mathrm{d} \ln \nu} = -\gamma^{\rm{CIB}} \,, 
\end{equation}
 and $T_d(z)$ is the dust temperature at redshift $z$ that we parameterize as 
 \begin{equation}
      T_d(z) = T_0 (1+z)^{\alpha^{\rm{CIB}}}, 
 \end{equation}
 with $T_0$ and $\alpha^{\rm{CIB}}$ being free parameters.

To sum up, the \cite{Shang2012} CIB model presented in this work  has ten free parameters \{$L_0$, $\alpha^{\rm{CIB}}$, $\beta^{\mathrm{CIB}}$, $\gamma^{\mathrm{CIB}}$, $T_0$, $M_{\rm{eff}}$, $\sigma^2_{L-M}$, $\delta^{\rm{CIB}}$,  $M_{\rm{min}}^{\rm{CIB}}$, ($z_p$)\}. 

In CIB modeling and analysis, e.g., \cite{Shang2012, Planck:2013cib, McCarthy:2021lfp}, one usually implements a flux cut above which bright sources are detected and can be removed (thus suppressing the Poisson power associated with these objects). In our modeling, we follow this prescription and allow to remove all halos whose total luminosity is larger than the luminosity corresponding to a given flux cut value,  $ S_{\nu}$ defined as 
\begin{equation}
    S_{\nu} = \frac{L_{(1+z)\nu}(M,z)}{4\pi(1+z)\chi^2} \,.
    \label{eq:flux_cut}
\end{equation}
We present the flux cut values for selected \emph{Planck} and SO frequencies in Table~\ref{table:fluxcut}; for the \emph{Planck} frequencies the values are taken from Ref.~\cite{Planck:2013cib}, and for SO frequencies those in the SO forecast paper~\cite{Ade_2019}.

\begin{table}
\centering
\setlength{\tabcolsep}{10pt}
    \begin{tabular}{|c|c|c|c|c|c|c|c|c|c|} 
        \hline
         Frequency $\left[ \mathrm{GHz} \right]$ & 93 & 100 & 143 & 145 & 217 & 225 & 280 & 353 & 545 \\
         \hline
         Flux cut $\left[ \mathrm{mJy} \right]$ & 7 & 400 & 350 & 15 & 225 & 20 & 25 & 315 & 350 \\
          \hline
          
    \end{tabular}
    \caption{
    Point source flux cut values (in mJy) for CIB frequencies (in GHz) considered in this work. The \emph{Planck} frequency (100, 217, 353, 545 GHz) flux cut values come from Table~1 in Ref.~\cite{Planck:2013cib}, while for the SO frequencies (93, 145, 280 GHz), we use the flux cut values from the SO forecast paper \cite{Ade_2019}. The flux cut is implemented according to Eq.~\eqref{eq:flux_cut} for each frequency (in both auto- and cross-correlations involving the CIB).}
    \label{table:fluxcut}
\end{table}

Finally, putting all of the pieces together, the CIB multipole-space kernel $u_\ell^{\nu} (M,z)$ at frequency $\nu$ can be written as
\begin{equation}
    u_\ell^{\nu} (M,z) = W_{I_{\nu}}(z) \bar{j}_\nu^{-1} \frac{L_{(1+z)\nu}^{c} + L_{(1+z)\nu}^{s} u^{m}_{\ell}(M,z)}{4\pi} \,,
\end{equation}
where the CIB window function $W_{I_{\nu}} (z)$ is defined as 
\begin{equation}
       W_{I_{\nu}} (z) = a (z) \bar{j}_\nu(z) \,.
\end{equation}

\begin{figure*}
    \includegraphics[width=0.5\columnwidth]{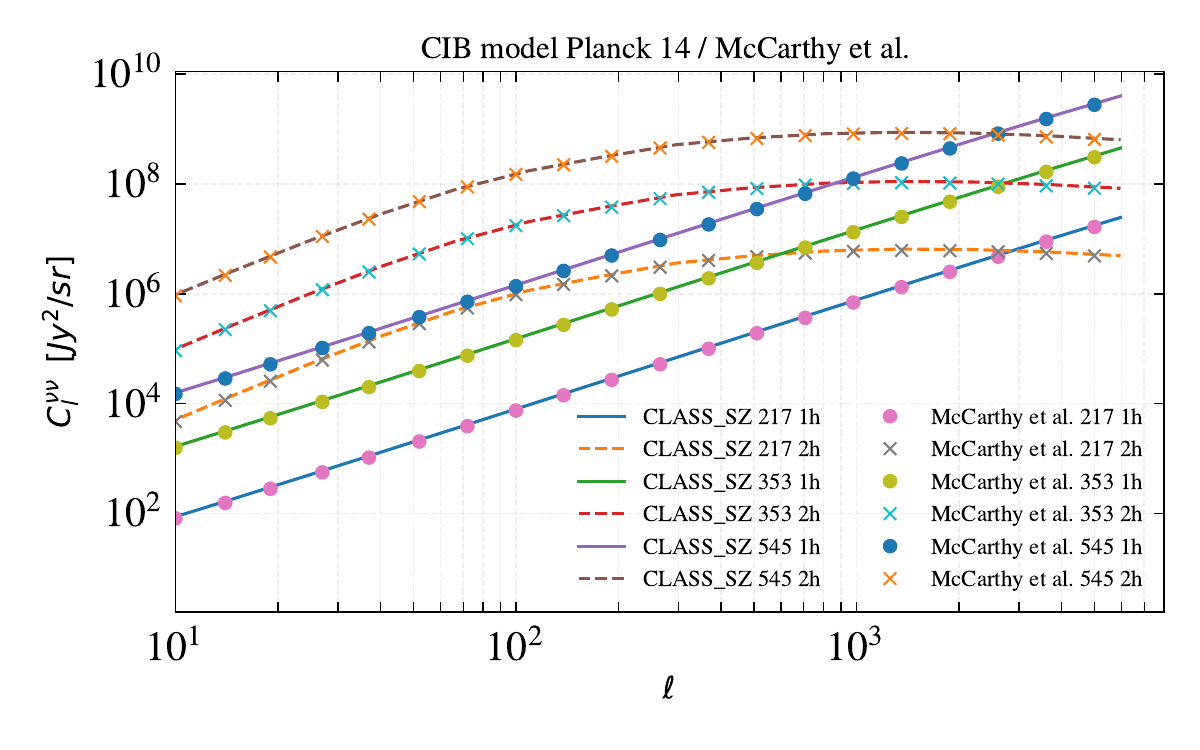}\includegraphics[width=0.5\columnwidth]{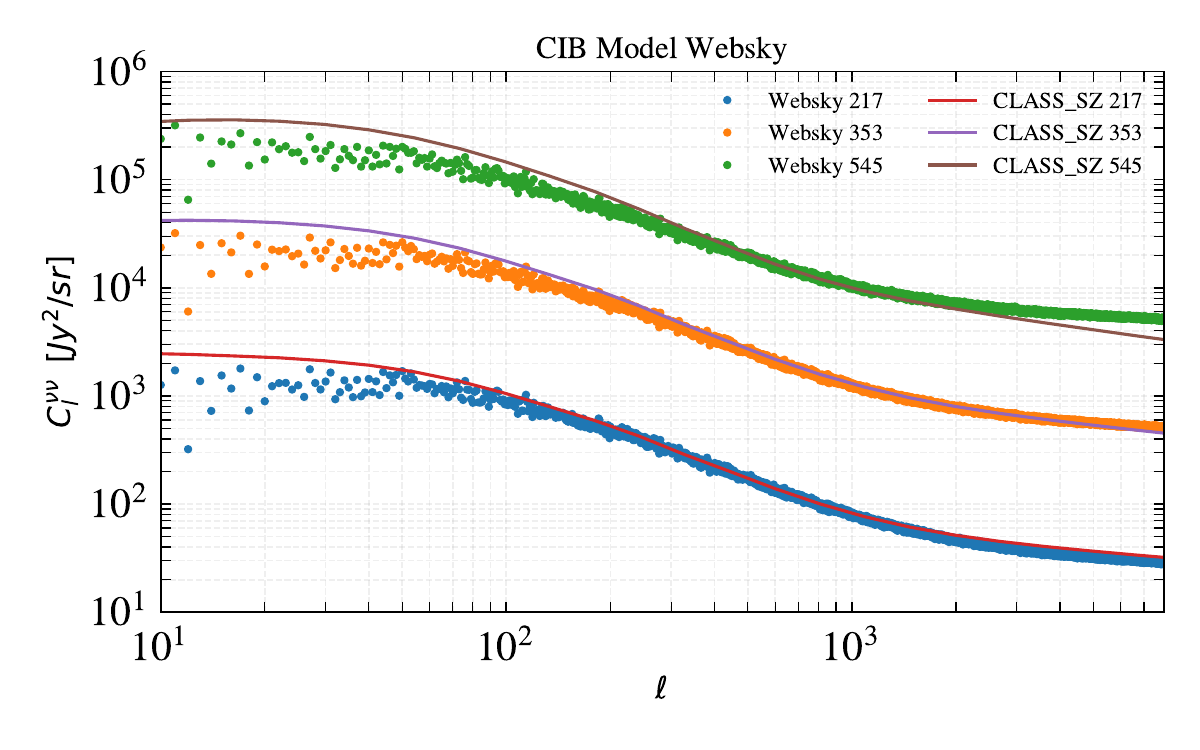}
    \vspace{-0.4cm}
    \caption{Benchmark against McCarthy and also websky model. For websky we added the shot noise values from Table 9 of the planck CIB paper.}
    \label{fig:shangmccarthy}
\end{figure*}

In Figure \ref{fig:shangmccarthy} we compare our CIB predictions with the ones from \cite{McCarthy:2020qjf} and \cite{websky_Stein_2020}.

{\bf{Maniyar et al. CIB model.}} A recent and popular CIB model, explicitely linked to star formation history is the \cite{Maniyar2021} model. We implemented it in \textsc{class\_sz} and validated against the original code from \cite{Maniyar2021}\footnote{\href{https://github.com/abhimaniyar/halomodel_cib_tsz_cibxtsz}{https://github.com/abhimaniyar/halomodel\_cib\_tsz\_cibxtsz}}. 

Within this model, the satellite galaxy luminosity is computed using the \cite{2010ApJ...719...88T} sub-halo mass function (see \ref{ss:subhmf}) with a galaxy luminosity give by
\begin{equation}
    L^{\rm{gal}}_{(1+z)\nu} = \mathrm{min}[\mathrm{SFRI},\mathrm{SFRII}]
\end{equation}
where 
\begin{equation}
    \mathrm{SFRI} = 4\pi S_\nu(m,z) \mathrm{SFR}(m_\mathrm{sub},z)
\end{equation}
and
\begin{equation}
    \mathrm{SFRII} = 4\pi S_\nu(m,z) \mathrm{SFR}(m,z) \frac{m_\mathrm{sub}}{m}
\end{equation}
where 
\begin{equation}
   \mathrm{SFR}(m,z)= 10^{10} \dot{\mathcal{M}} f_\mathrm{b} \mathrm{SFR}_c
\end{equation}
note that $10^{10}$ is the Kennicutt constant for a Chabrier IMF, where $f_\mathrm{b}$ is the baryon fraction and 
\begin{equation}
    \mathrm{SFR}_c \eta_\mathrm{max}\exp(-(\ln m- \ln m_\mathrm{eff})^2/2\sigma_{\ln m}^2 ) 
\end{equation}
with
\begin{equation}
    \sigma_{\ln m} = \sigma_{\ln m}^\star - \mathrm{H}(m_\mathrm{eff}-m) \tau_\mathrm{cib}(\mathrm{max}[0,z-z_\mathrm{cib}]).
\end{equation}

and 
\begin{equation}
    \dot{\mathcal{M}} = 46.1(1+1.11z) E(z) (m/10^{12}M_\mathrm{sun})^{1.1}
\end{equation}
This is also used in the central galaxy luminosity calculations, which is evaluated at $m(1-f_\mathrm{sub})$ in this model. 

% \subsection{Profile two-lhalo terms}\label{ss:prof2h}

\subsection{Angular Power Spectra}\label{ss:autospec}

With the Limber approximation in flat-sky \citep[e.g., Appendix A of  ][and references therein]{HillPajer2013}, angular anisotropy power spectra are obtained by integrating the 3D power spectra, evaluated at $k_\ell=(\ell+1/2)/\chi$, over comoving volume 
\begin{equation}
    C_\ell^{XY}=\int\mathrm{d}\mathrm{v}  W^X(\chi)W^Y(\chi)P_{XY}^\mathrm{hm}\left(k_\ell,z\right)\label{eq:clxy}
\end{equation}
where $W^X$ and $W^Y$ are redshift dependent projection kernels (see Eq.~\ref{eq:zint}). Here, $P_{XY}^\mathrm{hm}$ is the 3d power spectrum of tracers $X$ and $Y$ evaluated in the halo model (see \ref{ss:ps}). 

\subsubsection{Auto-powerspectra}
To compute the angular power spectra of the tracers listed above, we use the Fourier transforms of the radial profiles that we explicitely gave and specify the redshift dependent kernels. All we need to do here is specifying the redshift dependent kernels for each tracer. 

For CMB lensing, the redshift dependent kernel is given in Eq.~\eqref{eq:wkcmb}. For galaxy-lensing, the redshift dependent kernel is given in Eq.~\eqref{eq:wkg} and for galaxy counts the kernel is given in Eq.~\eqref{eq:wg}.

\begin{figure*}
    \includegraphics[width=1.\columnwidth]{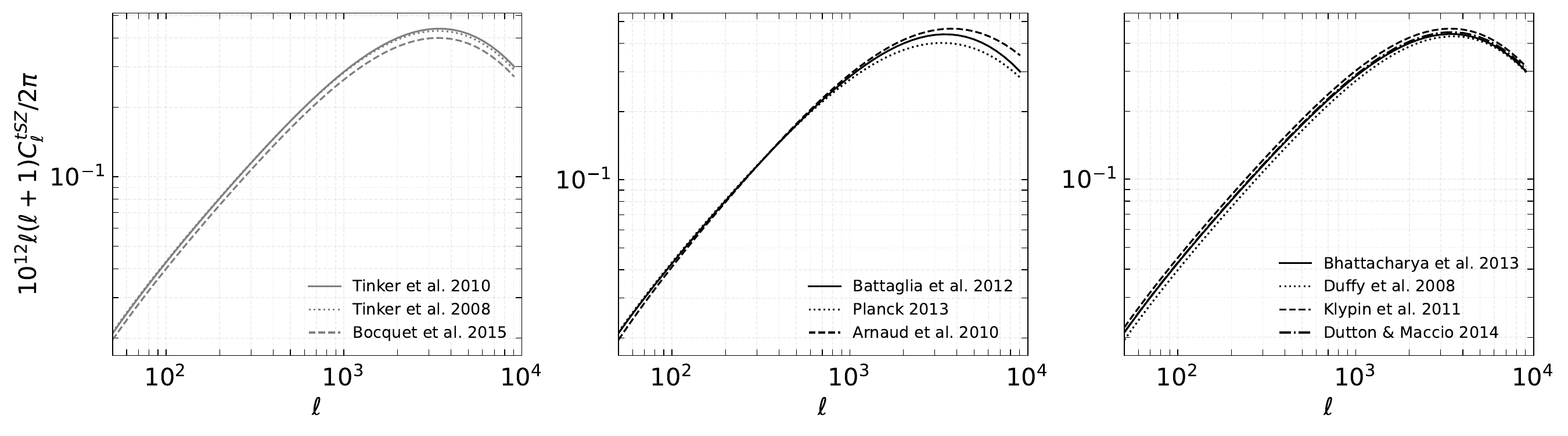}
    \vspace{-0.2cm}
    \caption{tSZ power spectrum as a function of several halo mass functions, pressure profiles and concentration parameters that are implemented in \textsc{class\_sz}. The baseline model in these example comparisons is plotted with a solid curve. For the purposes of illustration, we fix the HSE mass bias here to $B=(1-b)^{-1}=1.7$ and \cite{Battaglia_2012} pressure profile amplitude $P_{0}=11.$ Here we show the total halo model power spectrum, which is the sum of the one and two-halo terms. }
    \label{fig:tSZ_hmf_pp_conc}
\end{figure*}

For pressure we % \bb{galaxy}
show the total tSZ power spectrum computed by \textsc{class\_sz} in Figure \ref{fig:tSZ_hmf_pp_conc}. This computation is available in a tutorial notebook\footnote{\href{https://github.com/borisbolliet/class_sz/blob/master/notebooks/class_sz_tutorial_notebooks/class_sz_cltSZ.ipynb}{class\_sz\_tutorial\_notebooks/class\_sz\_cltSZ.ipynb}}. In the notebook we also show the one- and two-halo terms separately, and how to change the mass bounds and other things. 
The angular tSZ power spectrum from \textsc{class\_sz} has been used mutiple times in previous publication. The first work was \cite{Bolliet:2017lha} where we obtained bounds on the dark energy equation of state using the \textit{Planck} thermal SZ map. In \cite{Bolliet:2019zuz} we described how to consistently include neutrino masses and derived constraints on it. In \cite{Remazeilles:2018laq} we computed the electron-temperature power spectrum, which is based on the tSZ power spectrum. In \cite{Rotti:2020rdl} we modified the calculation to model the tSZ power spectrum after masking SZ selected clusters. In \cite{Sabyr_2022} we used to estimate the anisotropy of the SZ effect on the CIB.  In \cite{Kusiak:2023hrz} it was used for investigating novel ways of improving component separation for CMB maps. Moreover, the tSZ power spectrum calculations from \textsc{class\_sz} served as a benchmark for the \textsc{ccl} implementation. 

\begin{figure*}
    \includegraphics[width=1.\columnwidth]{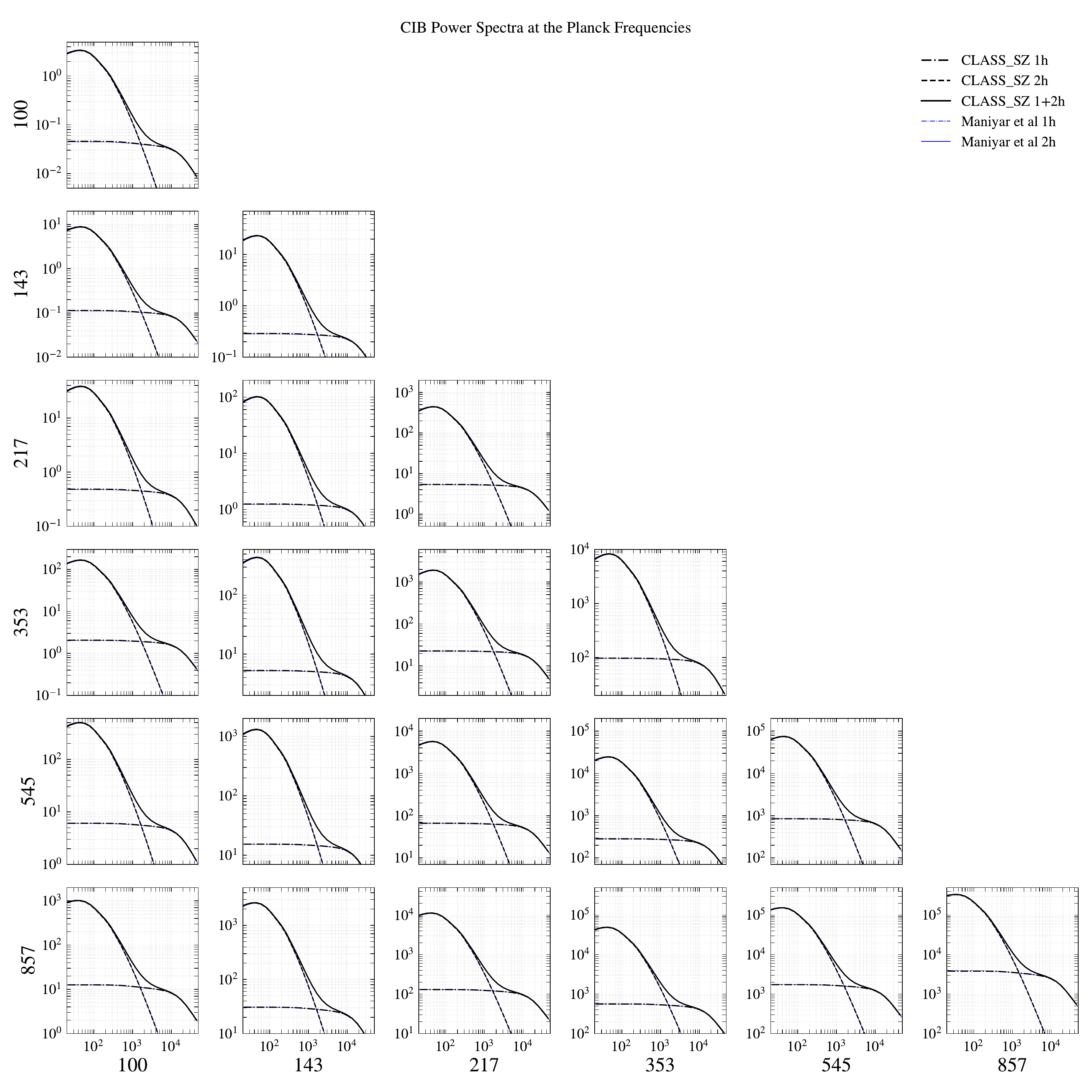}
    \vspace{-0.2cm}
    \caption{Benchmark against \cite{Maniyar2021}. For comparison purposes, this use fixed concentration $c=5$ and the \cite{Tinker2010} mass function as implemented in \textsc{class\_sz}.}
    \label{fig:maniyar}
\end{figure*}

Similarly, the CIB angular power spectra calculations have been carefully checked against previously published work, including \cite{McCarthy:2020qjf,Maniyar2021} and \cite{websky_Stein_2020}. In figure \ref{fig:maniyar} we show the reproduction of the \cite{Maniyar2021} model, for six planck frequencies between 100 and 857 GHz. We computed the auto and cross-frequency power spectra. The notebook to replicate this calculation is available\footnote{\href{https://github.com/borisbolliet/class_sz/blob/master/notebooks/class_sz_tutorial_notebooks/class_sz_CIB_maniyar.ipynb}{class\_sz\_tutorial\_notebooks/class\_sz\_CIB\_maniyar.ipynb}}. As can be seen on the figure, the \textsc{class\_sz} and \cite{Maniyar2021} spectra are in excellent agreement. The difference at low $\ell$ for the 1-halo term is because we switched on the damping in the \textsc{class\_sz} calculations (see Eq.~\ref{eq:damp}). In Figure \ref{fig:shangmccarthy} we show the comparison between the \cite{McCarthy:2020qjf}, \cite{websky_Stein_2020} and \textsc{class\_sz} calculation, again showing an excellent agreement. For the comparison with \cite{websky_Stein_2020} the small difference at low-$\ell$ can be attributed to sample variance while the one at high $\ell$ can be attributed to shot-noise slighlt different shot-noise level. (Note that the code to generate these \cite{websky_Stein_2020} results is not public, but future \textsc{Websky} releases will be and will be systematically compared with \textsc{class\_sz}).
We also note that the CIB angular power spectra calculations from \textsc{class\_sz} were used in \cite{Kusiak:2023hrz} for component separation purposes. 

The lensing auto-powerspectrum can be computed in the halo model. We made a notebook dedicated to this. And the result is also shown in our \textit{combo} plot in Fig~\ref{fig:combo} (top panel). The halo model lensing power spectrum is however not the most accurate approximation, because it fails in the 1-to-2-halo transition. Using the Limber integral of Eq.~\ref{eq:clkk}, with a matter power spectrum fitted to N-body simulations is a better approximation.  

The galaxy-lensing auto-power spectrum, from which one can compute the shear (see Subsection \ref{ss:realspace}) is shown in our \textit{combo} plot in Fig~\ref{fig:combo} (bottom right panel).

%\bb{ksz}

% \bb{cib}

\begin{figure*}
\includegraphics[width=1\columnwidth]{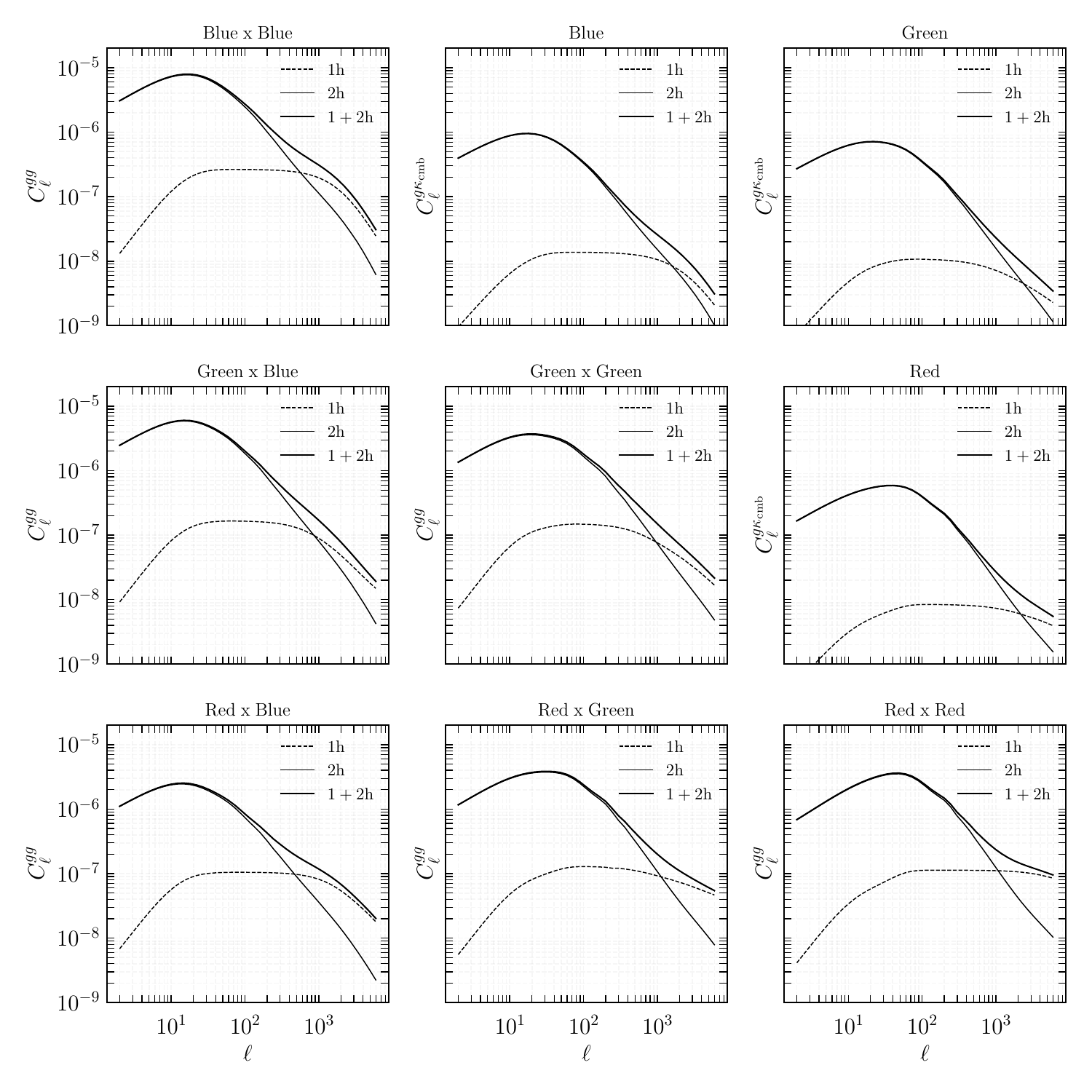}
    \vspace{-0.2cm}
    \caption{Auto- and cross-correlations for the unWISE galaxy samples, blue, green, and red (denoted as $g$) and CMB lensing ($\kappa_{\rm{CMB}}$) computed in the halo model. We assume the best-fit unWISE HOD from \citep{Kusiak_2022}. %\ok{maybe remove [units] from the plot ( both unitless), and instead of x, do $\times$, + typo in the title}
    }
    \label{fig:unwise_hod}
\end{figure*}

\subsubsection{Cross-correlations}

%\bb{effective galaxy bias.}

\begin{figure*}
\includegraphics[scale=0.6]{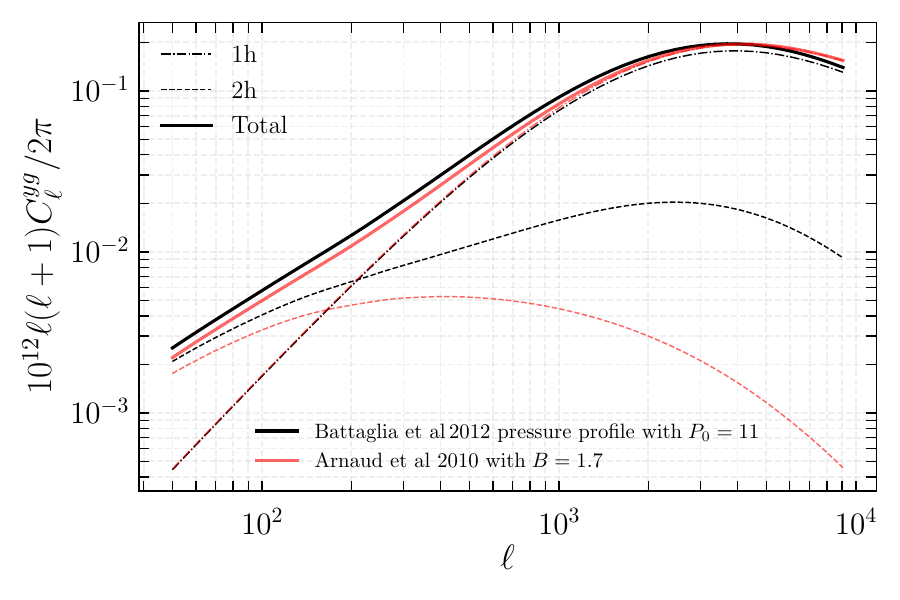}\includegraphics[scale=0.6]{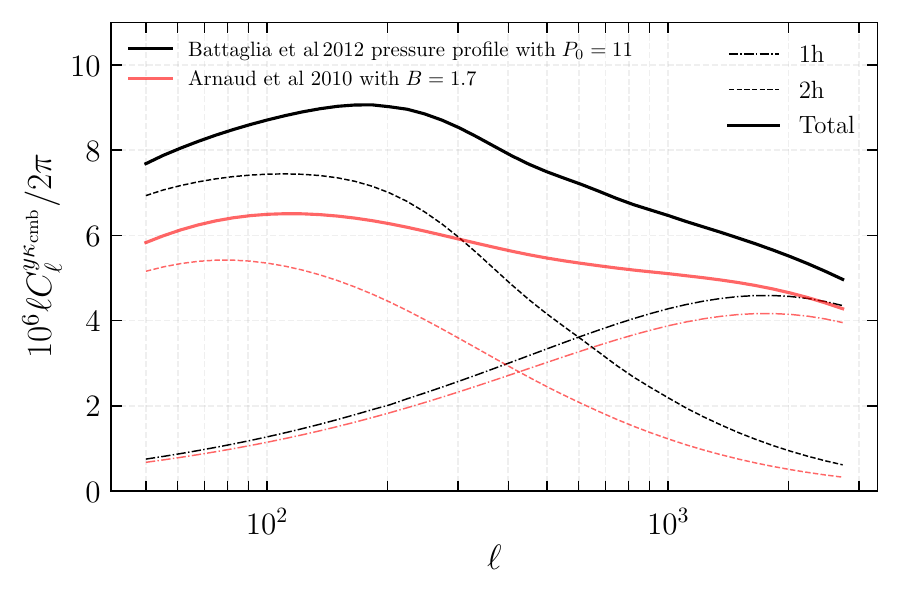}
    \centering
    % \vspace{-0.2cm}
    \caption{Cross power spectrum for Compton-$y$ $\times$ galaxy for the unWISE blue galaxy sample computed for two pressure profiles: B12 (solid blue) and A10 (dashed light blue).  We assume the best-fit unWISE HOD from \citep{Kusiak_2022} to compute the observable. Cross power spectrum for Compton-$y$ $\times$ $\kappa^{\mathrm{CMB}}$  for two pressure profiles: B12 (black) and A10 (red).}
    \label{fig:yg}
\end{figure*}

% \bb{ola: cmb lensing x galaxy; tsz x galaxy} 
Cross-correlation Compton-$y$ $\times$ galaxy in Fig.~\ref{fig:yg}.  Many other cross-correlations are shown in Fig.~\ref{fig:combo}.

% \begin{equation}
    
% \end{equation}

% \bb{boris: galaxy x galaxy-lensing}

% \bb{fiona: tsz x kappa; yogesh: cib x lensing; tsz x cib}
\begin{figure*}
\includegraphics[scale=0.65]{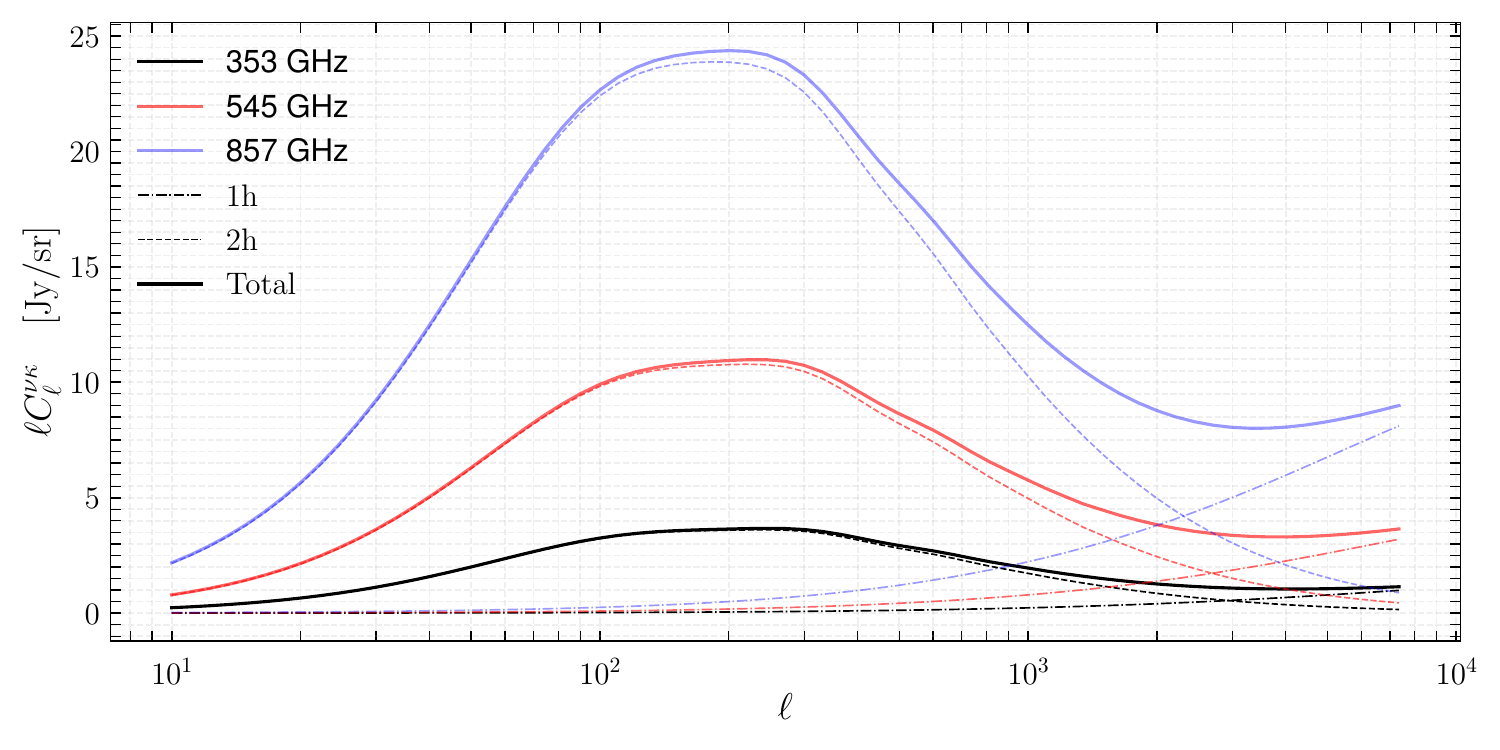}
    \centering
    \vspace{-0.2cm}
    \caption{CIB x $\kappa$ at three \textit{Planck} frequencies. The solid line shows the total cross spectrum, and the dashed and dashed-dotted lines correspond to the 2- and 1-halo terms, respectively. The flux cuts (see Table~\ref{table:fluxcut}) and fiducial CIB model parameters are from \cite{Planck:2013cib}.}
    \label{fig:cibxkappa}
\end{figure*}

% \bb{boris: ksz2-galaxy}

\subsection{Configuration-space}\label{ss:realspace}

From the angular power spectra we get the angular 2-point correlation functions (2PCF) as
\begin{equation}
    \xi^{XY}(\theta) = \frac{1}{2\pi}\int \mathrm{d}\ell \ell J_i(\ell\theta)C_\ell^{XY}\label{eq:xi}
\end{equation}
where $J_i$ is the $i$th  order Bessel function of the first kind and $i$ depends on the spin of the field. For instance, $i=2$ for  $X=\delta_\mathrm{g}$ and $Y=\kappa_\mathrm{g}$. In this case, the angular 2PCF is the so-called galaxy tangential shear $\gamma_t(\theta)$.  For $X=Y=\delta_\mathrm{g}$, we have $i=0$ and the angular 2PCF is the galaxy clustering correlation function, often denoted $w(\theta)$. For $X=Y=\kappa_\mathrm{g}$ we have $i=0/4$ and the angular 2PCF is the so-called shear $\xi_{+/-}(\theta)$  \citep[see,e.g.,][and references therein]{DES12018}. Numerically, the integral in Eq. \eqref{eq:xi} can be evaluated efficiently with FFTLog routines, for instance with the \textsc{mcfit} package. 

Analytical covariance of configuration-space statistics are also easily obtained. See eg \cite{DESI:2022kbk} for a summary of the formulas. 

We recall that spherical bessel and bessel functions are related as
\begin{equation}
    j_n(x)=\sqrt{\frac{\pi}{2x}}J_{n+\frac{1}{2}}(x)
\end{equation}

\begin{figure*}
\includegraphics[width=1.\columnwidth]{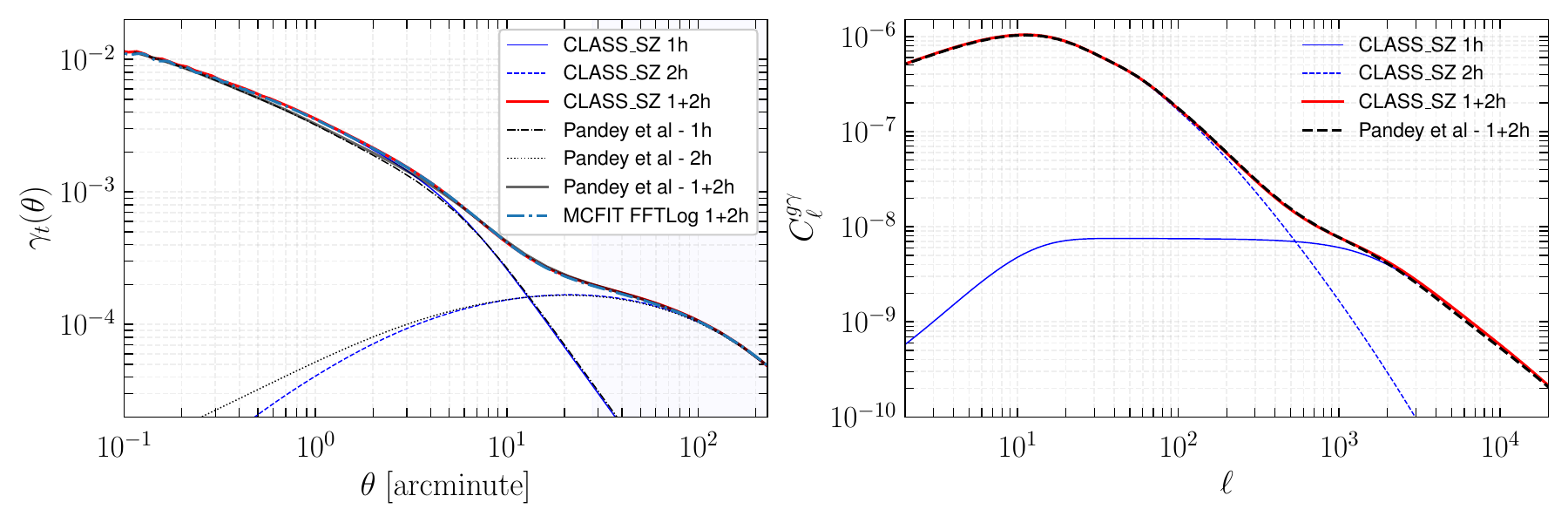}
    \centering
    \vspace{-0.2cm}
    \caption{DES benchmark. See notebook. See DES paper. This reproduces top left figure 4 of the DES paper https://arxiv.org/pdf/2106.08438.pdf modulo some diffs that can be explained by slightly different pk models/treatment. }
    \label{fig:gammat}
\end{figure*}

\subsection{Bispectra and Higher Order Statistics}

 Let $X,Y,Z$ be three LSS tracers. Their bispectrum is defined by 
\begin{equation}
    \langle X(\bm{k}_1) Y(\bm{k}_2) Z(\bm{k}_3)\rangle=(2\pi)^3\delta(\bm{k}_1+\bm{k}_2+\bm{k}_3)B(k_1,k_2,k_3).
\end{equation}
Its halo model expression is the sum of three terms,  $B^\mathrm{hm} = B^\mathrm{1h}+B^\mathrm{2h}+B^\mathrm{3h}$, associated with correlations between triplets within 1,2 and 3 halos, respectrively. The halo model terms expressions are \citep{Scoccimarro:2000gm,Valageas_2011,Lazanu_2016}:
\begin{align}
         B^\mathrm{1h} &= \langle \hat{u}_{k_1}^X\hat{u}_{k_2}^Y\hat{u}_{k_3}^Z\rangle_n +\,\mathrm{perm}(X,Y,Z)\label{eq:bispec1h}\\
     B^\mathrm{2h} &= \langle \hat{u}_{k_1}^X\hat{u}_{k_2}^Y\rangle_n\langle \hat{u}_{k_3}^Z\rangle_n P_{L}(k_3)+\langle \hat{u}_{k_3}^X\hat{u}_{k_1}^Y\rangle_n\langle \hat{u}_{k_2}^Z\rangle_n P_{L}(k_2)+\langle \hat{u}_{k_2}^X\hat{u}_{k_3}^Y\rangle_n\langle \hat{u}_{k_1}^Z\rangle_n P_{L}(k_1)+\,\mathrm{perm}(X,Y,Z)\label{eq:bispec2h}\\
     B^\mathrm{3h}&= 2\langle b^{(1)}\hat{u}_{k_1}^X\rangle_n P_{L}(k_1)\langle b^{(1)}\hat{u}_{k_2}^Y\rangle_n P_{L}(k_2)\langle b^{(1)} \hat{u}_{k_3}^Z\rangle_nF_2(k_1,k_2,k_3)\nonumber+2\mathrm{cyc}\\
     &+\langle b^{(1)}\hat{u}_{k_1}^X\rangle_n P_{L}(k_1)\langle b^{(1)}(m_2)\hat{u}_{k_2}^Y \rangle_n P_{L}(k_2)\langle b^{(2)}(m_3) \hat{u}_{k_3}^Z\rangle_n+2\mathrm{cyc}\\
     &+\mathrm{perm}(X,Y,Z)
     \label{eq:bispec3h}
\end{align}
where $F_2$ is given in Eq. \eqref{eq:f2s}
and $b^{(2)}$ is the second order halo bias (see Subsection \ref{ssec:bias}).

In a case where $X=Y=\delta_\mathrm{e}$ and where $Z$ is always evaluated at the scale $k_3$ for the 1-halo term, there is only one permutation to evaluate. For the 2-halo term there are three permutations (those of Eq.~\ref{eq:bispec2h} where $Z$ and $k_3$ are together). Similarly, for the 3-halo term there are three permutations proportional to $F_2$ and three other permutations proportional to $b^{(2)}$. In \cite{Hill:2018ypf}, the 2-halo term of the kSZ-kSZ-ISW bispectrum was computed including the nine  permutations (see their Eq.~30).

\begin{figure*}
\includegraphics[width=0.5\columnwidth]{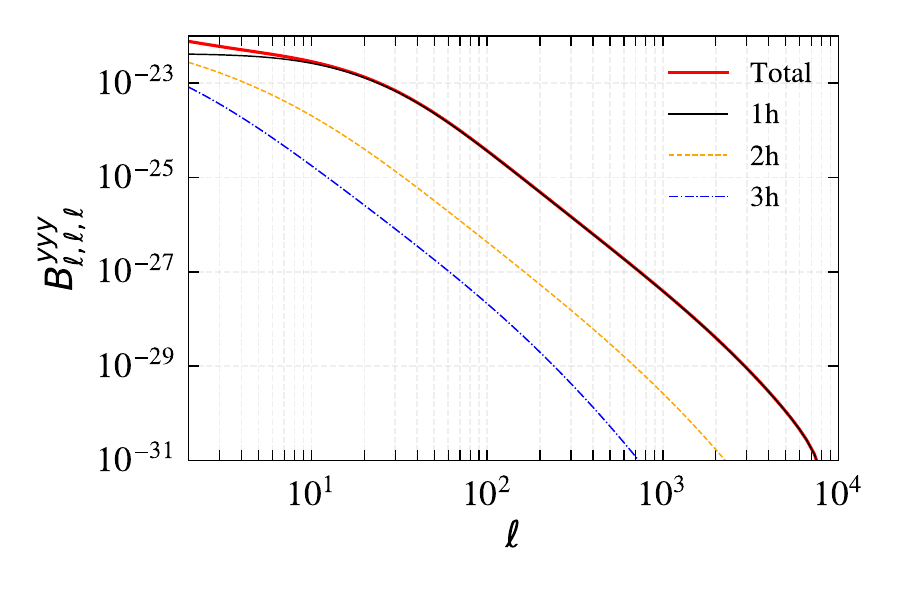}\includegraphics[width=0.5\columnwidth]{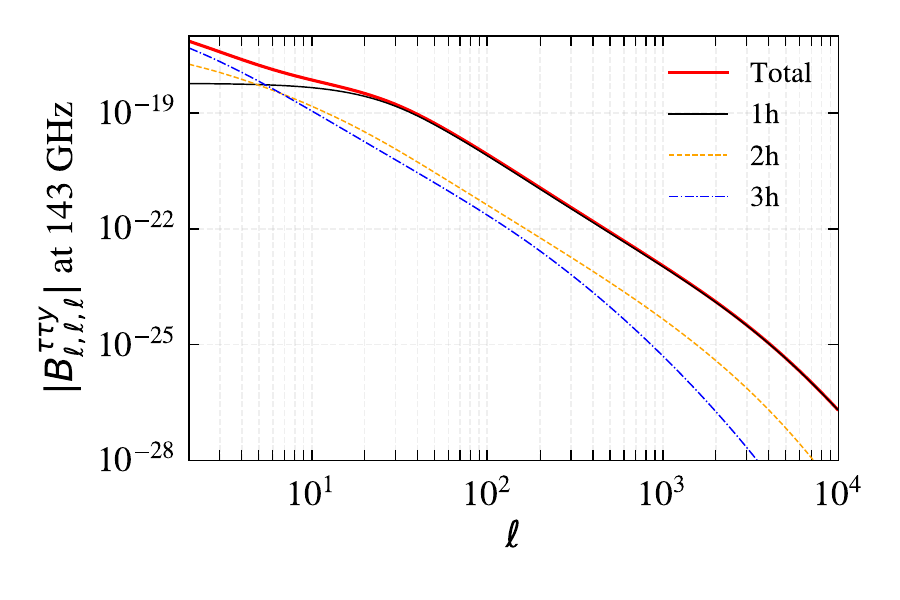}
    \centering
    \vspace{-0.2cm}
    \caption{Example of bispectra}
    \label{fig:szbispectra}
\end{figure*}

\begin{figure*}
\includegraphics[width=0.8\columnwidth]{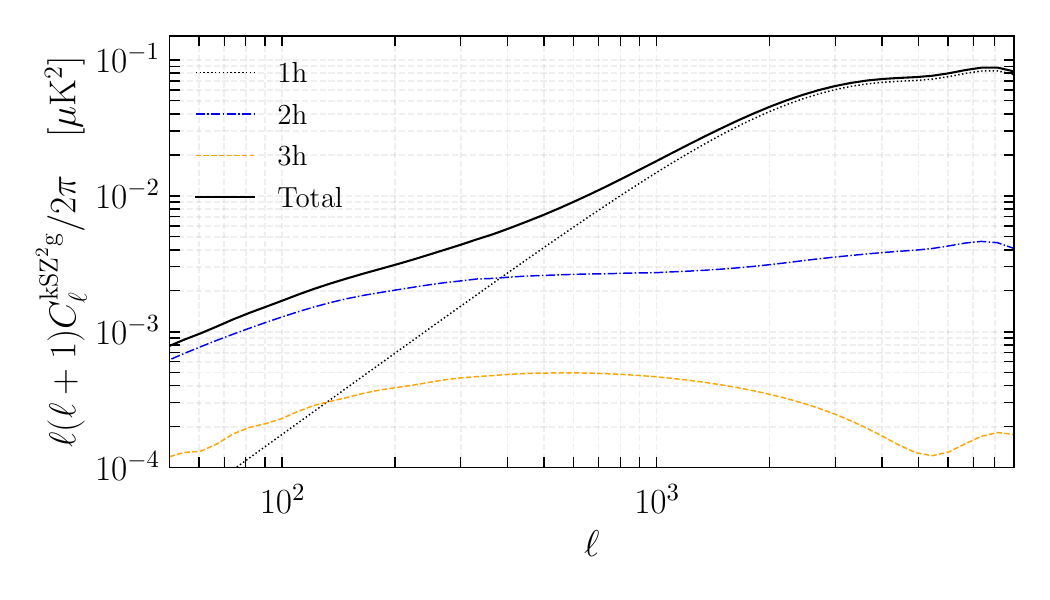}
    \centering
    \vspace{-0.2cm}
    \caption{Projected-field ksz power spectrum}
    \label{fig:ksz2g}
\end{figure*}

\subsection{Beyond Limber}\label{ssec:beyondlimber}

There are a few quantities that can be computed without Limber approximation (e.g., galaxy-galaxy spectra) within class\_sz. However, the current implementation is much slower than in ccl. As of July 2025, there is no plan to work on this.

\section{SZ Cluster Counts}\label{sec:szcounts}

We can compute cluster count likelihoods and predictions. Both binned and unbinned. Nevertheless, our implementations are superseded by \cite{Zubeldia:2024lke}.
% \bb{describe act cluster counts likelihood from hasselfield too}

% \begin{figure*}
% \includegraphics[width=0.5\columnwidth]{dndz.pdf}\includegraphics[width=0.5\columnwidth]{dndq.pdf}
%     \centering
%     \vspace{-0.2cm}
%     \caption{dndz and dndq comparison with cnc}
%     \label{fig:dndz_and_dndq}
% \end{figure*}

\begin{figure*}
\includegraphics[width=1\columnwidth]{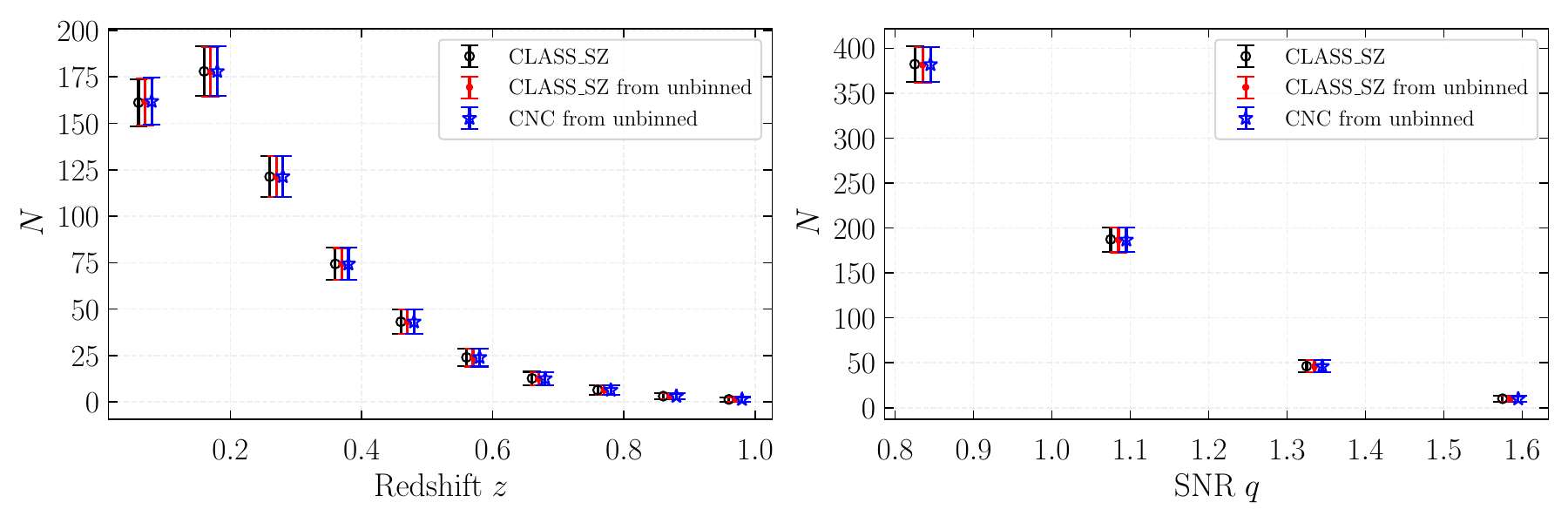}
    \centering
    \vspace{-0.2cm}
    \caption{cluster counts likelihood validation on \textit{Planck}.}
    \label{fig:dndzq_and_binned}
\end{figure*}

% \bb{Present binned and unbinned calculations}
% \bb{discussed masked ymap power spectrum calculations}

\section{Interface with Cosmological Likelihoods}\label{sec:lkls}

\subsection{Cobaya}

To run \textsc{class\_sz} using its fast mode in \texttt{cobaya} \citep{Torrado:2020dgo}, i.e., with CMB and Pk emulators, one just has to pass the following parameter in the theory block of the yaml file: use\_class\_sz\_fast\_mode=1. There are example in the class\_sz\_mcmcs/input\_files directory. If we want to run class\_sz, without the emulators for CMB and Pk, which is what we do when studying models not covered by the emulators, we just switch this parameter to 0.

\subsection{Cosmosis}
We use a wrapper file for \textsc{cosmosis} \citep{Zuntz:2014csq}, in order to run class\_sz in its fast mode. It can be found inside the classy\_szfast directory, and is named cosmosis\_classy\_szfast\_interface.py. An example parameter file is inside the class\_sz\_mcmc directory. To run it, one has to change directory and run the command \$ cosmosis $<$command$>$ from inside the cosmosis-standard-library directory.

We make the wrapper\footnote{\href{https://github.com/borisbolliet/cosmosis-standard-library/blob/dev-classy_szfast/boltzmann/class_sz/class_sz_interface.py}{https://github.com/borisbolliet/cosmosis-standard-library/blob/dev-classy\_szfast/boltzmann/class\_sz/class\_sz\_interface.py}} and a tutorial notebook\footnote{\href{https://github.com/CLASS-SZ/notebooks/blob/main/mcmcs/cosmosis/notebooks/desy3.ipynb}{https://github.com/CLASS-SZ/notebooks/blob/main/mcmcs/cosmosis/notebooks/desy3.ipynb}} available online.

\subsection{MontePython}

It is simple to use \textsc{class\_sz} with \textsc{MontePython} \citep{Audren:2012wb,Brinckmann:2018cvx}. It works in the same way with \texttt{class}.

\section{Modifying \textsc{class\_sz} and Adding Your Own Tracer}\label{sec:modify}

You can modify class\_sz to add what you need. There is a \texttt{custom\_tracer} module where you can define your own tracer and bias models.

\section{Conclusion}\label{sec:conclusion}

class\_sz continues to serve for benchmarking of many codes in cosmology and for parameter inference in cross-correlation analyses. Its wrapper for emulators has been used in many papers, including for parameter constraints based on ACT DR6 data releases. 

Looking ahead, the code will not be actively maintained. Instead, we are working towards integrating all class\_sz calculations in upcoming Jax packages and agentic frameworks such as \texttt{cmbagent}\footnote{\href{https://github.com/CMBAgents/cmbagent}{https://github.com/CMBAgents/cmbagent}}.

% \begin{equation}
%     M_{\ell\ell^\prime}=\frac{1}{(2\ell+1)f_\mathrm{sky}}\{C_\ell^{g^2g^2}(C_\ell^{\kappa\kappa}+N_\ell^{\kappa\kappa})+(C_\ell^{\mathrm{g}^2 \kappa})^2\}\delta_{\ell\ell^\prime},
% \end{equation}
%  where 
% \begin{equation}
% C_\ell^{g^2g^2} = \frac{1}{2\pi^2}\int\mathrm{d}^2\bm{\ell}^\prime C_{\ell^\prime}^{gg}C_{|\bm{\ell}-\bm{\ell}^\prime|}^{gg }.\label{eq:cltftf}
% \end{equation}

% \bb{summary of what we did.}
% Compared to \verb|CosmicNet| \citep{Gunther:2022pto},
% our emulators extend to much higher multipoles and achieve much faster evaluation. \asm{and we are fully differentiable -- however I don't think we should compare against other emulators, this was done already in the original CP paper}

\section*{Acknowledgements}
We thank David Alonso, Eiichiro Komatsu and Mathew Madhavacheril. JCH acknowledges support from NSF grant AST-2108536, NASA grant 21-ATP21-0129, DOE grant DE-SC00233966, the Sloan Foundation, and the Simons Foundation.  BS and BB acknowledges  support from the European Research Council (ERC) under the European Union’s Horizon 2020 research and innovation programme (Grant agreement No. 851274). AK and JCH acknowledge support from NSF grant AST-2108536. This material is based upon work supported by the National Science Foundation Graduate Research Fellowship Program under Grant No.~DGE 2036197 (KMS). 

%%%%%%%%%%%%%%%%%%%%%%%%%%%%%%%%%%%%%%%%%%%%%%%%%%
\section*{Data Availability}

\textsc{class\_sz} is public.



% \bibliographystyle{mnras}
\bibliographystyle{aasjournal}
\bibliography{example} % if your bibtex file is called example.bib

%\appendix

%%%%%%%%%%%%%%%%%%%%%%%%%%%%%%%%%%%%%%%%%%%%%%%%%%

% Don't change these lines
% \bsp	% typesetting comment
% \label{lastpage}
\end{document}